%% file: StochaAttractor.tex
\documentclass[a4paper,11pt]{article}
\pdfoutput=1 % to allow pdflatex compilation in JCAP
\usepackage{jcappub}
\usepackage{amsmath}
\usepackage{graphicx}
\usepackage{latexsym}
\usepackage{xspace}
\usepackage{color}
\usepackage{hyperref} 
\usepackage{bm}
\usepackage{relsize}
\usepackage{tabularx}
\usepackage{multirow}
\usepackage{amssymb}
\usepackage[table]{xcolor}
\usepackage{slashbox}
\usepackage{braket}
\usepackage{comment}
\usepackage[utf8]{inputenc}
\makeatletter
\gdef\@fpheader{}
\g@addto@macro\bfseries{\boldmath}
\makeatother

\input{newcommands}

\subheader{}

\title{Stochastic inflation in phase space:\\ Is slow roll a stochastic attractor?}

\author[a]{Julien Grain,}
\affiliation[a]{Institut d'Astrophysique Spatiale, UMR8617, CNRS, Univ. Paris Sud, Universit\'e Paris-Saclay, Bt. 121, Orsay, France, F-91405}
\emailAdd{julien.grain@ias.u-psud.fr}

\author[b]{Vincent Vennin}
\affiliation[b]{Institute of Cosmology \& Gravitation, University of Portsmouth, Dennis Sciama Building, Burnaby Road, Portsmouth, PO13FX, United Kingdom}
\emailAdd{vincent.vennin@port.ac.uk}

\date{today}

\begin{document}
\sloppy

\abstract{An appealing feature of inflationary cosmology is the presence of a phase-space attractor, ``slow roll'', which washes out the dependence on initial field velocities. We investigate the robustness of this property under backreaction from quantum fluctuations using the stochastic inflation formalism in the phase-space approach. A Hamiltonian formulation of stochastic inflation is presented, where it is shown that the coarse-graining procedure -- where wavelengths smaller than the Hubble radius are integrated out -- preserves the canonical structure of free fields. This means that different sets of canonical variables give rise to the same probability distribution which clarifies the literature with respect to this issue. The role played by the quantum-to-classical transition is also analysed and is shown to constrain the coarse-graining scale. In the case of free fields, we find that quantum diffusion is aligned in phase space with the slow-roll direction. This implies that the classical slow-roll attractor is immune to stochastic effects and thus generalises to a stochastic attractor regardless of initial conditions, with a relaxation time at least as short as in the classical system. For non-test fields or for test fields with non-linear self interactions however, quantum diffusion and the classical slow-roll flow are misaligned. We derive a condition on the coarse-graining scale so that observational corrections from this misalignment are negligible at leading order in slow roll.}


%\arxivnumber{16XX.XXXXX}

\maketitle

\flushbottom

\section{Introduction}
\label{sec:intro}
Inflation~\cite{Starobinsky:1980te, Sato:1980yn, Guth:1980zm, Linde:1981mu, Albrecht:1982wi, Linde:1983gd} is the leading paradigm to describe the physical conditions that prevail in the very early Universe. During this accelerated expansion epoch, vacuum quantum fluctuations of the gravitational and matter fields are amplified to cosmological perturbations~\cite{Starobinsky:1979ty, Mukhanov:1981xt, Hawking:1982cz,  Starobinsky:1982ee, Guth:1982ec, Bardeen:1983qw}, that later seed the cosmic microwave background (CMB) anisotropies and the large-scale structure of our Universe.

An appealing feature of inflation is that when implemented with one or several scalar fields, the dependence on initial conditions in phase space is usually washed out by the existence of a dynamical attractor~\cite{Clesse:2009ur, Remmen:2013eja} dubbed ``slow roll''~\cite{Mukhanov:1985rz, Mukhanov:1988jd, Stewart:1993bc, Gong:2001he, Schwarz:2001vv, Leach:2002ar} and characterised by small deviations from de-Sitter space time. Not only have these deviations been confirmed to be small by cosmological observations, but the existence of an attractor also makes inflation more predictive since less dependent on initial field velocities. It is therefore an important feature of this theory.

Slow-roll attractors have been shown to exist for classical homogeneous scalar fields during inflation. However, as inflation proceeds, small wavelength perturbations are excited out of the vacuum quantum fluctuations and stretched out to scales larger than the Hubble radius, backreacting on the background evolution. This effect is usually constrained to be small inside the observational window, but the erasure of initial field velocities has to take place before the scales probed in cosmological experiments cross the Hubble radius, at a time when it can a priori be larger. Besides, even within the observational window, whether backreaction from quantum fluctuations alters slow-roll attractors is an important question since the very high accuracy of the data now calls for high precision calculations where all corrective effects must be incorporated or checked to be negligible indeed. This is why in this paper, we study how the classical slow-roll attractors behave in the presence of backreaction from quantum fluctuations, modelled through the formalism of stochastic inflation.

This paper is organised as follows. In the rest of this section, we explain why slow roll is an attractor of the background dynamics of scalar fields during inflation and how backreaction from quantum fluctuations can be incorporated in the stochastic inflation formalism. In \Sec{sec:HamiltonianStochastic}, we then present a Hamiltonian formulation of stochastic inflation in order to describe the stochastic dynamics in the full phase space. The case of a free field is studied in \Sec{sec:testField}, since the linearity of the equations allow us to solve exactly the phase-space stochastic dynamics in this case. The quantum-to-classical transition is analysed, and we show that the coarse-graining procedure, where wavelengths smaller than the Hubble radius are integrated out, preserves the canonical structure, clarifying the literature with respect to this issue. In \Sec{sec:SRattractor}, it is shown that the classical slow-roll attractor is immune to stochastic effects in this same case, due to the fact that field perturbations follow the same equations of motion as the classical background on large scales. For non-free fields however, this stops being the case and quantum diffusion occurs in a different direction from classical slow roll. The observational consequences of this misalignment are discussed in \Sec{sec:BeyondFreeFields}. We present our main conclusions in \Sec{sec:concl}, and we end the paper with a few appendices containing various technical aspects.
\subsection{Slow-roll inflation and phase-space attractors}
\label{sec:SR}
In the slow-roll regime of inflation, the Hubble factor $H=\dot{a}/a$, where $a$ is the scale factor of the Friedmann-Lema\^itre-Robertson-Walker metric and a dot denotes a derivation with respect to cosmic time $t$, is almost a constant. This can be quantified by requiring that the Hubble-flow parameters $\epsilon_n$, defined by the flow equations \cite{Schwarz:2001vv,Schwarz:2004tz} 
\beq
\label{eq:epsFlow}
\epsilon_{n+1}=\frac{\dd\ln\vert\epsilon_n\vert}{\dd N_e} ,
\eeq
are parametrically small. In \Eq{eq:epsFlow}, the hierarchy is started at $\epsilon_0\equiv H_\uin/H$, and $N_e\equiv \ln a$ is the number of $e$-folds. With this definition, all the $\epsilon_n$ are typically of the same order of magnitude. The slow-roll regime of inflation is characterised by $\vert\epsilon_n\vert\ll 1$, for all $n>0$, while since $\epsilon_1=-\dot{H}/H^2=1-\ddot{a}/(aH^2)$ (where a dot denotes differentiation with respect to cosmic time), inflation ($\ddot{a}>0$) takes place provided $\epsilon_1<1$. 

Let us first see why the assumption that $\vert\epsilon_n\vert\ll 1$ removes the dependence on initial conditions in phase space before showing that such a regime is indeed an attractor of inflationary dynamics. If inflation is driven by a single scalar field $\phi$ with potential $V(\phi)$, its dynamics is determined by the Friedmann equation
\bea
\label{eq:Friedman}
H^2=\frac{V(\phi)+\dot{\phi}^2/2}{3\Mp^2} ,
\eea
which relates the expansion rate $H$ to the energy density contained in the inflaton field $\phi$ and the reduced Planck mass $\Mp$, and the Klein-Gordon equation
\bea
\label{eq:KG}
\ddot{\phi}+3H\dot{\phi}+V_{,\phi}=0 ,
\eea
which is the equation of motion for $\phi$. Hereafter, a subscript ``$,\phi$'' denotes derivative with respect to $\phi$. Inserting the Klein-Gordon equation~(\ref{eq:KG}) in the time derivative of the Friedman equation~(\ref{eq:Friedman}), one obtains $\dot{H}=-\dot{\phi}^2/(2\Mp^2)$, hence 
\bea
\label{eq:eps1exact}
\epsilon_1=-\frac{\dot{H}}{H^2}=3\frac{\dot{\phi}^2/2}{V(\phi)+\dot{\phi}^2/2} .
\eea
The condition $\epsilon_1\ll 1$ thus implies that the kinetic energy of the inflaton is much smaller than its potential energy, namely $\dot{\phi}^2/2\ll V(\phi)$. Under this condition, the Friedmann equation~(\ref{eq:Friedman}) simplifies and gives, at leading order in slow roll, $H^2\simeq V/(3\Mp^2)$. Moving on to $\epsilon_2$, one can insert the Klein-Gordon equation~(\ref{eq:KG}) in the time derivative of the relation $\dot{H}=-\dot{\phi}^2/(2\Mp^2)$ previously obtained, and get $\ddot{H}=3H\dot{\phi}^2/\Mp^2+\dot{\phi}V_{,\phi}/\Mp^2$, hence
\bea
\label{eq:eps2exact}
\epsilon_2=\frac{\ddot{H}}{H\dot{H}}-2\frac{\dot{H}}{H^2}=
6\left(\frac{\epsilon_1}{3}-\frac{V_{,\phi}}{3H\dot{\phi}}-1\right) .
\eea
The condition $\epsilon_2\ll 1$ thus implies that, at leading order in slow roll, $\dot{\phi}\simeq-V_{,\phi}/(3H)$, which means that the acceleration term $\ddot{\phi}$ can be neglected in the Klein-Gordon equation~(\ref{eq:KG}). This lowers by $1$ the order of the differential equation satisfied by $\phi$, and as a consequence, removes the dependence on the initial conditions by singling out a specific trajectory in phase space.
\begin{figure}[t]
\begin{center}
\includegraphics[width=0.49\textwidth]{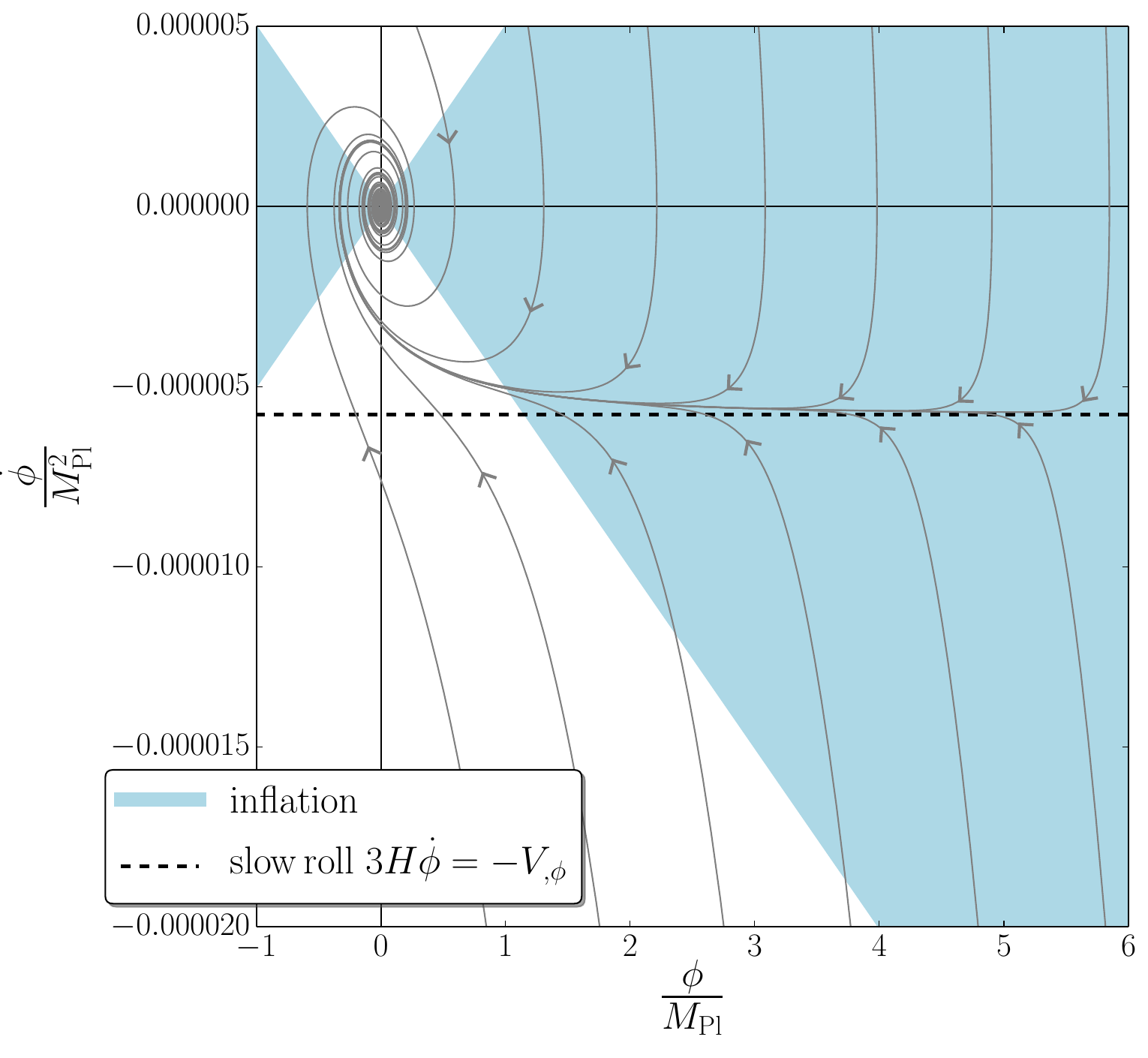}
\includegraphics[width=0.49\textwidth]{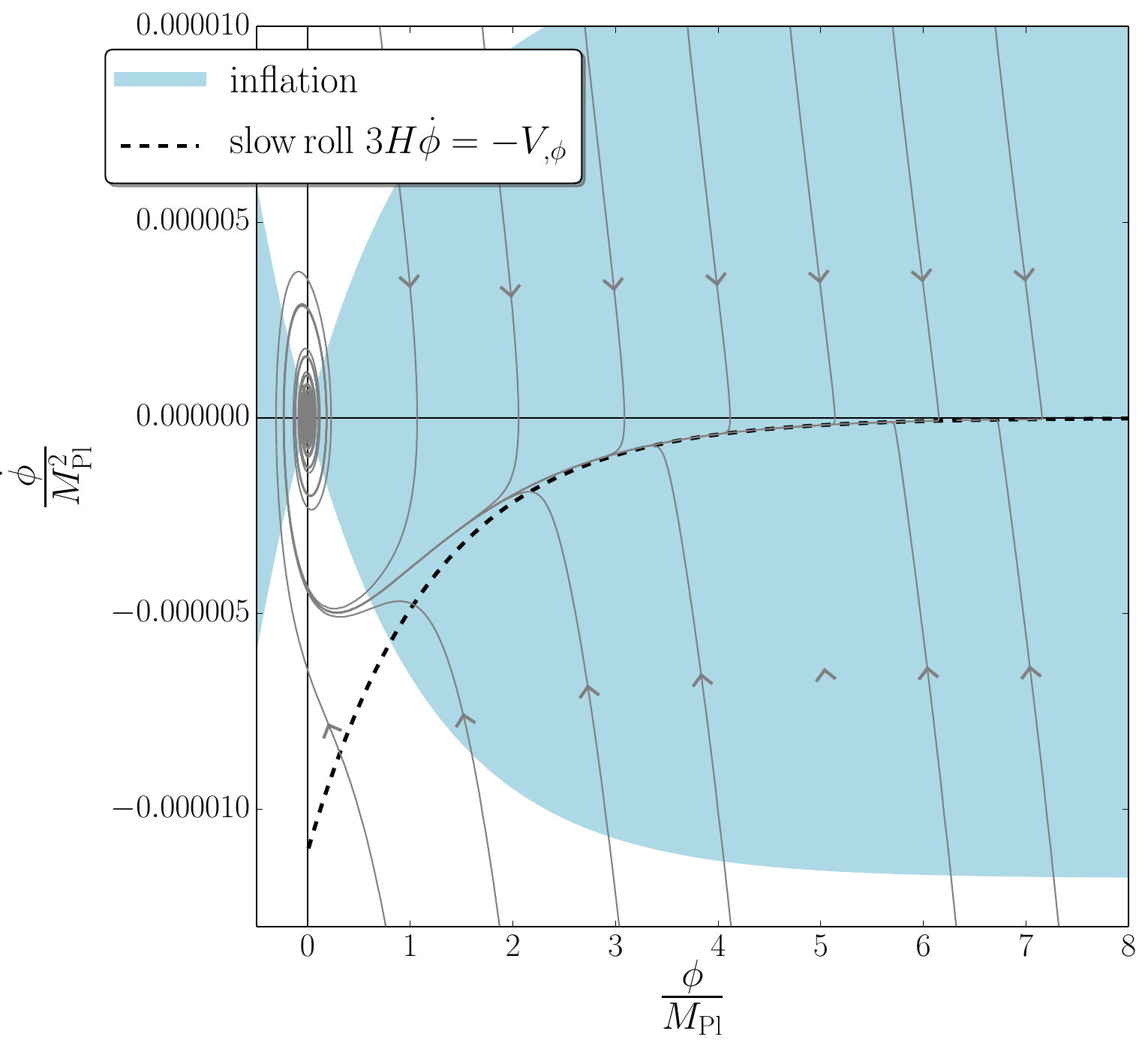}
\caption{Solutions of the the dynamical system~(\ref{eq:Friedman})-(\ref{eq:KG}) displayed in the phase space $(\phi,\dot{\phi})$. The blue shaded area is where $\epsilon_1<1$, with $\epsilon_1$ given in \Eq{eq:eps1exact}, and corresponds to the region where inflation proceeds. The black dashed line stands for the slow-roll solution $3H\dot{\phi}=-V_{,\phi}$ (at leading order in slow roll), which is an attractor of the phase-space dynamics. In the left panel, the inflaton potential is $V(\phi)=m^2\phi^2/2$, where $m=7\times 10^{-6}$ in order to match the scalar power spectrum amplitude, and the right panel corresponds to the $f(R)\propto R+R^2$ Starobinsky model \cite{Starobinsky:1980te} $V(\phi)=M^4 (1-\ee^{-\sqrt{2/3}\phi/\Mp} )^2$, where $M=0.0034\Mp$ to match the power spectrum amplitude as well.}
\label{fig:classical_phase_space}
\end{center}
\end{figure}

Let us now show that this regime is indeed an attractor of the inflationary dynamics. In \Fig{fig:classical_phase_space}, the phase portrait of the dynamical system~(\ref{eq:Friedman})-(\ref{eq:KG}) is displayed for two potentials, $V(\phi)=m^2\phi^2/2$ in the left panel and the $f(R)\propto R+R^2$ Starobinsky model \cite{Starobinsky:1980te} $V(\phi)=M^4 (1-\ee^{-\sqrt{2/3}\phi/\Mp} )^2$ in the right panel. The blue regions correspond to where inflation proceeds and the black dashed line stands for the (leading order) slow-roll solution $3H\dot{\phi} \simeq -V_{,\phi}(\phi)$. One can see that it corresponds indeed to an attractor of the phase-space dynamics, quickly attained from an extended basin of possible initial conditions. More precisely, if initial conditions are such that the kinetic term initially dominates the energy budget of the inflaton field, and that $\vert\dot{\phi}_\uin\vert\gg \vert V_{,\phi}(\phi_\uin)\vert/(3H)$, the Klein-Gordon equation~(\ref{eq:KG}) in this ``fast-roll'' limit implies that $\vert \dot{\phi} \vert\propto\ee^{-3N_e}$. Remembering that inflation starts when $\dot{\phi}^2<V$, this means that the velocity of the inflaton is damped to the slow-roll one within a few $\ee$-folds at most, of the order of $\ln\vert 3H\dot{\phi}_\uin/V_{,\phi}\vert /3$. Let us also note that the leading-order slow-roll trajectory, $\dot{\phi}\simeq-V_{,\phi}/(3H)$, allows one to approximate the slow-roll parameters by expressions that depend on the potential and its derivatives only
\bea
\label{eq:eps1:appr}
\epsilon_1&\simeq&\frac{\Mp^2}{2}\left(\frac{V_{,\phi}}{V}\right)^2 ,\\
\epsilon_2&\simeq& 2\Mp^2  \left[\left(\frac{V_{,\phi}}{V}\right)^2 - \frac{V_{,\phi\phi}}{V} \right] ,
\label{eq:eps2:appr}
\eea
which translate the slow-roll condition $\vert \epsilon_n \vert \ll 1$ into conditions on the potential $V$.

The notion of slow-roll dynamics can in fact be extended to all scalar fields during inflation. For a scalar field $\phi$ with potential $V(\phi)$, that may or may not substantially contribute to the total energy budget of the Universe, the fractional energy density contained in the kinetic term can still be quantified by a first ``slow-roll'' parameter
\bea
\label{eq:eps1:testfield}
\epsilon_{1}^{\phi}= 3\frac{\dot{\phi}^2/2}{V(\phi)+\dot{\phi}^2/2} ,
\eea
even if there is no relationship between $\epsilon_{1}^{\phi}$ and $H$ as in \Eq{eq:eps1exact}. By using the Klein-Gordon equation~(\ref{eq:KG}) only, that still holds for test fields (contrary to the Friedmann equation), the second slow-roll parameter $\epsilon_{2}^{\phi}=\dd\ln\epsilon_{1}^{\phi}/\dd N_e$ is given by
\bea
\label{eq:eps2:testfield}
\epsilon_{2}^{\phi}=6\left(\frac{\epsilon_{1}^{\phi}}{3}-\frac{V_{,\phi}}{3H\dot{\phi}}-1\right) ,
\eea  
even if, here again, there is no relationship between $\epsilon_{2}^{\phi}$ and $H$ as in \Eq{eq:eps2exact}. Similarly, the full hierarchy of slow-roll parameters $\epsilon_{n+1}^{\phi}=\dd\ln\vert \epsilon_{n}^{\phi} \vert/\dd N_e$ can be constructed, and if one defines ``slow roll'' for a generic scalar field as being the regime where $\vert \epsilon_n^\phi \vert \ll 1$ for all $n>0$, \Eq{eq:eps1:testfield} implies that the kinetic energy of a generic slowly rolling field is negligible compared to its potential energy, and \Eq{eq:eps2:testfield} means that its phase-space trajectory is $\dot{\phi}\simeq -V_{,\phi}/(3H)$ at leading order. Making use of this phase-space trajectory allows one to derive approximate expressions for the slow-roll parameters in terms of the potential $V$,
\bea
\label{eq:eps1:appr:test}
\epsilon_1^\phi&\simeq&\frac{V_{,\phi}^2}{6H^2V} ,\\
\epsilon_2^\phi&\simeq& 2\epsilon_{1}+\frac{V_{,\phi}^2}{3H^2V}-\frac{2V_{,\phi\phi}}{3H^2} .
\label{eq:eps2:appr:test}
\eea
In the case where $\phi$ is the inflaton, by plugging the Friedmann equation~(\ref{eq:Friedman}) into \Eqs{eq:eps1:appr:test} and~(\ref{eq:eps2:appr:test}), one recovers \Eqs{eq:eps1:appr} and~(\ref{eq:eps2:appr}), but the expressions derived here are more generic. Again, they translate the slow-roll conditions into conditions on the potential.

The slow-roll regime can also be shown to provide an attractor of test scalar fields during inflation. For instance, let us consider the case of a light scalar field $\phi$ with potential $V(\phi)=m^2\phi^2/2$ in a de-Sitter background with $H\gg m$. The Klein-Gordon equation~(\ref{eq:KG}) is in this case linear and has two independent solutions given by $\phi \propto \exp\{-3/2[1\pm\sqrt{1-4m^2/(9H^2)}]Ht\}$. The solution with the minus sign is denoted ``slow roll'' ($\mathrm{SR}$) and the solution with the plus sign ``non slow roll'' ($\mathrm{NSR}$) for the following reason. Along the SR solution, one has
\bea
\label{eq:phidotdotSR}
\frac{\ddot{\phi}_\sr}{3H\dot{\phi}_\sr}= -\frac{1}{2}\left(1-\sqrt{1-\frac{4}{9}\frac{m^2}{H^2}}\right)\simeq  -\frac{m^2}{9H^2} ,
\eea
where the second equality holds for $m\ll H$. In this limit, the acceleration term $\ddot{\phi}$ in the Klein-Gordon equation~(\ref{eq:KG}) is suppressed by $m^2/H^2$ compared to the two other ones. In fact, one can show that successive time differentiations obey the relation $\dd^n\phi_\sr/\dd N_e^n\simeq -m^2/(3H^2) \dd^{n-1}\phi_\sr/\dd N_e^{n-1}$, so that the time derivatives of $\phi_\sr$ are suppressed by higher and higher powers of $m^2/H^2$ and the field is indeed ``slowly'' rolling. Along the NSR solution on the other hand, one has
\bea
\label{eq:phidotdotNSR}
\frac{\ddot{\phi}_\nsr}{3H\dot{\phi}_\nsr} = -\frac{1}{2}\left(1+\sqrt{1-\frac{4}{9}\frac{m^2}{H^2}}\right)\simeq  - 1 ,
\eea
where the second equality corresponds again to $m\ll H$. Along this branch, the potential gradient term $V_{,\phi}$ is the one being suppressed by $m^2/H^2$ compared to the two other ones in the Klein-Gordon equation~(\ref{eq:KG}). For this reason, such a solution is sometimes referred to as ``ultra slow roll''~\cite{Kinney:2005vj, Martin:2012pe}. It satisfies $\dd^n\phi_\nsr/\dd N_e^n\simeq - 3 \dd^{n-1}\phi_\nsr/\dd N_e^{n-1}$, so that all time derivatives are of the same order of magnitude and the field is not ``slowly'' rolling. At late time, the SR branch of the solution dominates over the NSR branch, hence the SR solution is a dynamical attractor of the system.
\subsection{Stochastic inflation}
\label{sec:Stochastic_Inflation}
Let us now introduce perturbations about the homogeneous and isotropic expanding background. During inflation, scalar field perturbations are placed in two-mode squeezed states~\cite{Grishchuk:1990bj,Grishchuk:1992tw}, which undergo quantum-to-classical transitions~\cite{Polarski:1995jg,Lesgourgues:1996jc,Kiefer:2008ku,Martin:2012pea} in the sense that on super-Hubble scales, the non-commutative parts of the fields become small compared to their anti-commutative parts (see \Sec{sec:quantum_to_classical}). It gives rise to the stochastic inflation formalism~\cite{Starobinsky:1982ee, Starobinsky:1986fx, Nambu:1987ef, Nambu:1988je, Kandrup:1988sc, Nakao:1988yi, Nambu:1989uf, Mollerach:1990zf, Linde:1993xx, Starobinsky:1994bd, Finelli:2008zg, Finelli:2010sh}, consisting of an effective theory for the long-wavelength parts of the quantum fields, which are ``coarse grained'' at a fixed physical scale (\ie non-expanding), larger than the Hubble radius during the whole inflationary period. In this framework, the small wavelength fluctuations behave as a classical noise acting on the dynamics of the super-Hubble scales as they cross the coarse-graining scale, and the coarse-grained fields can thus be described by a stochastic classical theory, following Langevin equations that will be explicitly derived in \Sec{ssec:cosmo}.

The stochastic formalism accounts for the quantum modification of the super-Hubble scales dynamics, and allows us to study how quantum effects modify inflationary observable predictions~\cite{Vennin:2015hra,Kawasaki:2015ppx,Assadullahi:2016gkk,Vennin:2016wnk}. In the present work, we investigate whether stochastic effects induce a deviation of the background dynamics from the slow-roll attractor.
\section{Hamiltonian formulation of stochastic inflation}
\label{sec:HamiltonianStochastic}
The investigation of quantum diffusion in phase space first requires to derive a Hamiltonian formulation of stochastic inflation. 
%In this section, we present such a derivation and highlight how the quantum-to-classical transition occurs in phase space.
%
\subsection{Cosmology in the Hamiltonian framework}
\label{ssec:cosmo}
Let us first review the Hamiltonian framework for studying the dynamics of a scalar field $\phi$ minimally coupled to gravity in a 4-dimensional curved space-time with metric $g_{\mu\nu}$, described by the action
\bea
	S=\displaystyle\int \dd^4x\sqrt{-g}\left[\frac{\Mp^2}{2}R-\frac{1}{2}g^{\mu\nu}\partial_\mu\phi\partial_\nu\phi-V(\phi)\right] .
	\label{eq:action}
\eea
In this expression, $R$ is the scalar curvature and $V(\phi)$ is the potential of the scalar field. The Hamiltonian formulation is obtained through the ADM formalism~\cite{thieman_book,Langlois:1994ec}, which provides a foliation of 4-dimensional space-times into a set of 3-dimensional space-like hypersurfaces. The foliation is determined by the lapse function $N(\tau,x^i)$ and the shift vector $N^i(\tau,x^i)$, which enter the covariant line element as
\bea
\label{eq:line_element}
	\dd s^2=-N^2\dd \tau^2+\gamma_{ij}\left(N^i\dd \tau+\dd x^i\right)\left(N^j\dd \tau+\dd x^j\right),
\eea
where $\gamma_{ij}(t,x^i)$ is the induced metric on the 3-dimensional space-like hypersurfaces. 

The canonical variables for the gravitational sector are $\gamma_{ij}$ and $\pi^{ij}=\delta S/\delta\dot{\gamma}_{ij}$, where a dot means a derivation with respect to the time variable $\tau$. Their associated Poisson bracket is $\left\{\gamma_{ij}(\vec{x}),\pi^{kl}(\vec{y})\right\}= (\delta^{k}_i\delta^{l}_j+\delta^{l}_i\delta^{k}_j ) \delta^3(\vec{x}-\vec{y})/2$. Similarly, for the scalar field, the canonical variables are $\phi$ and $\pi_\phi=\delta S/\delta\dot{\phi}$, and their Poisson bracket reads $\left\{\phi(\vec{x}),\pi_\phi(\vec{y})\right\}=\delta^3(\vec{x}-\vec{y})$. The dynamics is thus described by the total Hamiltonian
\bea
\label{eq:total_Hamiltonian}
	C=\displaystyle\int \dd^3x\left[N\left(\mathcal{C}_G+\mathcal{C}_\phi\right)+N^i\left(\mathcal{C}^G_i+\mathcal{C}^\phi_i\right)\right].
\eea
In this expression, $G$ and $\phi$ stand for the gravitational and the scalar field sectors respectively, $\mathcal{C}=\mathcal{C}_G+\mathcal{C}_\phi$ is the scalar constraint and $\mathcal{C}_i=\mathcal{C}^G_i+\mathcal{C}^\phi_i$ is the spatial-diffeomorphism constraint. For the scalar field, they read
\bea
\label{eq:Cphi}
	\mathcal{C}_\phi&=&\frac{1}{2\sqrt{\gamma}}\pi^2_\phi+\frac{\sqrt{\gamma}}{2}\gamma^{ij}\partial_i\phi\partial_j\phi+\sqrt{\gamma}V(\phi) , \\
	\mathcal{C}^\phi_i&=&\pi_\phi\partial_i\phi ,
\label{eq:Ciphi}
\eea
where $\gamma$ denotes the determinant of $\gamma_{ij}$, and similar expressions can be found for the gravitational sector~\cite{Langlois:1994ec}. Any function $F$ of the phase-space variables then evolves under the Hamilton equations
\bea
\label{eq:Hamilton:constraint}
	\dot{F}(\gamma_{ij},\pi^{kl},\phi,\pi_\phi)=\left\{F,C\right\}.
\eea
Finally, variations of the action with respect to the lapse function and the shift vector show that both $\mathcal{C}$ and $\mathcal{C}_i$ are constrained to be zero.

Let us now study the case of a test scalar field, for which the gravitational part of the Hamiltonian does not depend on $\phi$ and $\pi_\phi$~\cite{Langlois:1994ec}. The Hamilton equations  $\dot\phi=\left\{\phi,C\right\}$ and $\dot\pi_\phi=\left\{\pi_\phi,C\right\}$ give rise to
\bea
	\dot\phi(\vec{x})&=&\displaystyle\int \dd^3y \left[N(\vec{y})\left\{\phi(\vec{x}),\mathcal{C}_\phi(\vec{y})\right\}+N^i(\vec{y})\left\{\phi(\vec{x}),\mathcal{C}^\phi_i(\vec{y})\right\}\right] , \\
	\dot\pi_\phi(\vec{x})&=&\displaystyle\int \dd^3y \left[N(\vec{y})\left\{\pi_\phi(\vec{x}),\mathcal{C}_\phi(\vec{y})\right\}+N^i(\vec{y})\left\{\pi_\phi(\vec{x}),\mathcal{C}^\phi_i(\vec{y})\right\}\right] ,
\eea	
where the time-dependence is made implicit for display convenience. Making use of \Eqs{eq:Cphi} and~(\ref{eq:Ciphi}), one obtains the local evolution equations
\bea
\label{eq:dotphigen}
	\dot\phi&=&\frac{N}{\sqrt{\gamma}}\pi_\phi+N^i\partial_i\phi ,  \\
	\dot\pi_\phi&=&-N\sqrt{\gamma}V_{,\phi}+\partial_i\left(N\sqrt{\gamma}\gamma^{ij}\partial_j\phi\right)
+N^i\partial_i\pi_\phi	
	+\pi_\phi\partial_iN^i ,
\label{eq:dotpgen}
\eea
where the space dependence is made implicit for clarity and where the last three terms in \Eq{eq:dotpgen} are obtained by integration by parts.

For simplicity, let us assume that $\phi$ is a test field sufficiently decoupled from the metric and other fields perturbations that the latter can be ignored~\cite{Langlois:1994ec, Lyth:2001nq} (this assumption will be relaxed in \Sec{sec:BeyondFreeFields}). One can drop out perturbations from the line element~(\ref{eq:line_element}), and in spatially flat universes, it is given by
\bea
\label{eq:line_element:flat}
	\dd s^2=-N^2(\tau)\dd\tau^2+p(\tau)\delta_{ij}\dd x^i\dd x^j ,
\eea
where the lapse function $N$ and $p\equiv a^2$ depend on time only. In this expression, the shift vector is zero and choosing a lapse function simply means choosing a time variable. For example, $N=1$ corresponds to working with cosmic time,  $N=a$ with conformal time, and $N=1/H$  with the number of $e$-folds. In absence of a shift vector, since \Eq{eq:line_element:flat} also gives rise to $\gamma_{ij}=p\delta_{ij}$,  \Eqs{eq:dotphigen} and~(\ref{eq:dotpgen}) simplify to
\bea
\label{eq:dotphiflrw}
	\dot\phi&=&\frac{N}{p^{3/2}}\pi_\phi ,  \\
	\dot\pi_\phi&=&-Np^{3/2}V_{,\phi}+Np^{1/2}\Delta\phi , 
\label{eq:dotpflrw}
\eea
where $\Delta\equiv \delta^{ij}\partial_i\partial_j$ is the 3-dimensional flat Laplace operator. 
\subsection{Langevin equation in phase space}
\label{sec:Langevin}
The strategy of the stochastic inflation formalism consists in deriving an effective theory for the long wavelength part of the scalar field $\phi$ by integrating out the small wavelengths. This requires to introduce a time-dependent cut-off in Fourier space
\bea
\label{eq:ksigma}
k_\sigma=\sigma aH ,
\eea
where $\sigma$ is the ratio between the Hubble radius and the cut-off wavelength. It disappears from all physical quantities in the limit $\sigma\ll 1$ under conditions that will be carefully discussed in \Secs{sec:quantum_to_classical} and~\ref{sec:BeyondFreeFields}. As explained in \Sec{sec:Stochastic_Inflation}, on super-Hubble scales, the quantum state of the field is such that it can be treated as a stochastic classical process. In practice, the dynamics of the long-wavelength part of the field can be described by a Langevin equation that we now derive, in which the small-wavelength part of the field provides the noise term as modes continuously cross $k_\sigma$.

In the Hamiltonian formalism, the coarse-graining is performed in phase space through the decomposition $\phi=\bar\phi+\phi_Q$ and $\pi_\phi=\bar\pi+\pi_Q$, where
\bea
\label{eq:phiq}
	\phi_Q&=&\displaystyle\int_{\mathbb{R}^3}\frac{\dd^3k}{(2\pi)^{3/2}}W\left(\frac{k}{k_\sigma}\right)\left[a_{\vec{k}}~\phi_{\vec{k}}(\tau)e^{-i\vec{k}\cdot\vec{x}}+a^\dag_{\vec{k}}~\phi^\star_{\vec{k}}(\tau)e^{i\vec{k}\cdot\vec{x}}\right] \\
	\pi_Q&=&\displaystyle\int_{\mathbb{R}^3}\frac{\dd^3k}{(2\pi)^{3/2}}W\left(\frac{k}{k_\sigma}\right)\left[a_{\vec{k}}~\pi_{\vec{k}}(\tau)e^{-i\vec{k}\cdot\vec{x}}+a^\dag_{\vec{k}}~\pi^\star_{\vec{k}}(\tau)e^{i\vec{k}\cdot\vec{x}}\right]
 \label{eq:piq}
\eea
are the small-wavelength parts of $\phi$ and $\pi_\phi$ defined through the window function $W$ such that $W\simeq0$ for $k \ll k_\sigma$ and $W\simeq1$ for $k \gg k_\sigma$, and $\bar{\phi}$ and $\bar{\pi}$ are the long-wavelength, or coarse-grained, parts of $\phi$ and $\pi_\phi$. In \Eqs{eq:phiq} and~(\ref{eq:piq}), $a_{\vec{k}}$ and $a^\dag_{\vec{k}}$ are annihilation and creation operators satisfying the usual commutation relations $[a_{\vec{k}},a^\dag_{\vec{k}'}]=\delta^3(\vec{k}-\vec{k}')$ and $[a_{\vec{k}},a_{\vec{k}'}]=[a^\dag_{\vec{k}},a^\dag_{\vec{k}'}]=0$. They are time independent, contrary to the mode functions $\phi_{\vec{k}}$ and $\pi_{\vec{k}}$ that are solutions of the linearised \Eqs{eq:dotphiflrw} and~(\ref{eq:dotpflrw}), which in spatial Fourier space read
\bea
\label{eq:eomphiq}
	\dot\phi_k&=&\frac{N}{p^{3/2}}\pi_k ,  \\
	\dot\pi_k&=&-Np^{3/2}V_{,\phi\phi}(\phi)\phi_k-Np^{1/2}k^2\phi_k . 
\label{eq:eompiq}
\eea
We note that if the initial state is statistically isotropic, because \Eqs{eq:eomphiq} and~(\ref{eq:eompiq}) only involve the norm of the wavenumber $k$, the mode functions depend only on $k$ as well. This is why hereafter, $\phi_{\vec{k}}$ and $\pi_{\vec{k}}$ are simply written $\phi_k$ and $\pi_k$. The canonical quantisation of the short-wavelength modes is made using the Klein-Gordon product as an inner product~\cite{birrell1982}. They are thus normalised according to\footnote{This normalisation is equivalent to the one performed in the Lagrangian approach. In the Hamiltonian formalism indeed, one has ${\Pi}_{\vec{k}}=\sqrt{\gamma}(\partial_\tau{\Phi}_{\vec{k}}-N^i\partial_i{\Phi}_{\vec{k}})/N$. By plugging this expression into $i\int_{\Sigma_\tau}\dd^3x({\Phi}_{\vec{k}}{\Pi}^\star_{\vec{k}'}-{\Pi}_{\vec{k}}{\Phi}^\star_{\vec{k}'})$, the standard Klein-Gordon product is obtained, $i\int_{\Sigma_\tau}\dd^3x\sqrt{\gamma}n^\mu ({\Phi}_{\vec{k}}\partial_\mu{\Phi}^\star_{\vec{k}'}-{\Phi}^\star_{\vec{k}'}\partial_\mu{\Phi}_{\vec{k}})$, where $n^\mu=(1/N,N^i/N)$ is a unit 4-vector orthogonal to $\Sigma_\tau$.}
$i\int_{\Sigma_\tau}\dd^3x({\Phi}_{\vec{k}}{\Pi}^\star_{\vec{k}'}-{\Pi}_{\vec{k}}{\Phi}^\star_{\vec{k}'})=\delta^3(\vec{k}-\vec{k'})$, where ${\Phi}_{\vec{k}}=\phi_k(\tau)e^{-i\vec{k}\cdot\vec{x}}$ and ${\Pi}_{\vec{k}}=\pi_k(\tau)e^{-i\vec{k}\cdot\vec{x}}$, and where $\Sigma_\tau$ is a space-like hypersurface of fixed time $\tau$. 

The Langevin equation for the long-wavelength part of the field is then obtained by plugging the decomposition $\phi=\bar\phi+\phi_Q$ and $\pi_\phi=\bar\pi+\pi_Q$ into \Eqs{eq:dotphiflrw} and~(\ref{eq:dotpflrw}). Linearising these equations in $\phi_Q$ and $\pi_Q$, one obtains
\bea
\label{eq:Hamilton:split:phi}
	\dot{\bar\phi}&=&\frac{N}{p^{3/2}}\bar\pi-\dot\phi_Q+\frac{N}{p^{3/2}}\pi_Q , \\
	\dot{\bar\pi}&=&-Np^{3/2}V_{,\phi}(\bar\phi)-\dot\pi_Q-Np^{3/2}V_{,\phi\phi}(\bar\phi)\phi_Q+Np^{1/2} \Delta\phi_Q .
\label{eq:Hamilton:split:pi}
\eea
In these expressions, the Laplacian of $\bar\phi$ has been dropped since it is suppressed by $\sigma$. Replacing $\phi_Q$ and $\pi_Q$ by \Eqs{eq:phiq} and~(\ref{eq:piq}), and making use of the fact that the mode functions $\phi_k$ and $\pi_k$ satisfy \Eqs{eq:eomphiq} and~(\ref{eq:eompiq}), the Hamilton equations for $\bar\phi$ and $\bar\pi$ can be written as~\cite{Nakao:1988yi, PhysRevD.46.2408, Rigopoulos:2005xx, Tolley:2008na, Weenink:2011dd}
\bea
	\dot{\bar\phi}&=&\frac{N}{p^{3/2}}\bar\pi+\xi_\phi(\tau), \label{eq:eombarphi} \\
	\dot{\bar\pi}&=&-Np^{3/2}V_{,\phi}(\bar\phi)+\xi_\pi(\tau), \label{eq:eombarpi}
\eea
where the quantum noises $\xi_\phi$ and $\xi_\pi$ are given by
\bea
	\xi_\phi&=&-\displaystyle\int_{\mathbb{R}^3}\frac{\dd^3k}{(2\pi)^{3/2}}\dot{W}\left(\frac{k}{k_\sigma}\right)\left[a_{\vec{k}}\phi_k(\tau)e^{-i\vec{k}\cdot\vec{x}}+a^\dag_{\vec{k}}\phi^\star_k(\tau)e^{i\vec{k}\cdot\vec{x}}\right] , \label{eq:noisephi}\\
	\xi_\pi&=&-\displaystyle\int_{\mathbb{R}^3}\frac{\dd^3k}{(2\pi)^{3/2}}\dot{W}\left(\frac{k}{k_\sigma}\right)\left[a_{\vec{k}}\pi_k(\tau)e^{-i\vec{k}\cdot\vec{x}}+a^\dag_{\vec{k}}\pi^\star_k(\tau)e^{i\vec{k}\cdot\vec{x}}\right] . \label{eq:noisepi}
\eea
We note that the above set of equations can alternatively be obtained by integrating out the small-wavelength degrees of freedom \cite{Morikawa:1989xz,Hu:1994dka,Matarrese:2003ye}. This gives rise to an effective action which can subsequently be Legendre transformed to get an effective Hamiltonian accounting for the influence of quantum fluctuations on the large-wavelength fields. The dynamical and stochastic equations derived from this effective Hamiltonian are \Eqs{eq:eombarphi} and~(\ref{eq:eombarpi}).\footnote{This can be shown as follows. Making use of the notations of \Refc{Matarrese:2003ye}, the effective action reads (using cosmic time)
\bea
	S_\mathrm{eff}=S[{\bar\phi}^+]-S[\bar\phi^-]+\displaystyle\int\dd t\dd^3x\left[\left(a^3\dot{\bar\phi}^\Delta\right)\widetilde\xi_\phi-\left(\bar\phi^\Delta\right)\widetilde\xi_\pi\right], \label{eq:seff}
\eea
where $S$ is the free action. The noises $\widetilde{\xi}_\phi$ and $\widetilde\xi_\pi$ are two Gaussian random variables whose covariance matrix reads
\bea
	\displaystyle\int\frac{\dd^3k}{(2\pi)^3}\dot{W}\left[\frac{k}{k_\sigma(t)}\right]\dot{W}\left[\frac{k}{k_\sigma(t')}\right]e^{-i\vec{k}\cdot(\vec{x}-\vec{x}')}\left(\begin{array}{cc}
		\phi_k(t)\phi^\star_k(t') & \phi_k(t)\left[a^3(t')\dot\phi^\star_k(t')\right] \\
		\left[a^3(t)\dot\phi_k(t)\right]\phi^\star_k(t')	 & \left[a^3(t)\dot\phi_k(t)\right]\left[a^3(t')\dot\phi^\star_k(t')\right]
	\end{array}\right). \label{eq:noisecoveff}
\eea
Performing a Legendre transform of the above leads to $\bar\pi^\Delta=a^3\dot{\bar\phi}^\Delta$. Similarly, a Legendre transform of the quantum fluctuations leads to $\pi_k=a^3(t)\dot\phi_k(t)$, and it is straightforward to show that the covariance matrix of the noises is
\bea
	\displaystyle\int\frac{\dd^3k}{(2\pi)^3}\dot{W}\left[\frac{k}{k_\sigma(t)}\right]\dot{W}\left[\frac{k}{k_\sigma(t')}\right]e^{-i\vec{k}\cdot(\vec{x}-\vec{x}')}\left(\begin{array}{cc}
		\phi_k(t)\phi^\star_k(t') & \phi_k(t)\pi^\star_k(t') \\
		\pi_k(t)\phi^\star_k(t')	 & \pi_k(t)\pi^\star_k(t')
	\end{array}\right).
\eea
One can therefore respectively identify $\widetilde\xi_\phi$ and $\widetilde\xi_\pi$ with the noises $\xi_\phi$ and $\xi_\pi$ involved in \Eqs{eq:eombarphi} and~(\ref{eq:eombarpi}) (or equivalently to $-\xi_\phi$ and $-\xi_\pi$) since that they have the same statistical properties (see \Sec{ssec:noise} for the statistical properties of the noises $\xi_\phi$ and $\xi_\pi$). The effective Hamiltonian is thus
\bea
	C_\mathrm{eff}=C[{\bar\phi}^+]-C[\bar\phi^-]+\displaystyle\int\dd t\dd^3x\left[\bar\pi^\Delta\xi_\phi-\bar\phi^\Delta\xi_\pi\right],
\eea
from which \Eqs{eq:eombarphi} and~(\ref{eq:eombarpi}) are easily retrieved (note that $C$ is just the free Hamiltonian for the homogeneous and isotropic part of the scalar field).

The key point is to prove \Eqs{eq:seff} and~(\ref{eq:noisecoveff}), which are the analog of Eqs. (2.10) and (2.12) of \cite{Matarrese:2003ye}. The last term in the effective action is derived in \Refc{Morikawa:1989xz} and in Appendix A of \Refc{Matarrese:2003ye}. Following this last reference, the part of the effective action encoding coupling of the large-wavelength fields to the quantum fluctuations is proportional to
\bea
	\displaystyle\iint\dd t\dd t' \bar\phi(t)\left\lbrace\int\frac{\dd^3k}{(2\pi)^3}\left[D_t\phi_k(t)\right]\left[D_{t'}\phi_k(t')\right]\right\rbrace\bar\phi(t'),
\eea
where we only write down the temporal dependence for simplicity. The differential operator introduced above is $D_t=a^3(\ddot{W}+3H\dot{W}+2\dot{W}\partial_t)$. It can be equivalently rewritten as $D_tf=\partial_t(a^3\dot{W}f)+a^3\dot{W}\partial_tf$. Using this new form for the differential operator $D_t$ and then performing an integration by part on its {\it first term only} (\ie on $\partial_t(a^3\dot{W}f)$ only), one can show that the part of the effective action encoding coupling of the large-wavelength fields to the quantum fluctuations can be written as
%\bea
%	\displaystyle\int\frac{\dd^3k}{(2\pi)^3}\iint\dd t\dd t'\left\lbrace \bar\phi(t) \dot{W}\left[k/k_\sigma(t)\right]a^3(t)\partial_t\phi_k(t)-a^3(t)\partial_t\bar\phi(t)\dot{W}\left[k/k_\sigma(t)\right]\phi_k(t)\right\rbrace \nonumber\\
%	\times\left\lbrace \bar\phi(t') \dot{W}\left[k/k_\sigma(t)\right]a^3(t')\partial_t\phi^\star_k(t')-a^3(t')\dot{\bar\phi}(t')\dot{W}\left[k/k_\sigma(t')\right]\phi^\star_k(t')\right\rbrace,
%\eea
\bea
	\displaystyle\int\frac{\dd^3k}{(2\pi)^3}\iint\dd t\dd t'\left\lbrace \bar\phi(t) \dot{W}\left[k/k_\sigma(t)\right]a^3(t)\dot\phi_k(t)-a^3(t)\dot{\bar\phi}(t)\dot{W}\left[k/k_\sigma(t)\right]\phi_k(t)\right\rbrace \nonumber\\
	\times\left\lbrace \bar\phi(t') \dot{W}\left[k/k_\sigma(t')\right]a^3(t')\dot\phi^\star_k(t')-a^3(t')\dot{\bar\phi}(t')\dot{W}\left[k/k_\sigma(t')\right]\phi^\star_k(t')\right\rbrace,
\eea
where the canonical momenta $\pi=a^3\dot\phi$ are made explicit. In vectorial form, the integrand reads 
\bea
	\left(\bar\phi(t),~\bar\pi(t)\right)\left(\begin{array}{c}
		-\dot{W}(t)\pi_k(t) \\
		\dot{W}(t)\phi_k(t)
	\end{array}\right)\left(-\dot{W}(t')\pi^\star_k(t'),~\dot{W}(t')\phi^\star_k(t')\right)\left(\begin{array}{c}
		\bar\phi(t') \\
		\bar\pi(t')
	\end{array}\right).
\eea
Apart from the integrations which have been omitted to lighten the expression, the above is the analog of Eq. (A.8) of \cite{Matarrese:2003ye} expressed here in the field configuration and field momentum basis. From this expression, and following Appendix A of \Refc{Matarrese:2003ye}, it is straightforward to get \Eqs{eq:seff} and~(\ref{eq:noisecoveff}).
}
\subsection{Statistical properties of the noise}
\label{ssec:noise}
Let us now assume that the field fluctuations $\phi_k$ and $\pi_k$ are placed in their vacuum state. Since we work at linear order in perturbation theory, they thus feature Gaussian statistics with vanishing mean. The statistical properties of the quantum noises $\xi_\phi$ and $\xi_\pi$ are therefore fully characterised by their two-points correlation matrix 
\bea
 \label{eq:xidef}
	\boldsymbol{\Xi} \left (\vec{x}_1,\tau_1;\vec{x}_2,\tau_2\right )=\left(\begin{array}{cc}
		\left<0\right|\xi_\phi(\vec{x}_1,\tau_1)\xi_\phi(\vec{x}_2,\tau_2)\left|0\right> & \left<0\right|\xi_\phi(\vec{x}_1,\tau_1)\xi_\pi(\vec{x}_2,\tau_2)\left|0\right> \\
		\left<0\right|\xi_\pi(\vec{x}_1,\tau_1)\xi_\phi(\vec{x}_2,\tau_2)\left|0\right> & \left<0\right|\xi_\pi(\vec{x}_1,\tau_1)\xi_\pi(\vec{x}_2,\tau_2)\left|0\right>
	\end{array}\right) .
\eea
Hereafter, bold notations denote vector or matrix quantities. Letting the annihilation and creation operators act on the vacuum state $|0\rangle$, the entries of this matrix read
\bea
	\Xi_{f_1,g_2}=\displaystyle\int_{\mathbb{R}^3}\frac{\dd^3k}{(2\pi)^3}\dot{W}\left[\frac{k}{k_\sigma(\tau_1)}\right]\dot{W}\left[\frac{k}{k_\sigma(\tau_2)}\right]f_k(\tau_1)g^\star_k(\tau_2)e^{i\vec{k}\cdot(\vec{x}_2-\vec{x}_1)} ,
\eea
where the notation $\Xi_{f_1,g_2}=\left<0\right|\xi_f(\vec{x}_1,\tau_1)\xi_g(\vec{x}_2,\tau_2)\left|0\right>$ has been introduced for display convenience, $f$ and $g$ being either $\phi$ or $\pi$. Note that, at this stage, the order of the subscripts $f$ and $g$ does matter as a result of the non-commutativity of $\xi_\phi$ and $\xi_\pi$. The angular integral over $\vec{k}/k$ can be performed easily since, as explained below  \Eqs{eq:eomphiq} and~(\ref{eq:eompiq}), the mode functions $\phi_k$ and $\pi_k$ only depend on the norm of $\vec{k}$. One obtains
\bea
	\Xi_{f_1,g_2}=\displaystyle\int_{\mathbb{R}^+}\frac{k^2\dd k}{2\pi^2}
	\dot{W}\left[\frac{k}{k_\sigma(\tau_1)}\right]\dot{W}\left[\frac{k}{k_\sigma(\tau_2)}\right]f_k(\tau_1)g^\star_k(\tau_2)
	\frac{\sin\left(k \vert \vec{x}_2-\vec{x}_1 \vert\right)}{k \vert \vec{x}_2-\vec{x}_1 \vert } .
\label{eq:Xif1g2:kintegrated}
\eea
We now need to specify the filter function $W$, which for simplicity we choose to be a Heaviside function $W(k/k_\sigma)=\Theta\left(k/k_{\sigma}-1\right)$. Its time derivative gives a Dirac distribution, and the integrand of \Eq{eq:Xif1g2:kintegrated} contains $\delta[k-k_\sigma(\tau_1)]\delta[k-k_\sigma(\tau_2)]$ which yields $\delta(\tau_1-\tau_2)$, meaning that the noises are white. One obtains
\bea
 \label{eq:noisecorrel}
	\Xi_{f_1,g_2}=\frac{1}{6\pi^2}\left.\frac{\dd k^3_\sigma(\tau)}{\dd\tau}\right|_{\tau_1}f_{k=k_\sigma(\tau_1)}g^\star_{k=k_\sigma(\tau_1)}\frac{\sin\left[k_\sigma(\tau_1) \vert \vec{x}_2-\vec{x}_1 \vert \right]}{k_\sigma(\tau_1)\vert \vec{x}_2-\vec{x}_1 \vert}\delta\left(\tau_1-\tau_2\right) .
\eea
In the following, we will be essentially interested in the autocorrelation of the noises, $\vec{x}_1=\vec{x}_2$, for which $\sin[k_\sigma(\tau_1) \vert \vec{x}_2-\vec{x}_1 \vert ]/[k_\sigma(\tau_1) \vert \vec{x}_2-\vec{x}_1 \vert ]=1$. The noises being white, the correlations are non-zero only at equal time. We will thus write the correlation matrix of the noise as $\Xi_{f_1,g_2}\equiv \Xi_{f,g}(\tau_1)\delta(\tau_1-\tau_2)$. The correlator $ \Xi_{f,g}(\tau)$ can be expressed in terms of the power spectrum of the quantum fluctuations
\begin{equation}
	 \mathcal{P}_{f,g}(k;\tau)=\frac{k^3}{2\pi^2}f_{k}(\tau)g^\star_{k}(\tau) ,
\end{equation}
which gives rise to
\begin{equation}
\label{eq:noisecorrel_Pk}
	\Xi_{f,g}(\tau)=\frac{\dd \ln\left[k_\sigma(\tau)\right]}{\dd\tau} \mathcal{P}_{f,g}\left[k_\sigma(\tau);\tau\right] . 
\end{equation}

Let us finally notice that the noises correlator is described by an hermitian matrix, i.e $\Xi^\star_{f,g}=\Xi_{g,f}$, the antisymmetric part of which is proportional to the Klein-Gordon product of the mode functions
\bea
	\Xi_{\phi,\pi}(\tau)-\Xi_{\pi,\phi}(\tau)&=&\frac{1}{6\pi^2}\frac{\dd k^3_\sigma(\tau)}{\dd\tau}\left[\phi_{k=k_\sigma(\tau)}\pi^\star_{k=k_\sigma(\tau)}-\pi_{k=k_\sigma(\tau)}\phi^\star_{k=k_\sigma(\tau)}\right]\\
	&=&\frac{-i}{6\pi^2}\frac{\dd k^3_\sigma(\tau)}{\dd\tau} .
\label{eq:commutator:normalised}
\eea
In this expression, the second equality is obtained by using canonical quantisation of the fluctuations which sets the Klein-Gordon product to $-i$, as explained below \Eq{eq:eompiq}.
\section{Generic solution for a free scalar field}
\label{sec:testField}
Let us now consider the case of a test scalar field with quadratic potential $V(\phi)=\Lambda^4+m^2\phi^2/2$ (terms linear in $\phi$ can always be reabsorbed by field shift symmetry). If $m^2>0$, the potential is convex and of the large-field type, if $m^2<0$ it is concave and of the hilltop type. For such a potential, \Eqs{eq:eombarphi} and~(\ref{eq:eombarpi}) form a linear differential system where the noises $\xi_\phi$ and $\xi_\pi$ do not depend on the phase-space variables of the coarse-grained field. This yields simplifications that allow one to analytically solve the full stochastic dynamics.
\subsection{Probability distribution in phase space}
\label{ssec:fokker}
Since the dynamics described by \Eqs{eq:eombarphi} and~(\ref{eq:eombarpi}) is linear, it is convenient to work with the vector notation
\bea
\boldsymbol{\Phi}=\left(\begin{array}{c}
	\bar\phi \\
	\bar\pi
\end{array}\right)~~~\mathrm{and}~~~\boldsymbol{\xi}=\left(\begin{array}{c}
	\xi_\phi \\
	\xi_\pi
\end{array}\right) .
\eea
In terms of these variables, \Eqs{eq:eombarphi} and~(\ref{eq:eombarpi}) can be written as the Langevin equation
\bea
 \label{eq:inhomopb}
	\dot{\boldsymbol{\Phi}}=\boldsymbol{A}(\tau)\boldsymbol{\Phi}+\boldsymbol{\xi}(\tau)~~~\mathrm{with}~~~\boldsymbol{A}(\tau)=\left(\begin{array}{cc}
	0 & N/p^{3/2} \\
	-m^2Np^{3/2} & 0						\end{array}\right) .
\eea
\subsubsection{Fokker-Planck equation}
This Langevin equation can be translated into a Fokker-Planck equation~\cite{risken1996fokker} for the probability density function (PDF hereafter) in phase space associated to the stochastic process~(\ref{eq:inhomopb}), given by
\bea
\label{eq:fokkerxi}
	\frac{\partial P(\boldsymbol{\Phi},\tau)}{\partial\tau}=-\displaystyle\sum_{i,j=1}^2\frac{\partial}{\partial\Phi}_i\left[A_{ij} \Phi_j P(\boldsymbol{\Phi},\tau)\right]+\frac{1}{2}\displaystyle\sum_{i,j=1}^2\frac{\partial^2}{\partial\Phi_i\partial\Phi_j}\left[\Xi_{ij}(\tau)P(\boldsymbol{\Phi},\tau)\right] . 
\eea
The first term in the right-hand side of \Eq{eq:fokkerxi} is called the drift term and traces the deterministic part of the dynamics, and the second term is the diffusive term that traces the stochastic component of the evolution. In the latter, $\Xi_{ij}(\tau)$ can be factored out of the phase-space differential operator since it does not depend on $\boldsymbol{\Phi}$. This term can therefore be written as $\mathrm{Tr}\left[\boldsymbol{H}\,\boldsymbol{\Xi}\right]/2$, where $\mathrm{Tr}$ is the trace operator and $H_{ij}\equiv \partial^2P(\boldsymbol{\Phi},\tau)/(\partial \Phi_i\partial \Phi_j)$ is the Hessian of the PDF.
\subsubsection{Losing the commutator}
\label{sec:lossing_commutator}
As noticed below \Eq{eq:noisecorrel_Pk}, the noise correlator matrix $\boldsymbol{\Xi}$ is hermitian and it can thus be decomposed on the basis $\{\boldsymbol{I},\boldsymbol{J}_x,\boldsymbol{J}_y,\boldsymbol{J}_z\}$, where $\boldsymbol{I}$ is the $2\times2$ identity matrix and the three following matrices are the Pauli matrices. The decomposition reads
\bea
\label{eq:Pauli}
	\boldsymbol{\Xi}=\frac{1}{2}\left(\Xi_{\phi,\phi}+\Xi_{\pi,\pi}\right)\boldsymbol{I}+\frac{1}{2}\left(\Xi_{\phi,\pi}+\Xi_{\pi,\phi}\right)\boldsymbol{J}_x+\frac{i}{2}\left(\Xi_{\phi,\pi}-\Xi_{\pi,\phi}\right)\boldsymbol{J}_y+\frac{1}{2}\left(\Xi_{\phi,\phi}-\Xi_{\pi,\pi}\right)\boldsymbol{J}_z .\quad
\eea
In this expression, the coefficient multiplying $\boldsymbol{J}_y$ is built from the commutator of the quantum fluctuations~(\ref{eq:commutator:normalised}) and thus traces the very quantum nature of the noise. However, its contribution to the Fokker-Planck equation vanishes. Indeed, the Hessian of the PDF, $H_{ij}$, is symmetric with respect to the indices $i$ and $j$ while the matrix $\boldsymbol{J}_y$ is antisymmetric. As a consequence, $\mathrm{Tr}\left[\boldsymbol{H}\boldsymbol{J}_y\right]=0$ and the quantum commutator disappears from \Eq{eq:fokkerxi}. This makes sense since it implies that, if one wants to describe the full quantum dynamics by a stochastic theory, one looses the information about the commutators. The reason why it provides a good approximation is because these commutators become negligible on large scales as mentioned in \Sec{sec:Stochastic_Inflation} and further explained below in \Sec{sec:quantum_to_classical}. As a consequence, the symmetric terms of $\boldsymbol{\Xi}$ remaining in the Fokker-Planck equation, although drawn out from a quantum state, can be equivalently described by a classical (though correlated) distribution. Defining the diffusion matrix $\boldsymbol{D}$ as the symmetric part of $\boldsymbol{\Xi}$, \ie $ \boldsymbol{D}=(\Xi_{\phi,\phi}+\Xi_{\pi,\pi})\boldsymbol{I}/2+(\Xi_{\phi,\pi}+\Xi_{\pi,\phi})\boldsymbol{J}_x/2+(\Xi_{\phi,\phi}-\Xi_{\pi,\pi})\boldsymbol{J}_z/2$, the Fokker-Planck equation is then given by
\bea
\label{eq:fokker}
	\frac{\partial P(\boldsymbol{\Phi},\tau)}{\partial\tau}=-\displaystyle\sum_{i,j=1}^2\frac{\partial}{\partial \Phi_i}\left[A_{ij} {\Phi}_jP(\boldsymbol{\Phi},\tau)\right]+\frac{1}{2}\displaystyle\sum_{i,j=1}^2 {D}_{ij}(\tau)\frac{\partial^2P(\boldsymbol{\Phi},\tau)}{\partial {\Phi}_i\partial {\Phi}_j} . 
\eea
\subsubsection{Green formalism}
Because the Langevin equation~(\ref{eq:inhomopb}), or equivalently the Fokker-Planck equation~(\ref{eq:fokker}), is linear, it can be analytically solved using the Green's matrix formalism. The Green's matrix $\boldsymbol{G}(\tau,\tau_0)$ is defined as the solution of the homogeneous (hence deterministic) problem associated to the stochastic dynamics of \Eq{eq:inhomopb}, $\partial \boldsymbol{G}(\tau,\tau_0)/\partial\tau=\boldsymbol{A}(\tau)\boldsymbol{G}(\tau,\tau_0)+\boldsymbol{I}\delta(\tau-\tau_0)$. In \App{app:green}, it is shown how such a matrix can be constructed in practice, why it always has unit determinant and why for any solution $\boldsymbol{\Phi}_{\mathrm{det}}$ of the homogenous problem one has 
\bea
\label{eq:Phidet}
\boldsymbol{\Phi}_{\mathrm{det}}(\tau)=\boldsymbol{G}(\tau,\tau_0)\boldsymbol{\Phi}_0 .
\eea 
Here, $\boldsymbol{\Phi}_{\mathrm{det}}$ is the deterministic trajectory that field variables would follow in the absence of the noises and starting from the initial state $\boldsymbol{\Phi}(\tau_0)=\boldsymbol{\Phi}_0$. Solutions of the Fokker-Planck equation~(\ref{eq:fokker}) can be formally obtained introducing the Green function $\mathcal{W}(\boldsymbol{\Phi},\tau|\boldsymbol{\Phi}_0,\tau_0)$,  giving the PDF in phase space at time $\tau$ if the field and its momentum are initially at $\boldsymbol{\Phi}(\tau_0)=\boldsymbol{\Phi}_0$, through 
\bea
P\left(\boldsymbol{\Phi},\tau\right)=\displaystyle\int \dd\boldsymbol{\Phi}_0\mathcal{W}(\boldsymbol{\Phi},\tau|\boldsymbol{\Phi}_0,\tau_0)P\left(\boldsymbol{\Phi}_0,\tau_0\right) .
\eea
For the Fokker-Planck equation~(\ref{eq:fokker}), the Green function is the gaussian distribution
\bea
\label{eq:Gaussian:Green}
	\mathcal{W}\left(\boldsymbol{\Phi},\tau|\boldsymbol{\Phi}_0,\tau_0\right)=\displaystyle\frac{1}{\sqrt{2\pi^2\det\left[\boldsymbol{\Sigma}(\tau)\right]}}\exp\left\lbrace-\frac{1}{2}\left[\boldsymbol{\Phi}-\boldsymbol{\Phi}_{\mathrm{det}}(\tau)\right]^\dag\boldsymbol{\Sigma}^{-1}(\tau)\left[\boldsymbol{\Phi}-\boldsymbol{\Phi}_{\mathrm{det}}(\tau)\right]\right\rbrace, 
\eea
where $\dag$ means the conjugate-transpose. From this expression, one can check that  $\left<\boldsymbol{\Phi}(\tau)\right>=\int\dd\boldsymbol{\Phi}~\boldsymbol{\Phi}\mathcal{W}(\boldsymbol{\Phi},\tau|\boldsymbol{\Phi}_0,\tau_0)$ is equal to $\boldsymbol{\Phi}_{\mathrm{det}}$, which means that the deterministic trajectory is also the average trajectory of the stochastic field since the noises have a vanishing mean, in agreement with Ehrenfest theorem. In \Eq{eq:Gaussian:Green}, $\boldsymbol{\Sigma}$ is the covariance matrix of the field variables that captures all the diffusive processes. It is obtained as the forward propagation of the diffusion matrix,
\bea
\label{eq:Sigma}
	\boldsymbol{\Sigma}(\tau)=\displaystyle\int^\tau_{\tau_0}\dd s~\boldsymbol{G}(\tau,s)\boldsymbol{D}(s)\boldsymbol{G}^\dag(\tau,s) ,
\eea	
and is related to the two-point correlation of the coarse-grained field through $\langle\left[\boldsymbol{\Phi}(\tau)-\left<\boldsymbol{\Phi}(\tau)\right>\right]\left[\boldsymbol{\Phi}(\tau)-\left<\boldsymbol{\Phi}(\tau)\right>\right]^\dag\rangle=\boldsymbol{\Sigma}(\tau)$. 
\subsection{Quantum-to-classical transition}
\label{sec:quantum_to_classical}
In \Sec{sec:lossing_commutator}, it was shown that a description of the full quantum dynamics in terms of a Fokker-Planck equation necessarily drops out the commutators of the theory, and can therefore only provide an approximation to the actual results. In this section, we show that this approximation becomes accurate in the limit $\sigma\ll 1$, illustrating the ``quantum-to-classical'' transition of inflationary perturbations on super-Hubble scales.

To this end we evaluate the moments of the coarse-grained scalar field with and without assuming the noises to be in a quasiclassical state, \ie with and without resorting to the Fokker-Planck equation. Let us consider the case where the coarse-grained field is in a classical state $\boldsymbol{\Phi}_0$ at initial time. Under \Eq{eq:inhomopb}, it becomes a mixture of a classical state and quantum operators a later time, through the contribution of the quantum noise $\boldsymbol{\xi}$. Under \Eq{eq:fokker} however, it simply becomes a random variable. The solution to \Eq{eq:inhomopb} is formally given by
\bea
\label{eq:quant_Sol}
	\boldsymbol{\Phi}_{\mathrm{quant}}(\tau)=\boldsymbol{G}(\tau,\tau_0)\boldsymbol{\Phi}_0+\displaystyle\int_{\tau_0}^\tau\dd s\boldsymbol{G}(\tau,s) \boldsymbol{\xi}(s) ,
\eea
where the subscript ``quant'' stresses that we are dealing with the solution of the full quantum-field theory, while the solution to \Eq{eq:fokker} has already be given in \Eqs{eq:Gaussian:Green}, (\ref{eq:Phidet}) and~(\ref{eq:Sigma}).
\subsubsection{Linear observables} 
Evaluating \Eq{eq:quant_Sol} on the vacuum, since $\langle 0 \vert \xi \vert 0 \rangle=0$, one obtains
\bea
	\left\langle 0 \left\vert \boldsymbol{\Phi}_{\mathrm{quant}} \right\vert 0 \right\rangle=\boldsymbol{G}\left(\tau,\tau_0\right)\boldsymbol{\Phi}_0=\boldsymbol{\Phi}_\mathrm{det} ,
\eea
as follows from \Eq{eq:Phidet}. This also corresponds to the solution for the mean stochastic path as noticed below \Eq{eq:Gaussian:Green}, so the full quantum and the stochastic theories match for linear observables.
\subsubsection{Quadratic observables} 
\label{sec:QuantumToClassical:Quadratic}
The initial coarse-grained state $\boldsymbol{\Phi}_0$ being classical, it commutes with the noise and evaluating the square of \Eq{eq:quant_Sol} on the vacuum gives rise to
\bea
\label{eq:quadmoment:quant}
	\left\langle 0 \left\vert  \boldsymbol{\Phi}_\mathrm{quant}\boldsymbol{\Phi}_\mathrm{quant}^\dag\right\vert 0 \right\rangle -
	\left\langle 0 \left\vert  \boldsymbol{\Phi}_\mathrm{quant}\right\vert 0 \right\rangle 
	\left\langle 0 \left\vert  \boldsymbol{\Phi}_\mathrm{quant}^\dag\right\vert 0 \right\rangle 
	= \displaystyle\int^\tau_{\tau_0}\dd s~\boldsymbol{G}(\tau,s)\boldsymbol{\Xi}(s)\boldsymbol{G}^\dag(\tau,s) .
\eea
On the other hand, the corresponding expression for the stochastic solution is given below \Eq{eq:Sigma}. It is identical to \Eq{eq:quadmoment:quant} except that $\boldsymbol{\Xi}$ is replaced by $\boldsymbol{D}$. The difference it yields can be thus quantified through
\bea
	\boldsymbol{\Delta}&\equiv&
	\left\langle 0 \left\vert  \boldsymbol{\Phi}_\mathrm{quant}\boldsymbol{\Phi}_\mathrm{quant}^\dag \right\vert 0 \right\rangle -
	\langle\boldsymbol{\Phi}\boldsymbol{\Phi}^\dag\rangle\\
	&=&\frac{i}{2}\displaystyle\int^\tau_{\tau_0}\dd s~\left[\Xi_{\phi,\pi}(s)-\Xi_{\pi,\phi}(s)\right]\boldsymbol{G}(\tau,s)\boldsymbol{J}_y\boldsymbol{G}^\dag(\tau,s),
\label{eq:quadratic:Delta}
\eea
where the decomposition of $\boldsymbol{\Xi}$ in terms of the Pauli matrices introduced in \Eq{eq:Pauli} has been used. Since $\boldsymbol{G}$ is a symplectic matrix, one has\footnote
{\label{footnote:symplectic}A symplectic matrix $M$ is a real matrix satisfying $\boldsymbol{M}^\mathrm{T}\boldsymbol{\Omega} \boldsymbol{M}=\boldsymbol{\Omega}$, where $T$ means transpose and 
\bea
\label{eq:def:Omega}
\boldsymbol{\Omega}\equiv \left(\begin{array}{cc} 0 & 1 \\ -1 & 0 \end{array}\right) .
\eea
In 2 dimensions, it is easy to show that the symplectic matrices are the real matrices with unit determinant, which is the case of the Green matrix as shown in \App{app:green}. More generally, evolution generated by quadratic Hamiltonians can always be viewed as the action of the symplectic group on the phase-space variables which also explains why $\boldsymbol{G}$ is symplectic. Since $\boldsymbol{J}_y=-i\boldsymbol{\Omega}$, this means that $\boldsymbol{G}\boldsymbol{J}_y\boldsymbol{G}^\dag=\boldsymbol{J}_y$.
}
$\boldsymbol{G}\boldsymbol{J}_y\boldsymbol{G}^\dag=\boldsymbol{J}_y$, which can be factored out of the integral as it does not depend on time. Making use of \Eq{eq:commutator:normalised}, the time integration can be performed and one obtains
\bea
\label{eq:Delta:quad}
	\boldsymbol{\Delta}=\frac{1}{12\pi^2}\left[k_\sigma^3(\tau)-k_\sigma^3(\tau_0)\right]\boldsymbol{J}_y
	=\sigma^3\frac{a^3(\tau)H^3(\tau)}{12\pi^2}\left[1-\frac{a^3(\tau_0)H^3(\tau_0)}{a^3(\tau)H^3(\tau)}\right]\boldsymbol{J}_y ,
\eea
where in the second equality one has made use of \Eq{eq:ksigma}. Since $\boldsymbol{J}_y$ is an off-diagonal matrix, this needs to be compared to the $\phi,\pi$ component of the covariance matrix. In \Sec{sec:massless}, by integrating the mode equations~(\ref{eq:eomphiq}) and~(\ref{eq:eompiq}), it is shown that for a massless field, at leading order in $\sigma$,  one has
\bea
\left.  {\Sigma}_{\phi,\pi}\right\vert_{m=0} = \sigma^2\frac{a^3(\tau)H^3(\tau)}{12\pi^2}\left[1-\frac{a^3(\tau_0)H^3(\tau_0)}{a^3(\tau)H^3(\tau)}\right] ,
\eea
while for a light test field with a non-vanishing mass $m\ll H$, it is found in \Sec{sec:light} that
\bea
 {\Sigma}_{\phi,\pi} = 
 \left\lbrace
 \begin{array}{ll}
\dfrac{m^2}{H^2} \dfrac{a^3(\tau)H^3(\tau)}{12\pi^2} \ln\left[\dfrac{a(\tau_0)H(\tau_0)}{a(\tau)H(\tau)}\right] &
\mathrm{if}\   \ln\left[\dfrac{a(\tau)H(\tau)}{a(\tau_0)H(\tau_0)}\right] \ll \dfrac{3H^2}{2m^2} \\
 \dfrac{ a^3(\tau) H^3(\tau)}{8\pi^2}& 
\mathrm{if}\   \ln\left[\dfrac{a(\tau)H(\tau)}{a(\tau_0)H(\tau_0)}\right] \gg \dfrac{3H^2}{2m^2} 
 \end{array}
 \right.
  .
\eea
In all cases, one can see that $ {\Delta}_{\phi,\pi}$ is suppressed by higher powers of $\sigma$ and can therefore be neglected if $\sigma\ll 1$, with the slightly stronger condition $\sigma\ll (m/H)^{2/3}$ for a light test field at early time.

In fact, the limit $\sigma\ll 1$ does not need to be invoked if one restricts correlators to observable operators. Indeed, observables are necessarily described in terms of hermitian operators, since the outcome of a quantum measurement can only be a real number. This is why instead of $\boldsymbol{\Phi}\boldsymbol{\Phi}^\dag$ in \Eq{eq:quadmoment:quant}, one should consider $(\boldsymbol{\Phi}\boldsymbol{\Phi}^\dag+\boldsymbol{\Phi}^\star\boldsymbol{\Phi}^\mathrm{T})/2$. Since $\boldsymbol{G}$ is real, this means that $\boldsymbol{\Xi}$ in \Eq{eq:quadmoment:quant} must be replaced with $(\boldsymbol{\Xi}+\boldsymbol{\Xi}^\star)/2$, and since $\boldsymbol{\Xi}$ is hermitian, this is identical to $(\boldsymbol{\Xi}+\boldsymbol{\Xi}^\mathrm{T})/2$. This precisely corresponds to $\boldsymbol{D}$, the symmetric part of $\boldsymbol{\Xi}$. In this case, one thus recovers the predictions of the stochastic theory, even without resorting to the large-scale limit.
\subsubsection{Quartic observables} 
In terms of observable correlators, the full quantum and the stochastic theories give the exact same results for linear and quadratic operators. Since cubic powers of Gaussian noises vanish, they also match for cubic correlators and one has to consider quartic observables to start probing observable deviations between the stochastic framework and the full quantum theory. When calculating such correlators, one has to evaluate
\bea
\left\langle 0 \left\vert \boldsymbol{\xi}(\tau_1)\boldsymbol{\xi}(\tau_2)\boldsymbol{\xi}(\tau_3)\boldsymbol{\xi}(\tau_4) \right\vert 0 \right\rangle = 3 \delta\left(\tau_1-\tau_2\right)\delta\left(\tau_1-\tau_3\right)\delta\left(\tau_1-\tau_4\right) \boldsymbol{\Xi}^2 ,
\eea
which can easily be derived from \Eqs{eq:noisephi} and~(\ref{eq:noisepi}) using a Heaviside window function as in \Sec{ssec:noise}, and which simply translates the Gaussian character of $\boldsymbol{\xi}$. Real correlators are therefore encoded in the real part of $\boldsymbol{\Xi}^2$, while in the stochastic framework they are given by $\boldsymbol{D}^2$. The difference between the two theories is therefore characterised by
\bea
\boldsymbol{\Xi}^2+\boldsymbol{\Xi}^{2\star} - \boldsymbol{D}^2 - \boldsymbol{D}^{2\star} &=& \frac{1}{2} \left(\boldsymbol{\Xi}-\boldsymbol{\Xi}^\mathrm{T}\right)^2 = -\frac{1}{2} \left(\Xi_{\phi,\pi}-\Xi_{\pi,\phi}\right)^2 \boldsymbol{I} \nonumber \\
&=& \frac{1}{72\pi^4}\left(\frac{\dd k_\sigma^3}{\dd\tau}\right)^2 \boldsymbol{I} = \frac{N^2a^6H^8}{8\pi^4}  \sigma^6\boldsymbol{I}   ,
\label{eq:diff:quartic}
\eea
where we have used that $\boldsymbol{D}=(\boldsymbol{\Xi}+\boldsymbol{\Xi}^\mathrm{T})/2$. Contrary to \Eq{eq:Delta:quad} where $\boldsymbol{J}_y$ is purely imaginary, this difference is real, hence observable. Unsurprisingly, it is proportional to the antisymmetric part of $\boldsymbol{\Xi}$, which can be evaluated using \Eq{eq:Pauli}, and where we have further used that $\boldsymbol{J}^2=\boldsymbol{I}$.  In the second line of \Eq{eq:diff:quartic}, the commutator $\Xi_{\phi,\pi}-\Xi_{\pi,\phi}$ is expressed using \Eqs{eq:commutator:normalised} and~(\ref{eq:ksigma}), which is evaluated in the de-Sitter case where $H$ is a constant for simplicity. Since this matrix is proportional to $\boldsymbol{I}$, it needs to be compared to $ {D}^2_{\ \phi,\phi}+ {D}^2_{\ \pi,\pi}$, which is the component of $ \boldsymbol{D}^2 + \boldsymbol{D}^{2*}$ along $\boldsymbol{I}$ according to the decomposition~(\ref{eq:Pauli}). Using the results of \Sec{sec:massless}, at leading order in $\sigma$, it is given by
\bea
\left. {D}^2_{\ \phi,\phi}+ {D}^2_{\ \pi,\pi}\right\vert_{m=0} = \frac{N^2a^6H^8}{4\pi^4}\sigma^4
\eea
for a purely massless field, and \Eq{eq:diff:quartic} is suppressed by higher powers of $\sigma$ hence can be neglected in the limit
\bea
\left.\sigma\right\vert_{m=0}\ll 1 .
\eea
Using the results of \Sec{sec:light}, for a light test field with mass $m\ll H$, one obtains
\bea
 {D}^2_{\ \phi,\phi}+ {D}^2_{\ \pi,\pi} = \frac{9N^2a^6H^8}{4\pi^4}\frac{m^4}{H^4} ,
\eea
which dominates over \Eq{eq:diff:quartic} if
\bea
\label{eq:cond:classicalTransition}
\sigma\ll \left(\frac{m}{H}\right)^{\frac{2}{3}} ,
\eea 
and the same conditions as in \Sec{sec:QuantumToClassical:Quadratic} are recovered. One concludes that even at quartic order where the quantum and the stochastic theories start giving different results for observable correlators, these differences are suppressed on large scales by $\sigma$ and can therefore be neglected if $\sigma$ is taken to be sufficiently small.
\subsubsection{Wigner function}
\label{sec:Wigner}
Finally, let us briefly mention that these results can be reformulated making use of the Wigner function~\cite{Graziani:1987uz, PhysRevD.46.2408, Burgess:2014eoa, Boyanovsky:2015tba, Martin:2015qta, Moss:2016uix}. From the two-mode squeezed state in which each scalar mode is placed during inflation, one can build a function in phase space, the Wigner function (defined as the Weyl transform of the density matrix), which has three remarkable properties. First, because the two-mode squeezed state is Gaussian, it is a Gaussian function, hence it is positive everywhere. Second, since the Hamiltonian~(\ref{eq:total_Hamiltonian}) of linear fluctuations is quadratic, it evolves under the classical Hamilton's equations of motion. Third, once the perturbations wavelength cross the Hubble radius, their quantum states become highly squeezed. In this limit, it can be shown that the expectation value of any observable can be well approximated by integrating over phase space the function associated to the observable multiplied by the Wigner function. In fact, this procedure is even exact for operators depending on $\phi_k$ or $\pi_k$ only, as well as for hermitian two-point correlators (see table 1 in \Refc{Martin:2015qta}). Otherwise, the difference is related to the commutators of the theory which become suppressed on super-Hubble scales, and one recovers the same results as above [for instance, Eq.~(91) of \Refc{Martin:2015qta} is identical to \Eq{eq:quadratic:Delta}]. In this limit, the Wigner function can therefore be seen as a PDF of classical processes in phase space, and the product of all Wigner functions with wavenumbers $k<k_\sigma$ follows the Fokker-Planck equation~(\ref{eq:fokker}).

Let us also notice that the reason why one is interested in the coarse-grained parts of the fields in the first place is that for inflationary backgrounds, the infrared sector provides the dominant contribution of all field correlators. It is a priori rather coincidental that in the same super-Hubble limit, a quantum-to-classical transition takes places that allows one to use stochastic techniques. In a matter contracting universe for instance, this unambiguous link between classicality and super-Hubble regime does not hold anymore. In this case also, defining initial vacuum state may be less unambiguous and this may lead to different types of classical transitions~\cite{Lesgourgues:1996jc}. The conditions under which a stochastic formalism can be derived in this context thus remain to be determined.
\subsection{Canonical transformations}
\label{sec:canonical}
In \Sec{sec:Langevin}, the coarse-graining procedure has been performed parametrising phase space in terms of the canonical variables $\phi_k$ and $\pi_k$. However, it is not obvious a priori that the coarse-graining procedure and the choice of canonical coordinates commute. In this section, we therefore want to check whether it is indeed equivalent to perform a change of canonical coordinates followed by coarse graining, or, conversely, to perform coarse graining followed by a change of the canonical coordinates.

Different sets of canonical variables $\boldsymbol{v}(\vec{x})$ and $\boldsymbol{z}(\vec{x})$ are related through linear (in order to keep quadratic Hamiltonians) transformations
\bea
\label{eq:canonical:transf}
\boldsymbol{v}(\vec{x})=\boldsymbol{M}(\tau)\boldsymbol{z}(\vec{x})
\eea
called canonical transformations, where $\boldsymbol{M}$ is a matrix which depends on time only. If canonical commutation relations are preserved, one has $[v_j,v_k]=[z_j,z_k]=i {\Omega}_{j,k}$, where $\boldsymbol{\Omega}$ has been defined in \Eq{eq:def:Omega}. This implies that $\boldsymbol{\Omega}=\boldsymbol{M}^\mathrm{T}\boldsymbol{\Omega}\boldsymbol{M}$, that is to say that $\boldsymbol{M}$ is a symplectic matrix (see footnote~\ref{footnote:symplectic}). In particular, it is invertible and has a unit determinant.

If one starts from the phase-space variables $\boldsymbol{z}(\vec{x})=\left[\phi(\vec{x}), \pi(\vec{x}) \right]$, in \Sec{ssec:fokker} the Fokker-Planck equation~(\ref{eq:fokker}) was obtained, which gives the PDF~(\ref{eq:Gaussian:Green}) for the coarse-grained variables $\boldsymbol{\Phi}$. Let us now start from a different set of canonical variables $\boldsymbol{v}$ related to $\boldsymbol{z}$ through \Eq{eq:canonical:transf}, and then apply the coarse-graining procedure to get the coarse-grained field $\widetilde{\boldsymbol{\Phi}}$ from $\boldsymbol{v}$. In \App{app:canonical}, it is shown that the PDF for $\widetilde{\boldsymbol{\Phi}}$ is also the solution of a Fokker-Planck equation, the associated Green function being given by
\bea
	\widetilde{\mathcal{W}}\left(\widetilde{\boldsymbol{\Phi}},\tau|\widetilde{\boldsymbol{\Phi}}_0,\tau_0\right)=\displaystyle\frac{1}{\sqrt{2\pi^2\det\left[\widetilde{\boldsymbol{\Sigma}}(\tau)\right]}}\exp\left\lbrace-\frac{1}{2}\left[\widetilde{\boldsymbol{\Phi}}-\widetilde{\boldsymbol{\Phi}}_{\mathrm{det}}(\tau)\right]^\dag\widetilde{\boldsymbol{\Sigma}}^{-1}(\tau)\left[\widetilde{\boldsymbol{\Phi}}-\widetilde{\boldsymbol{\Phi}}_{\mathrm{det}}(\tau)\right]\right\rbrace ,\nonumber \\ 
\eea
where one has defined $\widetilde{\boldsymbol{\Phi}}_{\mathrm{det}}=\widetilde{\boldsymbol{G}}(\tau,\tau_0)\widetilde{\boldsymbol{\Phi}}_0$ and $\widetilde{\boldsymbol{\Sigma}}=\int^\tau_{\tau_0}\dd s \widetilde{\boldsymbol{G}}(\tau,s)\widetilde{\boldsymbol{D}}(s)\widetilde{\boldsymbol{G}}^\dag(\tau,s)$, with $\widetilde{\boldsymbol{G}}(\tau,s)=\boldsymbol{M}(\tau)\boldsymbol{G}(\tau,s)\boldsymbol{M}^{-1}(s)$ and $\widetilde{\boldsymbol{D}}(\tau)=\boldsymbol{M}(\tau){\boldsymbol{D}}(\tau)\boldsymbol{M}^\dag(\tau)$.  If the initial conditions are properly mapped through the canonical transformations, $\widetilde{\boldsymbol{\Phi}}_0=\boldsymbol{M}(\tau_0){\boldsymbol{\Phi}}_0$, these relations show that
\bea
\label{eq:canonicalTransf:phidet}
	\widetilde{\boldsymbol{\Phi}}_{\mathrm{det}}=\boldsymbol{M}(\tau)\boldsymbol{\Phi}_{\mathrm{det}}
\eea
and that 
\bea
\label{eq:canonicalTransf:Sigma}
	\widetilde{\boldsymbol{\Sigma}}(\tau)=\boldsymbol{M}(\tau){\boldsymbol{\Sigma}}(\tau)\boldsymbol{M}^\dag(\tau) .
\eea
Since $\boldsymbol{M}$ has a unit determinant, this relation implies that $\det[\widetilde{\boldsymbol{\Sigma}}(\tau)]=\det[{\boldsymbol{\Sigma}}(\tau)]$. As a consequence, the gaussian Green function in the $\widetilde{\boldsymbol{\Phi}}$ phase space reads
\bea
	\widetilde{\mathcal{W}}\left(\widetilde{\boldsymbol{\Phi}},\tau|\widetilde{{\boldsymbol{\Phi}}}_0,\tau_0\right)=\displaystyle\frac{1}{\sqrt{2\pi^2\det\left[{\boldsymbol{\Sigma}}(\tau)\right]}}\exp\left\lbrace-\frac{1}{2}\left[{\boldsymbol{\Phi}}(\widetilde{\boldsymbol{\Phi}})-{\boldsymbol{\Phi}}_{\mathrm{det}}(\tau)\right]^\dag{\boldsymbol{\Sigma}}^{-1}(\tau)\left[{\boldsymbol{\Phi}}(\widetilde{\boldsymbol{\Phi}})-{\boldsymbol{\Phi}}_{\mathrm{det}}(\tau)\right]\right\rbrace ,\nonumber \kern-1.55em \\ 
\label{eq:canonical_transformation:PDF}
\eea
where ${\boldsymbol{\Phi}}(\widetilde{\boldsymbol{\Phi}})\equiv \boldsymbol{M}^{-1}(\tau)\widetilde{\boldsymbol{\Phi}}$. This implies that at each time, the Green function $\widetilde{\mathcal{W}}$ is mapped onto the Green function $\mathcal{W}$ simply by mapping the coarse-grained phase space coordinated by $\widetilde{\boldsymbol{\Phi}}$ onto the coarse-grained phase space coordinated by $\boldsymbol{\Phi}=\boldsymbol{M}^{-1}(\tau)\widetilde{\boldsymbol{\Phi}}$.

Let us finally determine how expectation values of functions defined on the coarse-grained phase space are related through canonical transformations. For analytic functions that can be written as convergent series of powers of the phase-space variables, one simply has to calculate the statistical moments of the phase-space variable. Using $\widetilde{\boldsymbol{\Phi}}$, these are given by integrals over the coarse-grained phase space of the form
\bea
	\left<\widetilde{ {\Phi}}_{i_1}\cdots\widetilde{ {\Phi}}_{i_n}\right>_{\widetilde{\mathcal{W}}}=\displaystyle\int\widetilde{ {\Phi}}_{i_1}\cdots\widetilde{ {\Phi}}_{i_n} \widetilde{\mathcal{W}}(\widetilde{\boldsymbol{\Phi}},\tau|{\boldsymbol{\Phi}}_0,\tau_0) \dd\widetilde{\boldsymbol{\Phi}},
\eea
where the subscript in the left-hand side specifies which PDF is used to perform the average. Under the canonical transformation~(\ref{eq:canonical_transformation:PDF}), this gives rise to
\bea
	\left<\widetilde{{\Phi}}_{i_1}\cdots\widetilde{ {\Phi}}_{i_n}\right>_{\widetilde{\mathcal{W}}}=\displaystyle\int\widetilde{ {\Phi}}_{i_1}\cdots\widetilde{ {\Phi}}_{i_n} {\mathcal{W}}({\boldsymbol{\Phi}},\tau|{\boldsymbol{\Phi}}_0,\tau_0) \dd{\boldsymbol{\Phi}},
\eea
where we have made use of $\dd\widetilde{\boldsymbol{\Phi}}=\det(\boldsymbol{M})\dd{\boldsymbol{\Phi}}$ with $\det(\boldsymbol{M})=1$, and where $\widetilde{\mathcal{W}}$ is mapped to $\mathcal{W}$ thanks to $\boldsymbol{\Phi}=\boldsymbol{M}^{-1}(\tau)\widetilde{\boldsymbol{\Phi}}$. The field variables are related through $\widetilde{ {\Phi}}_{i_1}\cdots\widetilde{ {\Phi}}_{i_n}=\sum_{j_1,\cdots j_n=1}^2 {M}^{-1}_{i_1j_1}\cdots {M}^{-1}_{i_nj_n}{ {\Phi}}_{i_1}\cdots{ {\Phi}}_{i_n}$. Because $\boldsymbol{M}$ is independent of the phase-space variables, it can be pulled out of the integral and one gets
\begin{align}
	\left\langle \widetilde{ {\Phi}}_{i_1}\cdots\widetilde{ {\Phi}}_{i_n}\right\rangle _{\widetilde{\mathcal{W}}}&=\displaystyle\sum_{j_1,\cdots j_n=1}^2 {M}^{-1}_{i_1j_1}\cdots {M}^{-1}_{i_nj_n}\displaystyle\int \Phi_{i_1}\cdots{ \Phi}_{i_n} {\mathcal{W}}({\boldsymbol{\Phi}},\tau \vert{\boldsymbol{\Phi}}_0,\tau_0)\dd{\boldsymbol{\Phi}} \nonumber \\
	&=\displaystyle\sum_{j_1,\cdots j_n=1}^2 {M}^{-1}_{i_1j_1}\cdots {M}^{-1}_{i_nj_n}\left\langle{ {\Phi}}_{i_1}\cdots{ {\Phi}}_{i_n}\right\rangle_{\mathcal{W}} .
\end{align}
This confirms that expectation values of phase-space variables functions are related through the same canonical transformations as the phase-space variables themselves.

These results show that the same predictions are obtained irrespectively of the choice of the canonical variables on which coarse graining is performed. This is at least the case for free fields and linear canonical transformations that are homogeneous in phase space. Let us note that this is at odd with the conclusions drawn in \Refc{PhysRevD.46.2408} where the field correlation functions are expanded in $\sigma$ and where it is noted that higher order terms in $\sigma$ depend on the choice of canonical variables. The reason is that the irrelevance of the choice of canonical variables is valid for the full, exact solutions of the Hamilton's equations. If one expands these solutions in terms of a small parameter, say $\sigma$, there is no guarantee that the correspondence will be verified order by order. More fundamentally anyway, in \Sec{sec:quantum_to_classical} we have shown that the stochastic formalism cannot be trusted beyond the leading order in $\sigma$, where  the full quantum theory must be used instead.  
\subsection{Explicit solution}
\label{sec:explicit}
Let us now solve the mode equations~(\ref{eq:eomphiq}) and~(\ref{eq:eompiq}) explicitly and carry out the computational program sketched in \Sec{ssec:fokker}.
\subsubsection{Massless field on a de-Sitter background}
\label{sec:massless}
We first consider the case of a massless test field evolving on a de-Sitter background where $a=-1/(H\eta)$. In terms of the conformal time $\eta$, the mode equations~(\ref{eq:eomphiq}) and~(\ref{eq:eompiq}) give rise to
\bea
\left( a \phi_k \right)^{\prime\prime}+\left(k^2-\frac{2}{\eta^2}\right)\left( a \phi_k \right) = 0 .
\eea
The solution to this equation satisfying the Klein-Gordon normalisation condition given below \Eq{eq:eompiq} reads\footnote{This solution is such that  in the remote past, \ie when $k\eta\rightarrow -\infty$, $a\phi_k=e^{ik\eta}/\sqrt{2k}$, which corresponds to the so-called Bunch-Davies vacuum.}
\bea
\phi_k = \frac{1}{a\sqrt{2k}} \left(1+\frac{i}{k\eta}\right)\ee^{i k \eta}  .
\eea 
From \Eq{eq:eomphiq},  the conjugated momentum is given by $\pi_\phi = a^2 \phi^\prime$, which leads to
\bea
\pi_k = a\sqrt{\frac{k}{2}} i \ee^{i k \eta} .
\eea
Making use of \Eqs{eq:ksigma} and~(\ref{eq:noisecorrel}), the symmetric part $\boldsymbol{D}$ of the noise correlator matrix $\boldsymbol{\Xi}$ is then given by
\bea
\label{eq:noisecorrelator:massless}
	\boldsymbol{D}=a\displaystyle\left(\begin{array}{cc}
		\dfrac{H^3\left(1+\sigma^2\right)}{4\pi^2} & \dfrac{H\sigma^2}{4\pi^2\eta^3} \\
		\dfrac{H\sigma^2}{4\pi^2\eta^3} & \dfrac{\sigma^4}{4H\pi^2\eta^6}
	\end{array}\right)
		 .
\eea
On the other hand, the homogeneous equation for the field is given by
\bea
\left( a \bar{\phi} \right)^{\prime\prime}-\frac{2}{\eta^2}\left( a \bar{\phi} \right) = 0
\eea
and has two independent solutions, $a \bar{\phi}^{(1)} \propto 1/\eta$ and $a \bar{\phi}^{(2)} \propto \eta^2 $. These solutions allow one to introduce the fundamental matrix $\boldsymbol{U}$ defined in \App{app:green},
\bea
\label{eq:U:massless}
\boldsymbol{U}(\eta)=
\left(\begin{array}{ccc}
	H &\ & \dfrac{H}{3}\eta^3  \\
	0 &\ & \dfrac{1}{H}
	\end{array}\right) ,
\eea
which then gives rise to the Green matrix
\bea
	\boldsymbol{G}(\eta,\eta_0)&=&
	\left(\begin{array}{ccc}
	1 & \ &\dfrac{H^2}{3}\left(\eta^3-\eta_0^3\right) \\
	0 &\  & 1
\end{array}\right) ,
\eea
see \Eq{eq:Green:fundamental}. From here, the covariance matrix~(\ref{eq:Sigma}) can be calculated, 
\bea
\boldsymbol{\Sigma}(\eta) = 
\left(\begin{array}{ccc}
\dfrac{H^2}{4\pi^2} \ln\left(\dfrac{\eta_0}{\eta}\right) &\ & \dfrac{\sigma^2}{12\pi^2}\left(\dfrac{1}{\eta^3}-\dfrac{1}{\eta_0^3}\right) \\
\dfrac{\sigma^2}{12\pi^2}\left(\dfrac{1}{\eta^3}-\dfrac{1}{\eta_0^3}\right) &\ &  \dfrac{\sigma^4}{24\pi^2 H^2}\left(\dfrac{1}{\eta^6}-\dfrac{1}{\eta_0^6}\right)
\end{array}\right) .
\label{eq:Sigma:massless}
\eea
%\bea
%\boldsymbol{\Sigma}_{\phi,\phi} &=& \frac{H^2}{4\pi^2} \left\lbrace\left(1+\frac{\sigma^2}{3}+\frac{\sigma^4}{9}\right)\ln\left(\frac{\eta_0}{\eta}\right)+\frac{\sigma^2}{54}\left(1-\frac{\eta^3}{\eta_0^3}\right)\left[12+\left(\frac{\eta^3}{\eta_0^3}-3\right)\sigma^2\right]\right\rbrace\\
%\boldsymbol{\Sigma}_{\phi,\pi} &=& \boldsymbol{\Sigma}_{\pi,\phi}
%= \frac{\sigma^2}{12\pi^2\eta_0^3}\left(\frac{\eta_0^3}{\eta^3}-1\right)
%\left[1+\left(\frac{\eta^3}{\eta_0^3}-1\right)\frac{\sigma^2}{6}\right]\\
%\boldsymbol{\Sigma}_{\pi,\pi} &=& \frac{\sigma^4}{24\pi^2 H^2}\left(\frac{1}{\eta^6}-\frac{1}{\eta_0^6}\right)
%\eea
In this expression, according to the considerations of \Sec{sec:quantum_to_classical}, only the leading order terms in $\sigma$ have been kept. In the limit $\sigma\rightarrow 0$, one has $ {\Sigma}_{\phi,\pi}= {\Sigma}_{\pi,\phi}\simeq  {\Sigma}_{\pi,\pi} \simeq 0$, and quantum diffusion takes place in the $\phi$ direction only, growing as the logarithm of the scale factor.
\subsubsection{Free field on a slow-roll background}
\label{sec:light}
Let us now see how these results generalise to a test field with mass $m$ on a slow-roll inflationary background, 
\bea
\label{eq:a:SR}
a=-\dfrac{1}{\eta H_*}\left[1+\epsilon_{1*}-\epsilon_{1*}\ln\left(\frac{\eta}{\eta_*}\right)\right] ,
\eea
where $\epsilon_1$ denotes the first slow-roll parameter introduced in \Sec{sec:SR} and a star denotes the time around which the slow-roll expansion is performed (for instance, one can take $\eta_*=\eta_0$). Note that \Eq{eq:a:SR} corresponds to a first-order expansion in slow roll of the background dynamics (see for instance Eq.~(2.9) of \Refc{Martin:2013uma}), but no assumption is made regarding the (slow-roll or non-slow-roll) dynamics of the test field, which will be further investigated in \Sec{sec:SRattractor}. The mode equations~(\ref{eq:eomphiq}) and~(\ref{eq:eompiq}) give rise to
\bea
\label{eq:mode:light}
	\left(a\phi_k\right)^{\prime\prime}+\left(k^2-\dfrac{2+3\epsilon_{1*}-\frac{m^2}{H_*^2}}{\eta^2}\right)\left(a\phi_k\right)=0 .
\eea
Requiring the Klein-Gordon product normalisation condition again, this equation is solved by
\bea
\label{eq:modesolution:light}
\phi_k = \dfrac{\sqrt{\pi}}{2 a \sqrt{k}}\sqrt{-k\eta}e^{i\frac{\pi}{4}+i\nu\frac{\pi}{2}}
H^{(1)}_\nu\left(-k\eta\right) ,
\eea
where $H^{(1)}_\nu$ is the Hankel function of the first kind with index 
\bea
\label{eq:nu:def}
\nu \equiv \frac{3}{2}\sqrt{1-\frac{4m^2}{9H_*^2}+\frac{4}{3}\epsilon_{1*}} .
\eea
One can check that at leading order in background slow roll, the only effect of $\epsilon_{1*}$ is to change the effective mass of the field perturbations through \Eq{eq:nu:def}. For the conjugated momentum, one obtains
\bea
\label{eq:conjugated_momentum:light}
\pi_k =  a \frac{\ee^{i\frac{\pi}{4}+i\nu\frac{\pi}{2}}}{4}\sqrt{\frac{\pi}{-\eta}}\left[2 k \eta H^{(1)}_{\nu-1}\left(-k\eta\right) - \left(3-2\nu+2\epsilon_{1*}\right) H^{(1)}_\nu\left(-k\eta\right) \right] .
\eea
Making use of \Eqs{eq:ksigma} and~(\ref{eq:noisecorrel}), the components of the symmetric part $\boldsymbol{D}$ of the noise correlator matrix $\boldsymbol{\Xi}$ can then be expanded in $\sigma\ll 1$. After a lengthy but straightforward calculation, one obtains
\bea
 {D}_{\phi,\phi}&=&
 \frac{a H^3 \Gamma^2(\nu)}{\pi^3 \left(1+3\epsilon_{1*}\right)}\left[\frac{\sigma}{2}\left(1+\epsilon_{1*}\right)\right]^{3-2\nu}
 \Bigg\lbrace
 1+\frac{2}{\nu-1}\left[\frac{\sigma}{2}\left(1+\epsilon_{1*}\right)\right]^2 
 \nonumber \\ & &
 -\frac{2\pi}{\nu \tan(\pi\nu) \Gamma^2(\nu)}\left[\frac{\sigma}{2}\left(1+\epsilon_{1*}\right)\right]^{2\nu}
 \nonumber \\ & & 
 +\frac{2\nu-3}{(\nu-2)(\nu-1)^2}\left[\frac{\sigma}{2}\left(1+\epsilon_{1*}\right)\right]^4
\Bigg\rbrace
 +\order{\sigma^5} ,
\label{eq:massive:Dphiphi}
\\
 {D}_{\pi,\pi}&=&
 \frac{a^7 H^5 \Gamma^2(\nu)}{4\pi^3\left(1+5\epsilon_{1*}\right)}\left[\frac{\sigma}{2}\left(1+\epsilon_{1*}\right)\right]^{3-2\nu}
 \Bigg\lbrace
 \left(3-2\nu+2\epsilon_{1*}\right)^2
   \nonumber \\ & &
  +\frac{2}{\nu-1}\left(2\nu-7-2\epsilon_{1*}\right)\left(2\nu-3-2\epsilon_{1*}\right)\left[\frac{\sigma}{2}\left(1+\epsilon_{1*}\right)\right]^2
   \nonumber \\ & & 
 +\frac{2\pi \left(2\nu-3-2\epsilon_{1*}\right)\left(2\nu+3+2\epsilon_{1*}\right)}{\nu\tan(\pi\nu)\Gamma^2(\nu)}\left[\frac{\sigma}{2}\left(1+\epsilon_{1*}\right)\right]^{2\nu}
 \label{eq:massive:Dpipi}
 \\ & & 
    +\frac{8\nu^3-\left(68+16\epsilon_{1*}\right)\nu^2+\left(166+80\epsilon_{1*}\right)\nu-131-84\epsilon_{1*}}{\left(\nu-1\right)^2\left(\nu-2\right)}\left[\frac{\sigma}{2}\left(1+\epsilon_{1*}\right)\right]^{4}
 \Bigg\rbrace
 +\order{\sigma^5} ,
 \nonumber 
\\
 {D}_{\phi,\pi}&=& {D}_{\pi,\phi} =
 \frac{a^4 H^4 \Gamma^2(\nu)}{2\pi^3 \left(1+4\epsilon_{1*}\right)}\left[\frac{\sigma}{2}\left(1+\epsilon_{1*}\right)\right]^{3-2\nu}
\Bigg\lbrace
 \left(2\nu-3-2\epsilon_{1*}\right)
   \nonumber \\ & & 
 +\frac{2}{\nu-1}\left(2\nu-5-2\epsilon_{1*}\right)\left[\frac{\sigma}{2}\left(1+\epsilon_{1*}\right)\right]^2
 +\frac{2\pi\left(3+2\epsilon_{1*}\right)}{\tan(\pi\nu)\Gamma^2(\nu)\nu}\left[\frac{\sigma}{2}\left(1+\epsilon_{1*}\right)\right]^{2\nu}
 \nonumber \\ & & 
 +\frac{\left(2\nu-3\right)\left(2\nu-7-2\epsilon_{1*}\right)}{\left(\nu-2\right)\left(\nu-1\right)^2}\left[\frac{\sigma}{2}\left(1+\epsilon_{1*}\right)\right]^4
\Bigg\rbrace
 +\order{\sigma^{5}} .
\label{eq:massive:Dphipi}
\eea
This expansion in $\sigma$ has been ordered under the assumption that $1<\nu<2$, which amounts to $ -7/4<m^2/H_*^2-3\epsilon_{1*}<5/4$. In practice, only the leading terms in $\sigma$ must be kept in order to be consistent with the stochastic classical approximation as explained in \Sec{sec:quantum_to_classical}. In \Eqs{eq:massive:Dphiphi}-(\ref{eq:massive:Dphipi}) however, the first four terms of the expansion are displayed to make clear that the massless de-Sitter case of \Sec{sec:massless} is recovered in the limit $\nu=3/2$ and $\epsilon_{1*}=0$ (the non-dominant terms will be dropped in what follows). In the noise correlators involving the conjugated momentum $\pi$ indeed, one can see that the leading-order contributions vanish when $\nu=3/2$ and $\epsilon_{1*}=0$. For example, in ${D}_{\pi,\pi}$ given by \Eq{eq:massive:Dpipi}, the first three terms vanish when $\nu=3/2$ and $\epsilon_{1*}=0$ and one has to go to fourth order to recover the $\pi,\pi$ component of \Eq{eq:noisecorrelator:massless}. As a consequence, the inclusion of a small mass or of a small departure from de Sitter does not only slightly modify the coefficients of the noise density matrix. It introduces new, lower order contributions in $\sigma$ that make all entries of $\boldsymbol{D}$ non-vanish in the limit $\sigma\rightarrow 0$, contrary to the massless de-Sitter case.

The homogeneous equation for the field is given by 
\bea
\label{eq:eom:classical:light}
	\left(a\bar{\phi}\right)^{\prime\prime}-\dfrac{2+3\epsilon_{1*}-\frac{m^2}{H_*^2}}{\eta^2}\left(a\bar{\phi}\right)=0
\eea
and has two independent solutions, $a\bar{\phi}^{(1)}\propto \left(-\eta\right)^{\frac{1}{2}-\nu}$ and $a\bar{\phi}^{(2)}\propto \left(-\eta\right)^{\frac{1}{2}+\nu}$, from which the fundamental matrix
\bea
\label{eq:U:massive}
\kern-2em \boldsymbol{U}(\eta)=
\left(\begin{array}{ccc}
	\dfrac{(-\eta)^{\frac{1}{2}-\nu}}{a}(-\eta_0)^{\nu-\frac{3}{2}}
	 &\ & 
	 \dfrac{(-\eta)^{\frac{1}{2}+\nu}}{a}(-\eta_0)^{-\nu-\frac{3}{2}}
	 \\
	a\left(\nu-\frac{3}{2}-\epsilon_{1*}\right)(-\eta)^{-\frac{1}{2}-\nu}(-\eta_0)^{\nu-\frac{3}{2}}
	 &\ &
	 -a\left(\nu+\frac{3}{2}+\epsilon_{1*}\right)(-\eta)^{-\frac{1}{2}+\nu}(-\eta_0)^{-\nu-\frac{3}{2}}
	\end{array}\right)
\eea
can be constructed as explained in \App{app:green}. This gives rise to the Green matrix
\bea
G_{\phi,\phi}(\eta,\eta_0) &=&\frac{3+2\nu+2\epsilon_{1*}}{4\nu}\left(\frac{\eta}{\eta_0}\right)^{\frac{3}{2}-\nu+\epsilon_{1*}}
-\frac{3-2\nu+2\epsilon_{1*}}{4\nu}\left(\frac{\eta}{\eta_0}\right)^{\frac{3}{2}+\nu+\epsilon_{1*}},\\
G_{\phi,\pi}(\eta,\eta_0) &=& \frac{-\eta_0}{2\nu a^2\left(\eta_0\right)}\left[\left(\frac{\eta}{\eta_0}\right)^{-\nu+\frac{3}{2}+\epsilon_{1*}}-\left(\frac{\eta}{\eta_0}\right)^{\nu+\frac{3}{2}+\epsilon_{1*}}\right],\\
G_{\pi,\phi}(\eta,\eta_0) &=& \frac{a^2\left(\eta_0\right)}{-8\nu\eta_0}\left(9+12\epsilon_{1*}-4\nu^2\right)\left[\left(\frac{\eta}{\eta_0}\right)^{\nu-\frac{3}{2}-\epsilon_{1*}}-\left(\frac{\eta}{\eta_0}\right)^{-\nu-\frac{3}{2}-\epsilon_{1*}}\right],\\
G_{\pi,\pi}(\eta,\eta_0) &=& \frac{3+2\nu+2\epsilon_{1*}}{4\nu}\left(\frac{\eta}{\eta_0}\right)^{\nu-\frac{3}{2}-\epsilon_{1*}}
-\frac{3-2\nu+2\epsilon_{1*}}{4\nu}\left(\frac{\eta}{\eta_0}\right)^{-\nu-\frac{3}{2}-\epsilon_{1*}},
\eea
from which the diffusion matrix~(\ref{eq:Sigma}) can be obtained,
\bea
\label{eq:Sigma:light}
\kern-2em
\boldsymbol{\Sigma}(\eta) =
\left(\frac{\sigma}{2}\right)^{3-2\nu}
\frac{\Gamma^2\left(\nu\right)}{\pi^3}
\left[1-\left(\frac{\eta}{\eta_0}\right)^{3-2\nu}\right]
	\left(\begin{array}{ccc}
		\frac{1+(1-2\nu)\epsilon_{1*}}{3-2\nu}H^2 &\ & 
		\frac{1+\left(\frac{2}{3-2\nu}+3-2\nu\right)\epsilon_{1*}}{2\eta^3} \\
		\frac{1+\left(\frac{2}{3-2\nu}+3-2\nu\right)\epsilon_{1*}}{2\eta^3} &\ & \frac{3-2\nu+\left(19-16\nu+4\nu^2\right)\epsilon_{1*}}{4 H^2\eta^6}
	\end{array}\right) .
\eea
In this expression, only the leading-order contributions in $\sigma$ have been kept since only these terms are expected to be correctly described in the stochastic classical approximation. The dependence on $\sigma$ only appears through the overall $(\sigma/2)^{3-2\nu}$ factor, which can be made $\sigma$-independent (and equal to one) if $3-2\nu$ is close enough to $0$, \ie if
\bea
\label{eq:condition:m_over_H}
\ee^{-\frac{1}{\left\vert 3-2\nu \right\vert }}\ll \dfrac{\sigma}{2}.
\eea
This condition matches Eq.~(81) of \Refc{Starobinsky:1994bd}. Let us also note that, from \Eq{eq:modesolution:light}, $3-2\nu$ is related to the spectral index $\nS -1 = \dd \ln \calP_{ a\delta\phi}/\dd \ln k$ of the power spectrum $\calP_{a \delta\phi} = k^3 \vert  a \delta\phi_k \vert^2/(2\pi^2)$ of the field fluctuations $a\delta\phi$ through $3-2\nu = \nS-1$.  The condition~(\ref{eq:condition:m_over_H}) therefore means that for a fixed comoving wavenumber, the amplitude of the field fluctuations should not vary much between the Hubble radius crossing time and the coarse-graining radius crossing time. It is compatible with the classical transition condition~(\ref{eq:cond:classicalTransition}), $\sigma\ll \vert 3-2\nu \vert^{1/3}$, if $\nu$ is sufficiently close to $3/2$ (for instance, if $\vert 3-2\nu\vert <0.1$, there are already 4 orders of magnitude between the two bounds). This will be further discussed in \Sec{sec:BeyondFreeFields}, but for now \Eq{eq:condition:m_over_H} allows one to expand \Eq{eq:Sigma:light} at leading order in $\nu-3/2\simeq \epsilon_{1*}-m^2/(3H_*^2)$, where one obtains
\bea
\label{eq:Sigma:light:lightlimit}
\boldsymbol{\Sigma}(\eta) = 
\frac{1-\left(\frac{\eta}{\eta_0}\right)^{3-2\nu}}{3-2\nu}
\dfrac{H^2}{4\pi^2}
	\left(\begin{array}{ccc}
		1 &\ & \dfrac{3-2\nu+2\epsilon_{1*}}{2H_*^2\eta^{3}} \\
		\dfrac{3-2\nu+2\epsilon_{1*}}{2H_*^2\eta^{3}}  &\ & \dfrac{(3-2\nu+2\epsilon_{1*})^2}{4H_*^4 \eta^6}
	\end{array}\right) .
\eea
It is interesting to notice that $\epsilon_{1*}$ disappears from the combination $3-2\nu+2\epsilon_{1*}$ when expanded at leading order. Therefore, at the order at which the calculation is performed, the direction along which the phase-space distribution is elongated is not affected by slow-roll corrections.
Compared to the massless case in de Sitter, one can see that the coarse-grained field now diffuses in the momentum $\pi$ direction as the cubic power of the scale factor. In the $\phi$ direction, at early time, when $\log(\eta_0/\eta)\ll 1/\vert 3-2\nu\vert$, one obtains $\Sigma_{\phi,\phi}\simeq H^2/(4\pi^2) \log(\eta_0/\eta)$ which coincides with the massless case~(\ref{eq:Sigma:massless}). At late time however, when $\log(\eta_0/\eta)\gg 1/\vert 3-2\nu \vert $, if $\nu<3/2$, it asymptotes to the equilibrium value $\Sigma_{\phi,\phi}\simeq H^2/[4\pi^2(3-2\nu)] \simeq 3H^4/(8\pi^2m^2)$ (where the second expression is valid for a light field with positive squared mass in de-Sitter~\cite{Starobinsky:1994bd}); while if $\nu>3/2$, it continues to increase as $\Sigma_{\phi,\phi}\simeq H^2/[4\pi^2(2\nu-3)] (\eta/\eta_0)^{3-2\nu}\simeq 3H^4/(8\pi^2\vert m^2\vert) (\eta_0/\eta)^{2\vert m^2\vert/(3H^2)}$ (where the second expression is valid for a light field with negative squared mass in de-Sitter).
\section{Is slow roll a stochastic attractor?}
\label{sec:SRattractor}
We are now in a position where we can address the main question raised in this article and investigate whether or not slow roll, \ie the classical attractor, is still an attractor of the stochastic theory. We first study the case of a free scalar field for which the phase-space PDF was obtained in \Sec{sec:testField}, before extending the discussion to other types of fields (including the inflaton field) in \Sec{sec:BeyondFreeFields}. 

For a free scalar field, in \Sec{sec:light}, it was shown that the equation of motion~(\ref{eq:eom:classical:light}) for the classical evolution of the homogeneous field has two independent solutions,
\bea
\label{eq:def:PhiSR}
	\boldsymbol{\Phi}_\sr=
	\left(\begin{array}{c}
		\phi_\sr \\
		\pi_\sr
	\end{array}\right)= (-\eta_0)^{\nu-\frac{3}{2}} \left(\begin{array}{c}
		\displaystyle \frac{(-\eta)^{\frac{1}{2}-\nu}}{a}  \\
		\displaystyle a \left(\nu-\frac{3}{2}-\epsilon_{1*}\right)(-\eta)^{-\frac{1}{2}-\nu}
			\end{array}\right)
\eea
and
\bea
\label{eq:def:PhiNSR}
	\boldsymbol{\Phi}_\nsr=\left(\begin{array}{c}
		\phi_\nsr \\
		\pi_\nsr
	\end{array}\right)= (-\eta_0)^{-\nu-\frac{3}{2}}
	\left(\begin{array}{c}
		\displaystyle  \frac{(-\eta)^{\frac{1}{2}+\nu}}{a}  \\
		\displaystyle -a \left(\nu+\frac{3}{2}+\epsilon_{1*}\right)(-\eta)^{-\frac{1}{2}+\nu} 
	\end{array}\right) ,
\eea
where the overall powers of $\eta_0$ are introduced for later convenience.
The subscripts ``SR'' and ``NSR'' stand for ``slow roll'' and ``non slow roll'' respectively since $\boldsymbol{\Phi}_\sr$ and $\boldsymbol{\Phi}_\nsr$ correspond to the solutions discussed around \Eqs{eq:phidotdotSR} and~(\ref{eq:phidotdotNSR}). The attractor direction of the classical phase-space dynamics therefore corresponds to the SR direction. The solutions $\boldsymbol{\Phi}_\sr$ and $\boldsymbol{\Phi}_\nsr$ also match the ones given below \Eq{eq:eom:classical:light} (where $\bar{\phi}^{(1)}$ needs to be identified with $\phi_\sr$ and $\bar{\phi}^{(2)}$ with $\phi_\nsr$), so that the fundamental matrix defined in \Eq{eq:U:massive} simply reads $\boldsymbol{U}(\eta)=\left(\boldsymbol{\Phi}_\sr,~\boldsymbol{\Phi}_\nsr\right)$. 
\subsection{Diffusion in the interaction picture}
\label{sec:InteractionPicture}
In the covariance matrix~(\ref{eq:Sigma:light:lightlimit}) of the free coarse-grained field, one can note that $ {\Sigma}_{\phi,\pi}=-\sqrt{ {\Sigma}_{\phi,\phi} {\Sigma}_{\pi,\pi}}$, implying that the noises in the $\phi$ and $\pi$ directions are totally anticorrelated. This also means that $\det(\boldsymbol{\Sigma})=0$, hence the covariance matrix has one non-zero eigenvalue and one vanishing eigenvalue, respectively defining a first direction where diffusion occurs, and a second non-diffusive direction. The question is how these two directions relate to the attractor (SR) and anti-attractor (NSR) directions. In order to establish this relation, let us formulate the stochastic dynamics in terms of canonical variables that are aligned with the SR and NSR solutions of the homogeneous problem, which we call the interaction picture.

Any solution of the homogeneous problem can be expressed as $\boldsymbol{\Phi}_\mathrm{det}=z_\sr\boldsymbol{\Phi}_\sr+z_\nsr \boldsymbol{\Phi}_\nsr$, where $z_\sr$ and $z_\nsr$ are two constants. If the solution satisfies the initial value problem $\boldsymbol{\Phi}_\mathrm{det}(\eta_0)=\boldsymbol{\Phi}_0$, these constants are given by $\boldsymbol{z}=(z_\sr,z_\nsr)^\dagger=\boldsymbol{U}^{-1}(\eta_0)\boldsymbol{\Phi}_0$. They are also formally obtained from the Wronskian
\bea
\label{eq:def:zsr}
	z_\sr=\boldsymbol{\Phi}^\dag\boldsymbol{\Omega}\boldsymbol{\Phi}_\nsr, \\
	z_\nsr=-\boldsymbol{\Phi}^\dag\boldsymbol{\Omega}\boldsymbol{\Phi}_\sr,
\label{eq:def:znsr}
\eea
where $\boldsymbol{\Omega}$ has been defined in \Eq{eq:def:Omega} and where the subscript ``det'' has been dropped to let $\boldsymbol{z}$ describe a generic parametrisation of phase space. Note that the SR constant is obtained by projecting the general solution on the NSR branch, and vice versa. This means that the set of classical solutions can be parametrised by the constants $\boldsymbol{z}$. Combining \Eqs{eq:Phidet} and~(\ref{eq:Green:fundamental}), the link between $\boldsymbol{z}$ and $\boldsymbol{\Phi}$ can also be written $\boldsymbol{\Phi}  = \boldsymbol{U}(\eta)\boldsymbol{z} $. Since $\boldsymbol{U}(\eta)$ is a $(2\times2)$ real matrix with unit determinant, as explained in footnote~\ref{footnote:symplectic}, it is symplectic. It thus defines a linear and homogeneous canonical transformation that allows us to make use of the results of \Sec{sec:canonical}. 

In the set of canonical variables $\boldsymbol{z}$, the classical dynamics is simply frozen,\footnote
{This can be shown using \Eqs{eq:canonicalTransf:phidet},~(\ref{eq:Phidet}) and~(\ref{eq:Green:fundamental}), which give rise to
 \bea
\boldsymbol{z}_\mathrm{det}(\eta)&=&\boldsymbol{U}^{-1}(\eta)\boldsymbol{\Phi}_\mathrm{det}(\eta)=\boldsymbol{U}^{-1}(\eta)\boldsymbol{G}(\eta,\eta_0)\boldsymbol{\Phi}_\mathrm{det}(\eta_0) 
 \nonumber \\&=&
\boldsymbol{U}^{-1}(\eta)\boldsymbol{U}(\eta)\boldsymbol{U}^{-1}(\eta_0)\boldsymbol{\Phi}_\mathrm{det}(\eta_0)=\boldsymbol{U}^{-1}(\eta_0)\boldsymbol{\Phi}_\mathrm{det}(\eta_0)=
\boldsymbol{z}_\mathrm{det}(\eta_0). 
\eea} 
\bea
\boldsymbol{z}_\mathrm{det}(\eta) = \boldsymbol{z}_\mathrm{det}(\eta_0) ,
\eea
and the averaged trajectory reduces to a single point in phase space. The deterministic part of the dynamics thus factors out and only diffusion remains, hence the name ``interaction picture''. Since the first and second entries of $\boldsymbol{z}$ correspond respectively to the attractor and anti-attractor branches,  the averaged trajectory asymptotes to the attractor solution unless $z_{\mathrm{det}}^{\sr}=0$, and the attractor branch dominates the dynamics at times $\eta\gg -\vert z_{\mathrm{det}}^{\sr}/z_{\mathrm{det}}^{\nsr} \vert^{1/(2\nu)}$. The covariance matrix is given by the integral over time of the diffusion matrix\footnote
{This can be shown combining \Eqs{eq:canonicalTransf:Sigma},~(\ref{eq:Sigma}),~(\ref{eq:Green:fundamental}) and the formula given above \Eq{eq:canonicalTransf:phidet} that relates $\boldsymbol{D}_{\boldsymbol{z}}$ and $\boldsymbol{D}$, which give rise to
\bea
\boldsymbol{\Sigma}_{\boldsymbol{z}}(\eta)&=&
\boldsymbol{U}^{-1}(\eta) \boldsymbol{\Sigma}(\eta) \left[\boldsymbol{U}^{-1}(\eta)\right]^{\dag}=
\boldsymbol{U}^{-1}(\eta) \displaystyle\int^{\eta}_{\eta_0}\dd s\, \boldsymbol{G}(\eta,s)\boldsymbol{D}(s)\boldsymbol{G}^\dag(\eta,s) \left[\boldsymbol{U}^{-1}(\eta)\right]^{\dag}\nonumber \\
&=&\boldsymbol{U}^{-1}(\eta)\displaystyle\int^{\eta}_{\eta_0}\dd s\, \boldsymbol{U}(\eta)\boldsymbol{U}^{-1}(s)\boldsymbol{D}(s)\left[\boldsymbol{U}(\eta)\boldsymbol{U}^{-1}(s)\right]^\dag\left[\boldsymbol{U}^{-1}(\eta)\right]^{\dag}\nonumber \\
&=&\displaystyle\int^{\eta}_{\eta_0}\dd s\,\boldsymbol{U}^{-1}(s)\boldsymbol{D}(s)\left[\boldsymbol{U}^{-1}(s)\right]^{\dag}=
\displaystyle\int^{\eta}_{\eta_0}\dd s\,\boldsymbol{D}_{\boldsymbol{z}}(s) .
\eea
}
\bea
\boldsymbol{\Sigma}_{\boldsymbol{z}}(\eta)=\displaystyle\int^{\eta}_{\eta_0}\dd s\,\boldsymbol{D}_{\boldsymbol{z}}(s) ,
\eea
\ie it is simply the noise power cumulated over time. Combining \Eqs{eq:canonicalTransf:Sigma} and~(\ref{eq:Sigma:light:lightlimit}), it reads
\bea
\label{eq:Sigma:z}
\boldsymbol{\Sigma}_{\boldsymbol{z}}(\eta)=
\frac{1-\left(\frac{\eta}{\eta_0}\right)^{2\nu-3}}{2\nu-3}\frac{1+2\epsilon_{1*}}{\left(2\pi\right)^2}
\left(\begin{array}{cc}
1 &0 \\
0  &0
\end{array}\right) .
\eea
The relationship between this expression and the diagonalisation of the covariance matrix~(\ref{eq:Sigma:light:lightlimit})  in standard phase-space coordinates is elaborated on in \App{app:orthonormalisation}. For now, it is important to note that \Eq{eq:Sigma:z} clearly entails that diffusion takes place in the SR direction only, and that the NSR direction remains deterministic. As will be carefully shown in \Sec{sec:attractiveness}, this implies that the classical attractor generalises to a stochastic attractor of the free-field theory.

One may also notice that the late-time behaviour of the covariance matrix depends on the sign of $2\nu-3$ in a way that is opposite to what was observed for $\boldsymbol{\Sigma}_{\phi\phi}$ in \Eq{eq:Sigma:light:lightlimit}. Indeed, $\boldsymbol{\Sigma}_{\boldsymbol{z}}$ asymptotes to a constant value when $\nu>3/2$ while it continues to increase if $\nu<3/2$. This reversal is due to the squeezing induced by the matrix $\boldsymbol{U}$ that maps the field picture to the interaction picture, and that adds up to the one induced by dynamical evolution in the field picture.

This result can be reformulated by noticing that, for free fields (and only for free fields), the equation of motion of field perturbations~(\ref{eq:mode:light}) on super-Hubble scales $k\ll -1/\eta$ is the same as the one for the background~(\ref{eq:eom:classical:light}). The use of the interaction picture is again convenient to make this argument, since it can be extended down to the quantum fluctuations. Using vectorial notation $\boldsymbol{\Phi}_k=(\phi_k,\pi_k)^\mathrm{T}$, their dynamics is given by \Eq{eq:mode:light}, namely
\bea
\label{eq:eom:pert:vectorial}
\boldsymbol{\Phi}'_k=\left(\boldsymbol{A}+\boldsymbol{V}_k\right)\boldsymbol{\Phi}_k ,
\eea
where $\boldsymbol{A}$ has been defined in \Eq{eq:inhomopb} and corresponds to the deterministic evolution of the coarse-grained field, to which the additional potential
\bea
	\boldsymbol{V}_k=\left(\begin{array}{cc}
	0 & 0 \\
	-a^2(\eta)k^2 & 0
	\end{array}\right)
\eea
is added. Under the canonical transformation $\boldsymbol{\Phi}_k=\boldsymbol{U}(\eta)\boldsymbol{z}_k$, making use of the relation $\boldsymbol{U}^\prime = \boldsymbol{AU}$ given below \Eq{eq:U:def}, \Eq{eq:eom:pert:vectorial} reads 
\bea
\label{eq:eom:pert:z}
\boldsymbol{z}'_k=\boldsymbol{U}^{-1}\boldsymbol{V}_k\boldsymbol{U}\boldsymbol{z}_k .
\eea
This shows that the canonical transformation sending to the interaction picture of the classical, deterministic dynamics is also sending to the interaction picture of the quantum fluctuations and hence to the interaction picture of the stochastic, coarse-grained dynamics.

In this interaction picture, the solution~(\ref{eq:modesolution:light}) of \Eq{eq:eom:pert:z} is such that, at leading order in the coarse-graining parameter $\sigma$, $\boldsymbol{z}_k\propto (1,0)^\dagger$. This implies that in the super-Hubble limit, quantum fluctuations are highly squeezed along the attractor branch of the classical theory. The reason is that on super-Hubble scales, the interaction potential $\boldsymbol{V}_k$ becomes subdominant so that the equation of motion of perturbations matches the one for the background. In the asymptotic future, perturbations therefore confine to the classical attractor and quantum diffusion takes place in phase space along this attractor direction only.
\subsection{Stochastic attractor of free fields}
\label{sec:attractiveness}
In \Sec{sec:InteractionPicture}, we have shown that quantum diffusion takes place in phase space along the slow-roll classical attractor  only. This implies that, if one starts on the attractor, the subsequent stochastic dynamics remains confined to the attractor at any later time. If initial conditions are displaced from the attractor however, the stochastic dynamics explores regions of phase space that are inaccessible to the classical dynamics, and that may lie outside the slow-roll domain. This raises two questions that we address in the section: for which initial conditions does slow roll generalise to a stochastic attractor; and when this is the case, how much time does it take to relax towards slow roll and how does it compare to the classical situation?

Before answering these two questions, let us formulate them in more quantitative terms. Since quantum diffusion takes place along the SR direction only, in the interaction picture, solutions of the stochastic dynamics can be written as
\bea
\label{eq:decom:stoch:interaction}
\boldsymbol{\Phi}(\eta)=\widehat{z}_\sr\left(\eta\right)\boldsymbol{\Phi}_\sr+z_\nsr^{(0)}\boldsymbol{\Phi}_\nsr .
\eea
In this expression, $\widehat{z}_\sr$ is a Gaussian random variable with mean equal to $z_\sr(\eta_0)$ that we denote $z_\sr^{(0)}$ for simplicity, and variance equal to the $(z_\sr,z_\sr)$ component of the covariance matrix~(\ref{eq:Sigma:z}) that we denote $\Sigma_\sr(\eta)$ for simplicity. The deterministic quantity $z_\nsr^{(0)}$ is set by initial conditions, and hereafter, stochastic quantities are denoted with a hat. The averaged trajectory of the coarse-grained field is $\langle \boldsymbol{\Phi} (\eta)\rangle =z_\sr^{(0)}\boldsymbol{\Phi}_\sr+z_\nsr^{(0)}\boldsymbol{\Phi}_\nsr= \boldsymbol{\Phi} (\eta_0) $, meaning that, as already stressed, the averaged coarse-grained field evolves according to the classical dynamics (which is also the most probable trajectory, the PDF being Gaussian) and reaches the slow-roll late-time attractor unless $z_\sr^{(0)}=0$. Making use of \Eq{eq:Gaussian:Green}, the field-space PDF is given by
\bea
\label{eq:PDF:interactionPic}
\mathcal{W}\left( \boldsymbol{z},\eta \left\vert \boldsymbol{z}^{(0)},\eta_0 \right. \right)=\delta\left(z_\nsr-z_\nsr^{(0)}\right)
\dfrac{\exp\left[-\dfrac{1}{2\Sigma_\sr}\left(z_\sr-z^{(0)}_\sr\right)^2\right]}{\sqrt{2\pi\Sigma_\sr}} .
\eea
In this expression, it is clear that if the coarse-grained field is initially set in the attractor branch, \ie if $z^{(0)}_\nsr=0$, it never leaves the attractor although it diffuses along the SR direction. If $z^{(0)}_\nsr \neq 0$ however, at any finite time, the PDF never lies exactly on the attractor branch, but the question is whether it gets sufficiently close to it. To answer it, we define the ``slow-roll'' region of phase space $(z_\sr,z_\nsr)$ as being the domain where\footnote
{As explained in \Sec{sec:SR}, the notion of slow roll for a test scalar field is defined by requiring that the parameters $\epsilon_{n}^\phi$ are small. Plugging the decomposition $\phi = z_\sr \phi_\sr +  z_\nsr \phi_\nsr$ into \Eqs{eq:eps1:testfield} and~(\ref{eq:eps2:testfield}), with  $\dot{\phi}_\sr/\phi_\sr = (\nu-3/2)/H$ and $\dot{\phi}_\nsr/\phi_\nsr = (\nu+3/2)/H $, on can express $\epsilon_1^\phi$ and $\epsilon_2^\phi$ as functions of the ratio $z_\nsr\phi_\nsr/(z_\sr \phi_\sr)$ only. This shows that $\epsilon_1^\phi$ and $\epsilon_2^\phi$ are small when this ratio is.
}
$\left\vert z_\nsr\phi_\nsr\right\vert<R \left\vert z_\sr\phi_\sr\right\vert$, with $R\ll 1$ a dimensionless parameter. In the following, the PDF will be said to have reached the slow-roll attractor if its overlap with this domain is close to total.
\subsubsection{Probability to enter slow roll}
\begin{figure}
\begin{center}
\includegraphics[width=0.99\textwidth]{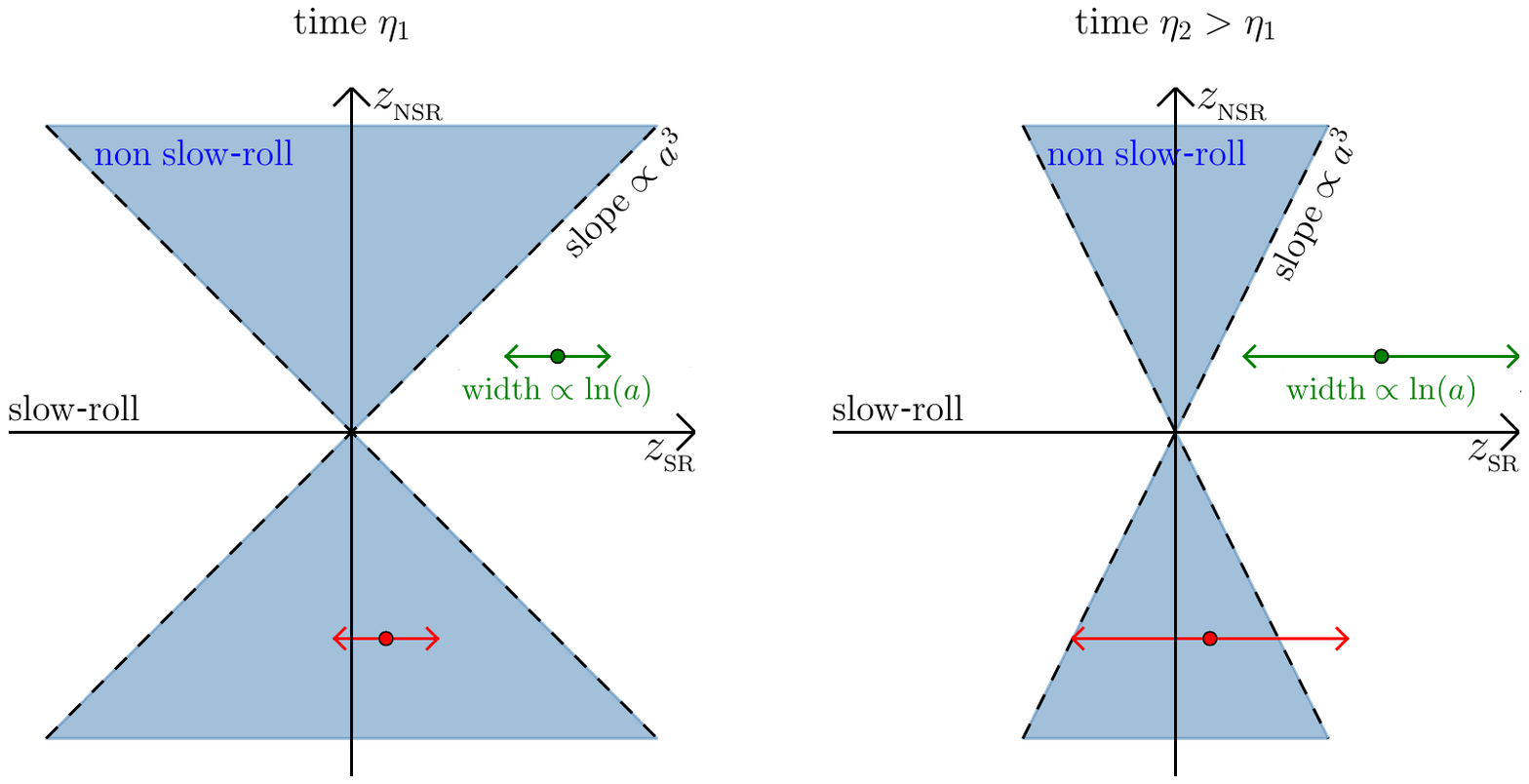} 
\caption{Schematic representation of the stochastic dynamics in phase space for a free scalar field. The right panel corresponds to a time $\eta_2$ greater than the time $\eta_1$ of the left panel. The shaded area corresponds to the non slow-roll region, which shrinks as $a^{-3}$ with time. The green and red arrows depict the width $\Sigma_\sr$ of the covariance matrix along the slow-roll direction (which first increases with time as $\ln(a)$ before either saturating to the value $3H^2/(2\pi^2 \vert m^2 \vert)$ if the squared mass is negative, or continuing to increase as $a^{2m^2/(3H^2)}$ if the squared mass is positive), for two different choices of the initial conditions, $z^{(0)}_\sr>z_R(\eta_0)$ in blue, and $z^{(0)}_\sr<z_R(\eta_0)$ in red (note that $\Sigma_\sr$ is independent of the averaged evolution of the field). In the interaction picture parametrised by the variable $\boldsymbol{z}$, the averaged dynamics is given by a single constant point, and quantum diffusion only takes place along the slow-roll direction.}
\label{fig:phsp}
\end{center}
\end{figure}
Unless $z_\sr^{(0)}=0$, the slow roll condition $\left\vert z_\nsr\phi_\nsr\right\vert<R \left\vert z_\sr\phi_\sr\right\vert$ is always satisfied at late time for the classical dynamics since $z_\sr$ and $z_\nsr$ are frozen in this case, and the ratio $\phi_\nsr/\phi_\sr$ decreases asymptotically to $0$. In the stochastic dynamics however, $z_\nsr$ is frozen but $\widehat{z}_\sr$ undergoes quantum diffusion, so that values of $z_\sr$ such that $\left\vert z_\nsr\phi_\nsr\right\vert > R \left\vert z_\sr\phi_\sr\right\vert$, \ie values of $\vert z_\sr \vert<z_R(\eta)$ with
\bea
\label{eq:def:zR}
z_R(\eta) = \frac{\left\vert z_\nsr^{(0)} \right\vert}{R}\left(\frac{\eta}{\eta_0}\right)^{2\nu} ,
\eea 
are not forbidden even at late time. Let us study how the probability for such values to be realised evolves in time.
\begin{figure}
\begin{center}
\includegraphics[width=0.99\textwidth]{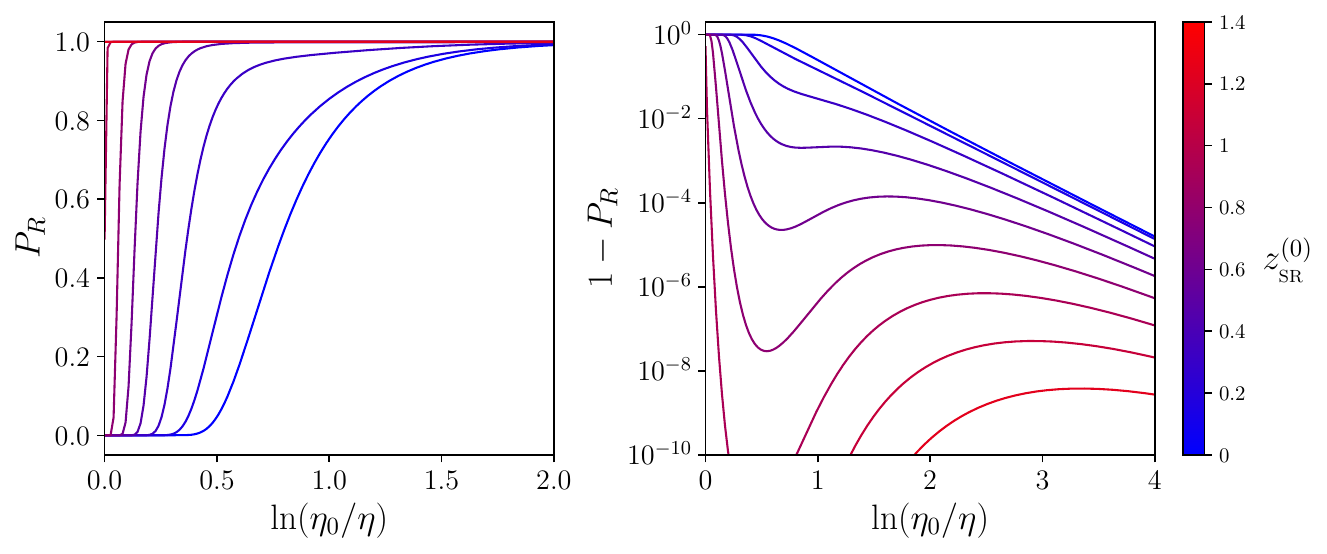} 
\caption{Probability to lie in the slow-roll region [$z_\sr>z_R(\eta)$, left panel] and probability to lie in the non slow-roll region [$z_\sr<z_R(\eta)$, right panel] as a function of conformal time for a free scalar field with mass $m/H=10^{-1}$ in de Sitter. Initial conditions are such that $z_\nsr^{(0)}=10^{-2}$ and different curves correspond to different values of $z_\sr^{(0)}$ given in the colour bar (from blue to red,  $z_\sr^{(0)}=0,\,0.2,\,0.4,\,0.6,\,0.8,\,1,\,1.2,\,1.4$). In the definition~(\ref{eq:def:zR}) of $z_R$, the parameter $R$ is taken to $R=10^{-2}$ and $\eta_0=-1$.}
\label{fig:probaSR}
\end{center}
\end{figure}

In \Fig{fig:phsp}, the slow-roll and non slow-roll regions of phase space are displayed, separated by the lines $z_\sr=\pm z_R$, and one can see that the slow-roll region expands as a power-law in conformal time. On the other hand, the width of the PDF along the SR direction is $\Sigma_\sr$ given in \Eq{eq:Sigma:z}. As noticed below \Eq{eq:Sigma:z}, at early time, when $\log(\eta_0/\eta)\ll 1/\vert 3-2\nu\vert$, it increases logarithmically in conformal time, $\Sigma_\sr \simeq \ln(\eta_0/\eta)/(2\pi)^2 $; while at late time, when $\log(\eta_0/\eta)\gg 1/\vert 3-2\nu \vert$, if $\nu>3/2$ then it asymptotes to the constant value $\Sigma_\sr\simeq 1/[(2\pi)^2(2\nu-3)]$, while if $\nu<3/2$, it increases as $\Sigma_\sr\simeq 1/[(2\pi)^2(3-2\nu)](\eta_0/\eta)^{3-2\nu}$. In either case the slow-roll region always expands faster than the width of the PDF, which suggests that the overlap between the PDF and the slow-roll region increases with time. This can be checked by calculating the probability $P_R$ to be in slow roll, that is to say the probability for $\vert \widehat{z}_\sr \vert$ to be larger than $z_R$,
\bea
P_{R}\left(\eta\left\vert \boldsymbol{z}^{(0)},\eta_0\right.\right)&=&1-\displaystyle\int^{z_R}_{-z_R} \dd z_\sr 
\int^{\infty}_{-\infty} \dd z_\nsr 
\mathcal{W}\left(\boldsymbol{z},\eta \left\vert \boldsymbol{z}^{(0)},\eta_0\right. \right)
\nonumber \\ &=&
1+\dfrac{1}{2}\erf\left[\dfrac{z^{(0)}_\sr-z_R(\eta)}{\sqrt{2\Sigma_\sr(\eta)}}\right]-\dfrac{1}{2}\erf\left[\dfrac{z^{(0)}_\sr+z_R(\eta)}{\sqrt{2\Sigma_\sr(\eta)}}\right] ,
\label{eq:probaSR}
\eea
where \Eq{eq:PDF:interactionPic} has been used in the second line and the function $z_R(\eta)$ given in \Eq{eq:def:zR} implicitly depends on $z_\nsr^{(0)}$. This probability is displayed as a function of time in \Fig{fig:probaSR} for a free field with mass $m/H=10^{-1}$ in de Sitter and taking $R=10^{-2}$. Initial conditions are set to $z_\nsr^{(0)}=10^{-2}$ and different curves correspond to different values of $z_\sr^{(0)}$. Initially, $z_R(\eta_0)=1$, so that if $z_\sr^{(0)}<1$, the initial PDF entirely lies in the non slow-roll region and $P_R(\eta_0)=0$. The slow-roll probability $P_R$ then increases and asymptotes to $1$ after a few \efolds. If $z_\sr^{(0)}>1$, the initial PDF entirely lies in the slow roll region and $P_R(\eta_0)=1$. Subsequently, $P_R$ slightly decreases (which can be better seen in the right panel) but never departs much from $1$ and increases back towards $1$ after a few \efolds. From \Eq{eq:probaSR}, one can see that this is generic and that $P_R$ always tends to $1$ at late time. The answer to the first question raised at the beginning of this section is therefore that for \emph{all} initial conditions, the classical slow roll attractor generalises to a stochastic attractor of free fields.
\subsubsection{Relaxation time towards slow roll}
The classical, deterministic dynamics is frozen in the interaction picture and thus reaches the slow-roll region of phase space when $z_R(\eta) < \vert z_\sr^{(0)}\vert$, \ie at a time $\eta>\eta_\sr^\mathrm{det}$ where
\bea
\label{eq:etasr:det}
\eta_\sr^\mathrm{det} = \eta_0\left(R\left\vert \frac{z_\sr^{(0)}}{z_\nsr^{(0)}} \right\vert\right)^{\frac{1}{2\nu}} .
\eea
Let us see how the relaxation time of the stochastic dynamics towards slow roll, defined as being the time $\eta_\sr^\mathrm{stoch}$ such that for $\eta>\eta_\sr^\mathrm{stoch}$, $P_R(\eta)>1/2$, compares with this value. A first remark is that since the classical trajectory coincides with the mean stochastic one, when $\eta>\eta_\sr^\mathrm{det}$, the center of the phase-space PDF lies in the slow-roll region. In fact, when $\eta=\eta_\sr^\mathrm{det}$, the slow-roll probability~(\ref{eq:probaSR}) reads $P_R(\eta_\sr^\mathrm{det})=1-1/2\erf[z_\sr^{(0)}\sqrt{2/\Sigma_\sr(\eta_\sr^\mathrm{det})}]$ and is larger than $1/2$ since the error function is always smaller than one. This shows that 
\bea
\eta_\sr^\mathrm{stoch}<\eta_\sr^\mathrm{det} ,
\eea
\ie relaxation towards slow roll is faster in the stochastic theory than in the classical one. This answers the second question raised at the beginning of this section. In some cases, the stochastic relaxation time can even be much smaller. For instance, let us consider the situation in which initial conditions are set on the anti-attractor branch, $z_\sr^{(0)}=0$. In this case, $\eta_\sr^\mathrm{det}$ given in \Eq{eq:etasr:det} vanishes, meaning that the classical trajectory never enters the slow-roll regime. In the stochastic theory however, in the regime $\log(\eta_0/\eta)\ll 1/(3-2\nu)$ where $\Sigma_\sr \simeq \ln(\eta_0/\eta)/(2\pi)^2$, one obtains
\bea
\eta_\sr^\mathrm{stoch} \simeq \eta_0\exp\left( -\frac{1}{4\nu} W_0\left\lbrace 8\nu\left[\frac{\pi}{\erf^{-1}\left(1/2\right)}\frac{z_\nsr^{(0)}}{R}\right]^2\right\rbrace \right)\, ,
\eea
where $W_0$ is the $0^\mathrm{th}$ branch of the Lambert function and $\erf^{-1}$ is the inverse error function. In the limit where the argument of the Lambert function is large, one obtains $\eta_\sr^\mathrm{stoch} \sim \vert R/z_\nsr^{(0)}\vert^{1/(2\nu)}$, which corresponds to $\eta_\sr^\mathrm{det}$ given in \Eq{eq:etasr:det} for $z_\sr^{(0)}\sim 1$. In the opposite regime where $\log(\eta_0/\eta)\gg 1/(3-2\nu)$  and $\Sigma_\sr\simeq (\eta_0/\eta)^{3-2\nu}/[4\pi^2(3-2\nu)]$ (here we assume $\nu<3/2$, so a positive squared mass), one has
\bea
\eta_\sr^\mathrm{stoch} \simeq \eta_0 \left[\frac{\erf^{-1}(1/2)}{\pi\sqrt{2}}\frac{R}{z_\nsr^{(0)}}\frac{1}{\sqrt{3-2\nu}}\right]^{\frac{1}{\nu+3/2}} .
\eea
In this case, even though the mean trajectory never reaches the slow-roll domain, the overlap between the PDF and the slow-roll region becomes close to total when $\eta>\eta_\sr^\mathrm{stoch}$. 

Let us finally mention that so far, the initial state of the coarse-grained field has been assumed to be known exactly, \ie the PDF at initial time $\eta_0$ has been taken to a Dirac function. For a generic initial PDF $P(\boldsymbol{z}_0,\eta_0)$, one can check that the average field is given by $\left<\boldsymbol{z}\right>=\left<\boldsymbol{z}^{(0)}\right>_0$, where $\langle\cdot\rangle_0$ denotes average over the initial PDF, \ie $\left<f(\boldsymbol{z}^{(0)})\right>_0\equiv \int\dd\boldsymbol{z}^{(0)}f(\boldsymbol{z}^{(0)})P(\boldsymbol{z}^{(0)},\eta_0)$. This implies that $\left<\boldsymbol{\Phi}\right>=\left<z^{(0)}_\sr\right>_0\boldsymbol{\Phi}_\sr+\left<z^{(0)}_\nsr\right>_0\boldsymbol{\Phi}_\nsr$ and means that in this case too, the average coarse-grained field evolves towards the slow-roll attractor unless $\left\langle z^{(0)}_\sr\right\rangle_0=0$, with a relaxation time still given by \Eq{eq:etasr:det} if one replaces $z^{(0)}_\sr$ by $\left\langle z^{(0)}_\sr\right\rangle_0$ and  $z^{(0)}_\nsr$ by $\left\langle z^{(0)}_\nsr\right\rangle_0$. For the full stochastic dynamics, the probability to be in the slow-roll region is given by $P_R(\eta)=\left\langle P_{R}\left(\eta|\boldsymbol{z}^{(0)},\eta_0\right) \right\rangle_0$, and from \Eq{eq:probaSR} one can check that it still asymptotes to $1$ at late time.
\section{Beyond free fields}
\label{sec:BeyondFreeFields}
So far, we have shown that for test fields with quadratic potentials, the classical slow-roll attractor generalises to a stochastic attractor. In \Sec{sec:InteractionPicture}, it was explained why this property is related to the fact that for such free fields, the equation of motion of field perturbations on large scales coincides with the one of the background, hence shares the same attractor. When the potential is not quadratic, or when the field is not a mere spectator and sources metric perturbations, this stops being the case and slow roll is not an exact stochastic attractor anymore. In this section, we quantify the deviation from slow roll induced by stochastic effects for non-free fields.

The stochastic phase-space dynamics of non-free fields is more challenging to study than the one of free fields for the two following reasons. First, the background equation of motion~(\ref{eq:KG}) is not linear anymore, so the ability to use Green formalism, Gaussian solutions of the Fokker-Planck equation and canonical transformations to the interaction picture is lost. Second, the equation of motion for the field perturbations~(\ref{eq:eomphiq}) and~(\ref{eq:eompiq}) now depends on the background value of $\phi$, through the term $V_{,\phi\phi}(\phi)$ in \Eq{eq:eompiq}. This means that \Eq{eq:mode:light} is still valid, but $m^2$ has to be replaced by an effective mass 
\bea
\label{eq:meff:def}
m_\ueff^2=V_{,\phi\phi}(\bar{\phi})
\eea
which depends on the background field $\bar{\phi}$. However $\bar{\phi}$ is stochastic and different for each realisation of the Langevin equation. This implies that, in principle, the diffusion matrix must be re-computed at every given time and for every given realisation of the Langevin equation by integrating the equation of motion of field perturbations sourced by this realisation. In practice, this prevents any conclusion to be drawn from analytical arguments only without resorting to some approximation.

For this reason, let us restrict the analysis to the phase-space region sufficiently close to the classical attractor so that the ``reference'' background solution about which the equation of motion of field perturbations is solved can be taken as the classical slow-roll solution $\phi_\sr$. This will allow us to assess how much stochastic effects alter the slow-roll dynamics \emph{if} one starts on the classical attractor.\footnote{This also corresponds to the first recursive level of \Refc{Levasseur:2013tja}.} Under this assumption, the effective mass term in \Eq{eq:mode:light} can be taken as 
\bea
\label{eq:meffSR:def}
m_\ueff^2=V_{,\phi\phi}(\phi_\sr)
\eea
and becomes an explicit, fixed function of time. 

In fact, this function of time can be expressed in terms of slow-roll parameters. Indeed, combining \Eqs{eq:eps1:appr:test} and~(\ref{eq:eps2:appr:test}), one obtains
\bea
\label{eq:meff:eps}
\frac{m_\ueff^2}{H^2}=3\epsilon_1+3\epsilon_1^\phi-\frac{3}{2}\epsilon_2^\phi .
\eea
If $\phi$ is the inflaton field, one can combine \Eqs{eq:eps1:appr} and~(\ref{eq:eps2:appr}) instead, and obtain $m_\ueff^2/H^2=6\epsilon_1-3\epsilon_2/2$, which is consistent with \Eq{eq:meff:eps} if one equates $\epsilon_1^\phi$ and $\epsilon_2^\phi$ to $\epsilon_1$ and $\epsilon_2$ respectively. This means that $m_\ueff^2/H^2$ does not substantially vary over the time scale of one \efold. Since the amplitude acquired by field perturbations is mostly determined by the background dynamics around the few \efolds~surrounding their Hubble exit time, the ratio $m_\ueff^2/H^2$ can thus be approximated as being constant during this time interval.  This implies that \Eqs{eq:modesolution:light} and~(\ref{eq:nu:def}) still provide an accurate solution to the equation of motion of field perturbations if one replaces $\nu$ by $\nu[\eta_*(k)]$, in which $\epsilon_1$ and the ratio $m_\ueff^2/H^2$ are evaluated at the time $\eta_*(k)$ when the mode $k$ crosses the Hubble radius. This adiabatic approximation is in fact the standard way cosmological perturbations are calculated in the slow-roll approximation~\cite{Stewart:1993bc}.
\begin{figure}
\begin{center}
\includegraphics[width=0.49\textwidth]{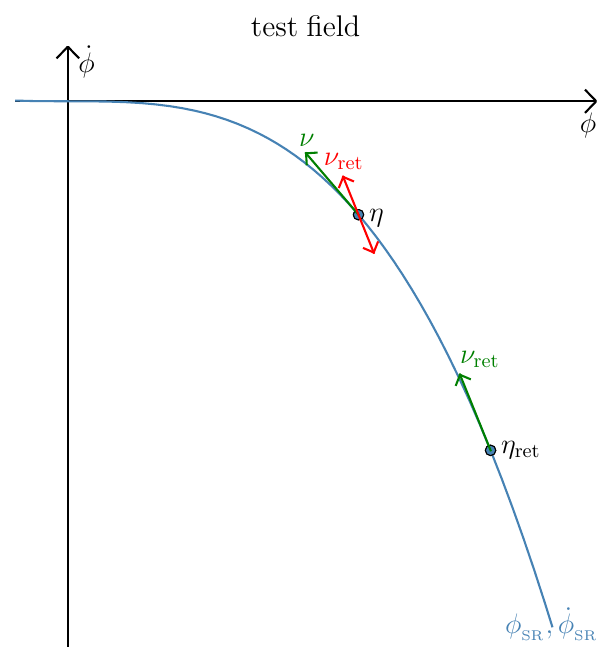} 
\includegraphics[width=0.49\textwidth]{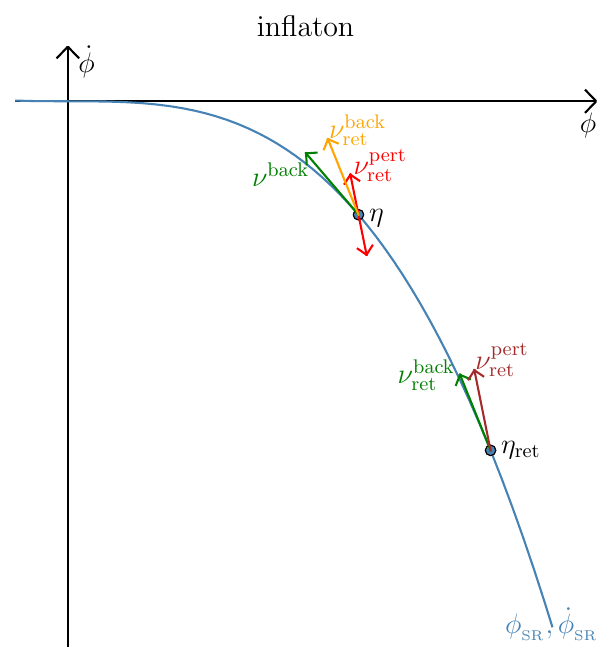} 
\caption{Quantum diffusion about the classical slow-roll trajectory $(\phi_\sr,\dot{\phi}_\sr)$ for a test field (left panel) and for the inflaton field (right panel). At time $\eta$, the direction to which the classical dynamics points (green arrow) is determined by $\nu^\mathrm{back}(\eta)$, while the direction along which quantum diffusion takes place (red double arrow) is determined by $\nu^\mathrm{pert}(\eta_\mathrm{ret})$, which corresponds to the field perturbations phase-space direction at the retarded time $\eta_\mathrm{ret}$ when the mode that crosses the coarse-graining radius at time $\eta$ crossed the Hubble radius. For a test field, the background and the perturbations share the same phase-space direction at a given time, $\nu^\mathrm{back}(\eta)=\nu^\mathrm{pert}(\eta)$, while this is not the case for the inflaton field that also couples to metric fluctuations. In both cases, quantum diffusion is not aligned with the classical flow and stochastic effects induce deviations from the classical attractor.}
\label{fig:retardednu}
\end{center}
\end{figure}
\subsection{Test fields}
In the case of test fields, this approximation implies that the diffusion matrix $\boldsymbol{D}(\eta)$ is still given by the leading-order terms in \Eqs{eq:massive:Dphiphi}-(\ref{eq:massive:Dphipi}), except that $\nu$ must now be evaluated at the retarded time $\eta_\mathrm{ret}(\eta)=\eta_*[k_\sigma(\eta)]$, \ie at the time when the mode $k_\sigma(\eta)$, that crosses the coarse-graining radius at time $\eta$, crosses the Hubble radius. This means that the Langevin equations~(\ref{eq:eombarphi}) and~(\ref{eq:eombarpi}), where $\xi_\phi$ and $\xi_\pi$ are totally anti-correlated, can be written as
\bea
\frac{\dd \phi}{\dd N_e} &=& \gamma + \frac{H}{2\pi}\xi\\
\frac{\dd \gamma}{\dd N_e} &=& \left(\epsilon_1-3\right)\gamma - \frac{V_{,\phi}(\phi)}{H^2} +\frac{H}{2\pi} \left[\nu_\mathrm{ret}(N_e)-\frac{3}{2}\right] \xi ,
\eea
where the number of \efolds~is used as the time variable for simplicity, phase space is parametrised by $\phi$ and $\gamma\equiv \pi/(H a^3)$, and $\xi$ is a normalised white Gaussian noise, such that $\langle \xi (N_{e}) \xi (N_{e}^\prime) \rangle = \delta(N_{e}-N_{e}^\prime)$.

However, the direction of the classical trajectory in phase space is still determined by the value of $\nu$ at current time $\eta$, since at leading order in slow roll, $\dot{\phi}_\sr\simeq -V_{,\phi}(\phi_\sr)/(3H)$ gives rise to
\bea
\frac{\dd}{\dd t} 
\left(\begin{array}{c}
	\phi_\sr  \\
	\dot{\phi}_\sr
\end{array}\right) 
\propto
\left(\begin{array}{c}
	1  \\
	H_*\epsilon_1-\frac{V_{,\phi\phi}(\phi_\sr)}{3H}
\end{array}\right)
=
\left(\begin{array}{c}
	1  \\
	H_* \left[\nu\left(\eta\right)-\frac{3}{2}\right]
\end{array}\right) ,
\label{eq:fieldshift:class}
\eea
where the expression~(\ref{eq:nu:def}) relating $\nu$ to the effective mass defined in \Eq{eq:meffSR:def} and to $\epsilon_1$ has been used. The situation is summarised in the left panel of \Fig{fig:retardednu}. Because $\nu(\eta)$ and $\nu(\eta_\mathrm{ret})$ are a priori different, the classical flow and the quantum diffusion do not point to the same direction in phase space and this is why stochastic effects induce a deviation from the classical attractor, related to 
\bea
\label{eq:nuMinusnuRet}
\nu(\eta)-\nu(\eta_\mathrm{ret}) &\simeq & \alphaS(\eta) \ln\left(\sigma\right) .
\eea
In this expression, $\alphaS=\dd \nS/\dd\ln k$ denotes the running of the spectral index of the field fluctuations power spectrum introduced below \Eq{eq:condition:m_over_H}. Since $\nS=1-2(\nu-3/2)$ indeed, $\nu-\nu_\mathrm{ret}\simeq \alphaS [N_e(\eta_\mathrm{ret})-N_e(\eta)]$. The comoving wavenumber that crosses the coarse-graining radius at time $\eta$ is $k=\sigma a(\eta) H(\eta)$, see \Eq{eq:ksigma}, so it crosses the Hubble radius at time $\eta_\mathrm{ret}$ such that $k= a(\eta_\mathrm{ret}) H(\eta_\mathrm{ret})$, and one finds that $N_e(\eta_\mathrm{ret})-N_e(\eta)  = \ln[a(\eta_\mathrm{ret})/a(\eta)]\simeq \ln(\sigma)$ at leading-order in slow roll, hence \Eq{eq:nuMinusnuRet}. 
%For a free scalar field, the running vanishes and $\nu(\eta)=\nu(\eta_\mathrm{ret})$.

Let us now assess the effect of the misalignment between the quantum noise and the classical slow-roll direction in phase space on the relative fluctuation $\delta\rho_\phi/\rho_\phi$ in the energy density contained in $\phi$. The reason is that this ratio corresponds to the contribution of the fluctuations in $\phi$ to the total curvature perturbation $\zeta$ measured \eg in the CMB. It is therefore directly related to observable quantities, for instance in the context of curvaton scenarios~\cite{Linde:1996gt, Enqvist:2001zp, Lyth:2001nq, Moroi:2001ct, Bartolo:2002vf, Vennin:2015vfa, Vennin:2015egh}. Fourier transforming the energy density field, the amplitude of perturbations at the comoving scale $k$ is given by $\delta\rho_\phi (k)/\rho_\phi (k)$, where $\rho_\phi (k)$ is the energy density contained in $\phi$ at the time when $k$ crosses the Hubble radius, and $\delta\rho_\phi (k)$ is the fluctuation in this quantity induced by quantum diffusion over one \efold~around this time. Since $\rho_\phi=V(\phi)+H^2\gamma^2/2$, one has $\delta \rho_\phi = V_{,\phi}(\phi) \delta\phi + H^2 \gamma \delta \gamma$. If quantum diffusion was aligned with the classical slow-roll direction, one would have $\delta\gamma=(\nu-3/2)\delta\phi$ and $\delta\rho_\phi=\delta\phi[V_{,\phi}+H^2\gamma(\nu-3/2)]$. However, quantum diffusion occurs along the retarded direction $\nu_\mathrm{ret}$, which gives rise to the corrected $\delta\gamma^\mathrm{corr}=(\nu_\mathrm{ret}-3/2)\delta\phi$ and $\delta\rho_\phi^\mathrm{corr}=\delta\phi[V_{,\phi}+H^2\gamma(\nu_\mathrm{ret}-3/2)]$. At leading order in slow roll, $\gamma\simeq -V_{,\phi}/(3H^2)$, and one has
\bea
\label{eq:deltarhoCorr:test}
\frac{\delta\rho_\phi^\mathrm{corr}-\delta\rho_\phi}{\delta\rho_\phi}=\frac{\nu-\nu_\mathrm{ret}}{3} .
\eea
Making use of \Eq{eq:nuMinusnuRet}, this yields a small correction if
\bea
\label{eq:conditions:sigma:test}
\sigma\gg \ee^{-\frac{1}{\vert \alphaS \vert}} ,
\eea
which has a similar form as the condition $\sigma\gg \ee^{-\frac{1}{\vert \nS-1 \vert}}$ derived in \Eq{eq:condition:m_over_H}. It implies that between the Hubble radius crossing time and the coarse-graining radius crossing time, not only the amplitude of the field fluctuations should not vary substantially but also their spectral index. Let us note that in the slow-roll approximation, $\vert \alphaS \vert \ll \vert \nS-1 \vert$, so \Eq{eq:conditions:sigma:test} is a weaker constraint than \Eq{eq:condition:m_over_H}.
\subsection{Inflaton field}
\label{sec:Inflaton}
In the case of the inflaton field, an additional subtlety is that even at the same fixed time, the slow-roll background flow and the perturbations dynamics have different directions $\nu^\mathrm{back}$ and $\nu^\mathrm{pert}$ in field space. In \Sec{ssec:cosmo} indeed, it was assumed that $\phi$ is a test field sufficiently decoupled from the metric perturbations such that the latter can be ignored. If $\phi$ is the inflaton field, this is not the case anymore. In \App{App:InflatonMetricFluctuations}, it is shown that the Langevin equations~(\ref{eq:eombarphi}) and~(\ref{eq:eombarpi}) with quantum noises~(\ref{eq:noisephi}) and~(\ref{eq:noisepi}) still apply for the inflaton field, but in the equation of motion~(\ref{eq:mode:light}) for the field perturbations,  the effective mass~(\ref{eq:meff:def}) receives a correction from gravitational coupling with metric fluctuations that reads~\cite{Langlois:1994ec, Lyth:2001nq}
\bea
\label{eq:meff:inflaton}
\widetilde{m}_{\mathrm{eff}}^2=V_{,\phi\phi}\left(\bar{\phi}\right)-\frac{1}{\Mp^2a^3}\frac{\dd}{\dd t}\left(\frac{a^3}{H}\dot{\bar{\phi}}^2\right) .
\eea
Making use of the Friedman equation~(\ref{eq:Friedman}) and of the Klein-Gordon equation~(\ref{eq:KG}) for the background field $\bar{\phi}$, this additional term can be written $-2H^2\epsilon_1(3+2\epsilon_2-\epsilon_1)$. At leading order in slow roll, the index $\nu$ given in \Eq{eq:nu:def} is then modified according to
\bea
\label{eq:nuPertMinusNuBack}
\nu^{\mathrm{pert}} = \nu^{\mathrm{back}}+2\epsilon_1 ,
\eea
where $\nu^\mathrm{back}=3/2-\epsilon_1+\epsilon_2/2$ is the value obtained in absence of gravitational coupling to metric fluctuations and corresponds to the direction of the slow-roll background dynamics. 

The situation is summarised in the right panel of \Fig{fig:retardednu}. At a given time $\eta$, not only the phase-space direction along which quantum diffusion takes place has to be evaluated at the retarded time $\eta_\mathrm{ret}$, but it is related to the dynamics of perturbations which occurs in a different direction than the one of the background. As before, let us assess the effect of this misalignment on the curvature perturbation $\zeta$, directly proportional to $\delta\rho/\rho$~\cite{Wands:2000dp}. Similarly to \Eq{eq:deltarhoCorr:test}, one has
\bea
\label{eq:deltarhoCorr:inflaton}
\frac{\delta\rho^\mathrm{corr}-\delta\rho}{\delta\rho}=\frac{\nu^\mathrm{back}-\nu^\mathrm{pert}_\mathrm{ret}}{3} .
\eea
Decomposing $\nu^\mathrm{back}-\nu_\mathrm{ret}^\mathrm{pert} =\nu^\mathrm{back}-\nu_\mathrm{ret}^\mathrm{back}   +     \nu_\mathrm{ret}^\mathrm{back}-\nu_\mathrm{ret}^\mathrm{pert} $, where $\nu^\mathrm{back}-\nu_\mathrm{ret}^\mathrm{back}$ is given by \Eq{eq:nuMinusnuRet} and $\nu_\mathrm{ret}^\mathrm{back}-\nu_\mathrm{ret}^\mathrm{pert} $ by \Eq{eq:nuPertMinusNuBack} (at retarded time $\eta_\mathrm{ret}$), one obtains the same condition~(\ref{eq:conditions:sigma:test}) as for a test field since $\epsilon_1\ll 1$. Therefore, in both cases, one finds that the observational effect of the phase-space misalignment between the classical homogeneous slow-roll dynamics and quantum diffusion remains small if the amplitude \emph{and} tilt of the field fluctuations at the Hubble radius crossing time and at the quantum-to-classical transition time are sufficiently close. 
\section{Conclusion}
\label{sec:concl}
The phase-space dynamics of homogeneous scalar fields during inflation is endowed with classical attractors along which their kinetic energy is parametrically small compared to their potential energy. These ``slow-roll'' solutions play an important role in the predictivity of the inflationary paradigm since they allow initial conditions in phase space to be erased after a few \efolds. When perturbations are included however, they alter the dynamics of the large-scale coarse-grained fields as they cross the coarse-graining radius, via quantum diffusion that can be modelled through the formalism of stochastic inflation. In this paper, we have quantified the departure from slow roll induced by these stochastic effects. 

We have found that for free fields, that is to say for test scalar fields with quadratic potentials, quantum diffusion takes place exactly along the slow-roll attractor. This is because in this case, once perturbations cross the Hubble radius, they follow an equation of motion that matches the one of the classical background, so that the ``growing mode'' of perturbations matches the slow-roll direction of the background. By the time perturbations cross the coarse-graining radius, they thus have reached the classical attractor. This result has been derived using a Hamiltonian formulation of stochastic inflation, in which the ability to perform canonical phase-space transformations to the interaction picture has played a crucial role. We also studied the quantum-to-classical transition in this context. We found that the stochastic dynamics relaxes towards the classical attractor regardless of initial conditions, at a rate at least as fast as in the classical picture, and sometimes even much faster. This reinforces the attractiveness (in both senses of the word) of slow-roll solutions.

For non-test scalar fields, this property is lost and quantum diffusion shifts the dynamics away from slow roll. This is because the statistical properties of the noise when some perturbations cross the coarse-graining radius depend on the field effective mass at the earlier time when those perturbations crossed the Hubble radius. Since the effective mass varies in time for non-test fields, this results into a misalignment between the classical flow and the direction of the stochastic noise. Moreover, when the field perturbations substantially source metric fluctuations (as for the inflaton field), the effective mass of perturbations differs from the one of the homogeneous component of the field by a slow-roll suppressed factor that further contributes to this misalignment. We found that, if the physical coarse-graining radius $(\sigma H)^{-1}$ is chosen sufficiently close to the Hubble radius $H^{-1}$ so that the amplitude and tilt of the field fluctuations does not vary much between these two scales (a necessary condition for the theory to be $\sigma$-invariant anyway), then these two phenomenons have a minor effect on the energy density fluctuations, which are the observables of the system. Combined with the requirement that quantum fluctuations must have completed their quantum-to-classical transition when they join the coarse-grained field (that cannot be described by a stochastic theory  otherwise), the set of conditions on the coarse-graining scale we obtained reads
\bea
	&&\ee^{-\frac{1}{\left\vert \nS-1 \right\vert }} \ll \sigma \ll \left\vert \nS-1 \right\vert^{\frac{1}{3}}, \\
	&&\ee^{-\frac{1}{\left\vert \alphaS \right\vert }} \ll \sigma,
\eea
where the first lower bound -- involving the tilt $\nS$ -- means that the amplitude of the field fluctuations does not vary much between Hubble radius crossing and coarse-graining radius crossing, while the second -- involving the running $\alphaS$ -- means that the tilt does not vary much. The above set of conditions is easily met in slow-roll inflation since the tilt and its running are both small numbers scaling with the slow-roll parameters (note that this simply reduces to $\sigma\ll 1$ in the singular case of a massless field in de Sitter).

Our main conclusion is therefore that classical attractors appear to be immune to quantum diffusion. This can be understood, and maybe summarised, by noticing that along the slow-roll solution, all quantities of physical interest change over time scales much larger than one \efold. Since the relaxation time to slow roll is of the order a few \efolds~(at least in the classical picture), this hierarchy in time scales allows for the existence of an adiabatic regime in which the effective mass of the field can be considered as constant over the time it takes for the system to relax towards the attractor, and the field can be seen as being ``locally free'' in that sense. This is to be contrasted with the phase-space late-time distribution, which typically requires thousands of \efolds~or even more [for instance, for a quadratic test scalar field with mass $m$, the relaxation number of \efolds~is of order $H^2/m^2$, see discussion below \Eq{eq:Sigma:light:lightlimit}] to relax to the stationary distribution. Even in slow roll, the background can vary more rapidly than this, in which case no adiabatic regime exists~\cite{Hardwick:2017fjo}. Therefore, we see that the equilibration time scales towards slow-roll and towards stationarity can be very different, which makes the phase-space dynamics of fields during inflation rather non trivial.

Finally, the formalism we derived may be relevant in other contexts where the full phase-space dynamics needs to be resolved. In contracting cosmologies for instance, the inflow of quantum perturbations towards the large-scale sector of the theory also occurs and is expected to modify the classical dynamics. Since there is no slow-roll attractor in this case, a description of the entire phase space is required and we plan to investigate this issue in the future.
\section*{Acknowledgements}
It is a pleasure to thank Jibril Ben Achour for interesting and enjoyable discussions and David Wands for interesting comments on the manuscript. V.V. acknowledges financial support from STFC grants ST/K00090X/1 and ST/N000668/1. 
We would like to thank Lucas Pinol for pointing out a typo in a previous version of the manuscript, which led to several minor modifications made after publication in the online arXiv version.
\appendix
\section{Green's matrices}
\label{app:green}
Let us consider the linear homogeneous system associated to the stochastic dynamics of \Eq{eq:inhomopb}, $\dot{\boldsymbol{\Phi}}=\boldsymbol{A}(\tau)\boldsymbol{\Phi}$, and let us assume that two independent solutions $(\bar\phi^{(1)},\bar\pi^{(1)})$ and $(\bar\phi^{(2)},\bar\pi^{(2)})$ are known. The so-called ``fundamental'' matrix of the system is defined as
\bea
\label{eq:U:def}
\boldsymbol{U}(\tau)=\left(\begin{array}{cc}
	\bar\phi^{(1)} & \bar\phi^{(2)} \\
	\bar\pi^{(1)} & \bar\pi^{(2)}
\end{array}\right).
\eea
By construction, one can check that $\dd \boldsymbol{U}(\tau)/\dd \tau =\boldsymbol{A}(\tau)\boldsymbol{U}(\tau)$. Let us also notice that since the two solutions are independent, $\det(\boldsymbol{U})\neq0$. The matrix $\boldsymbol{U}$ is then invertible and gives rise to the Green's matrix
\bea
\label{eq:Green:fundamental}
	\boldsymbol{G}(\tau,\tau_0)&=&\boldsymbol{U}(\tau)\left[\boldsymbol{U}(\tau_0)\right]^{-1}
	\Theta(\tau-\tau_0) ,
\eea
which satisfies $\partial \boldsymbol{G}(\tau,\tau_0)/\partial\tau=\boldsymbol{A}(\tau)\boldsymbol{G}(\tau,\tau_0)+\boldsymbol{I}\delta(\tau-\tau_0)$, where $\boldsymbol{I}$ is the identity matrix. One can also note that ${\dd}\det[\boldsymbol{U}(\tau)]/{\dd\tau}=\mathrm{Tr}\left[\boldsymbol{A}(\tau)\right]\det[\boldsymbol{U}(\tau)]$ with ``$\mathrm{Tr}$'' being the trace operation. The coefficients matrix $\boldsymbol{A}$, defined in \Eq{eq:inhomopb}, is traceless and $\det[\boldsymbol{U}(\tau)]$ is thus a conserved quantity. It is therefore sufficient to find two solutions such as $\det[\boldsymbol{U}(\tau_0)]\neq0$, and this ensures the Green's matrix to be properly defined throughout the evolution. In this case, one can always normalise the two independent solutions so that $\det[\boldsymbol{U}(\tau)]=1$. But even if this is not the case, the Green's matrix is such that $\det[\boldsymbol{G}(\tau,\tau_0)]=1$, which is easily derived from the fact that the determinant of $\boldsymbol{U}$ is conserved through evolution. 
\section{Coarse graining and canonical transformations for quadratic Hamiltonians}
\label{app:canonical}
Let us consider a quadratic local Hamiltonian of the form
\bea
	\mathcal{H}=\dfrac{1}{2}\displaystyle\int \dd^3x~\boldsymbol{z}^T(\vec{x})\boldsymbol{K}(\vec{x},\tau)\boldsymbol{z}(\vec{x}),
\eea
where $\boldsymbol{z}(\vec{x})=\left[\phi(\vec{x}) , \pi(\vec{x}) \right]$ is built from the canonical variables $\phi(\vec{x})$ and $\pi(\vec{x})$. In homogeneous and isotropic space-times, one can find a foliation such that $N^i=0$ and all the metric components depend on time only. One can thus write the kernel $\boldsymbol{K}$ as
\bea
	\boldsymbol{K}=\left(\begin{array}{ccc}
		f_\Delta(\tau)\delta^{ij}\overleftarrow{\partial_i}~\overrightarrow{\partial_j}+f_m(\tau)& {}^{} & f_\times(\tau) \\
		f_\times(\tau)&  & f_\pi(\tau)
	\end{array}\right) ,
\eea
where the notation $\overleftarrow{\partial_i}~\overrightarrow{\partial_j}$ is to be understood as $F(\vec{x})\overleftarrow{\partial_i}~\overrightarrow{\partial_j}G(\vec{x})\equiv \left(\partial_i F\right)\left(\partial_j G\right)$. For example, the Hamiltonian obtained from the action~(\ref{eq:action}) corresponds to $f_\Delta=-Np^{1/2}$, $f_m=Np^{3/2} V_{,\phi\phi}$, $f_\pi=N/p^{3/2}$ and $f_\times=0$. The resulting Hamilton equations are then given by
\bea
	\dot{\boldsymbol{z}}(\vec{x})=\boldsymbol{\Omega}{\boldsymbol{\mathcal{K}}}(\vec{x},\tau)\boldsymbol{z}(\vec{x}) ,
\eea
where
\bea
\label{eq:calk}
	\boldsymbol{\mathcal{K}}=\left(\begin{array}{ccc}
		f_\Delta(\tau)\Delta+f_m(\tau) & & f_\times(\tau) \\
		f_\times(\tau) & & f_\pi(\tau)
	\end{array}\right) , 
\eea
and $\boldsymbol{\Omega}$ has been defined in \Eq{eq:def:Omega}, $\Delta\equiv\delta^{ij}\partial_i\partial_j$ being the Laplace operator. In terms of these canonical variables, the coarse graining procedure explained in \Sec{sec:Langevin} gives rise to
\bea
	\dot{\bar{\boldsymbol{z}}}=\boldsymbol{\Omega}\boldsymbol{\mathcal{K}}_0\bar{\boldsymbol{z}}+\xi,
\eea
where $\bar{\boldsymbol{z}}$ is the coarse-grained field, $\boldsymbol{\mathcal{K}}_0$ refers to the ``homogeneous'' kernel $\boldsymbol{\mathcal{K}}$ where the Laplace operator has been dropped, and
\bea
\label{eq:noise:z}
\boldsymbol{\xi}= -\displaystyle\int_{\mathbb{R}^3}\dfrac{\dd^3k}{(2\pi)^{3/2}}\dot{W}\left(\frac{k}{k_\sigma}\right)\left[a_{\vec{k}}\boldsymbol{z}_k(\tau)e^{-i\vec{k}\cdot\vec{x}}+a^\dag_{\vec{k}}\boldsymbol{z}^\star_k(\tau)e^{i\vec{k}\cdot\vec{x}}\right]
\eea
is the noise. In this expression, the mode functions are solution of the differential system $\dot{\boldsymbol{z}}_k=\boldsymbol{\Omega}\boldsymbol{\mathcal{K}}_k\boldsymbol{z}_k$,  where $\boldsymbol{\mathcal{K}}_k$ is obtained by replacing $\Delta$ by $-k^2$ in the expression~(\ref{eq:calk}) for $\boldsymbol{\mathcal{K}}$. Their normalisation according to the Klein-Gordon product reads $-i\int_{\Sigma_\tau}\dd^3x\,\boldsymbol{z}^\dag_{\vec{k}}(\tau,\vec{x})\boldsymbol{\Omega}\boldsymbol{z}_{\vec{k}^\prime}(\tau,\vec{x})=\delta^3(\vec{k}-\vec{k^\prime})$, where $\boldsymbol{z}_{\vec{k}}(\tau,\vec{x})=\boldsymbol{z}_k(\tau)e^{-i\vec{k}\cdot\vec{x}}$. The correlation matrix of the noise is given by
\bea
	\boldsymbol{\Xi}(\tau_1)=\dfrac{1}{6\pi^2}\left.\dfrac{\dd k^3_\sigma(\tau)}{\dd\tau}\right|_{\tau_1}\boldsymbol{z}_{k_\sigma(\tau_1)}\boldsymbol{z}^\dag_{k_\sigma(\tau_1)}\delta\left(\tau_1-\tau_2\right)\left<0_{\boldsymbol{z}}\right|a_{\vec{k}}a^\dag_{\vec{k}}\left|0_{\boldsymbol{z}}\right> ,
\eea
where a Heaviside distribution for the window function has been taken. In this expression, the term $\left<0_{\boldsymbol{z}}\right|a_{\vec{k}}a^\dag_{\vec{k}}\left|0_{\boldsymbol{z}}\right>$ is kept explicit to remind that this correlation matrix is obtained from quantum expectation values of the vacuum state defined by the annihilation operator $a_{\vec{k}}$ .

Let us now perform a canonical transformation and introduce $\boldsymbol{v}$ as in \Eq{eq:canonical:transf}. The dynamics of these new variables is still described by a quadratic Hamiltonian leading to
\bea
	\dot{\boldsymbol{v}}(\vec{x})=\boldsymbol{\Omega}\widetilde{\boldsymbol{\mathcal{K}}}(\vec{x},\tau)~\boldsymbol{v}(\vec{x})
\eea
where the structure of $\widetilde{\boldsymbol{\mathcal{K}}}$ is similar\footnote{For example, the canonical transformation defined by $v(\vec{x})=a\phi(\vec{x})$ and $p(\vec{x})=a'\phi(\vec{x})+\pi(\vec{x})/a$ (where $a=\sqrt{p}$ the scale factor and where primes denote the derivative with respect to conformal time) leads to $\widetilde{\mathcal{K}}$ having the structure of \Eq{eq:calk} with $f_\Delta=-1$, $f_m=-(m^2a^2-a''/a)$, $f_\pi=1$ and $f_\times=0$.} to the one of $\mathcal{K}$ and is given by
\bea
\label{eq:ktok}
	\widetilde{\boldsymbol{\mathcal{K}}}=\left(\boldsymbol{M}^{-1}\right)^\dag\boldsymbol{\mathcal{K}}\boldsymbol{M}^{-1}-\boldsymbol{\Omega}\dot{\boldsymbol{M}}\boldsymbol{M}^{-1}. 
\eea
In this expression, since $\boldsymbol{M}$ is symplectic (see \Sec{sec:canonical}), one can replace $\boldsymbol{M}^{-1}=-\boldsymbol{\Omega}\boldsymbol{M}^\mathrm{T}\boldsymbol{\Omega}$ and $\left(\boldsymbol{M}^{-1}\right)^\dag=-\boldsymbol{\Omega}\boldsymbol{M}\boldsymbol{\Omega}$.
Performing the coarse-graining procedure using this new set of canonical variables leads to the following Langevin equation
\bea
	\dot{\bar{\boldsymbol{v}}}=\boldsymbol{\Omega}\widetilde{\boldsymbol{\mathcal{K}}}_0~\bar{\boldsymbol{v}}+\tilde{\boldsymbol{\xi}} ,
\eea
where the new noise is given by
\bea
\label{eq:canonical:noisetilda}
	\tilde{\boldsymbol{\xi}}=-\displaystyle\int_{\mathbb{R}^3}\dfrac{\dd^3k}{(2\pi)^{3/2}}\dot{W}\left(\frac{k}{k_\sigma}\right)\left[b_{\vec{k}}\boldsymbol{v}_k(\tau)e^{-i\vec{k}\cdot\vec{x}}+b^\dag_{\vec{k}}\boldsymbol{v}^\star_k(\tau)e^{i\vec{k}\cdot\vec{x}}\right],
\eea
In this expression, the mode functions $\boldsymbol{v}_k$ are solutions of $\dot{\boldsymbol{v}}_k=\boldsymbol{\Omega}\widetilde{\boldsymbol{\mathcal{K}}}_k\boldsymbol{v}_k$ and are subject to the Klein-Gordon normalisation $-i\int_{\Sigma_\tau}\dd^3x\,\boldsymbol{v}^\dag_{\vec{k}}(\tau,\vec{x})\boldsymbol{\Omega}\boldsymbol{v}_{\vec{k}^\prime}(\tau,\vec{x})=\delta^3(\vec{k}-\vec{k^\prime})$. We note that the creation and annihilation operators are a priori different from the operators introduced in \Eq{eq:noise:z} for the noise $\boldsymbol{\xi}$. Using a Heaviside distribution for the window function, the correlation matrix of the noise $\tilde{\boldsymbol{\xi}}$ is given by
\bea
	\widetilde{\boldsymbol{\Xi}}(\tau_1)=\dfrac{1}{6\pi^2}\left.\dfrac{\dd k^3_\sigma(\tau)}{\dd\tau}\right|_{\tau_1}\boldsymbol{v}_{k_\sigma(\tau_1)}\boldsymbol{v}^\dag_{k_\sigma(\tau_1)}\delta\left(\tau_1-\tau_2\right)\left<0_{\boldsymbol{v}}\right|b_{\vec{k}}b^\dag_{\vec{k}}\left|0_{\boldsymbol{v}}\right> .
\eea

The coarse-graining procedure splits the dynamics according to the spatial scales, and this splitting is not affected by canonical transformations~(\ref{eq:canonical:transf}) since they are space-independent. As a consequence, one can easily check that not only $\boldsymbol{\mathcal{K}}$ and $\widetilde{\boldsymbol{\mathcal{K}}}$ are related through \Eq{eq:ktok}, but also $\boldsymbol{\mathcal{K}}_0$ and $\boldsymbol{\mathcal{K}}_k$, \ie
\begin{align}
	 \widetilde{\boldsymbol{\mathcal{K}}}_0&=\left(\boldsymbol{M}^{-1}\right)^\dag\boldsymbol{\mathcal{K}}_0\boldsymbol{M}^{-1}-\boldsymbol{\Omega}\dot{\boldsymbol{M}}\boldsymbol{M}^{-1}, \\
	 \widetilde{\boldsymbol{\mathcal{K}}}_k&=\left(\boldsymbol{M}^{-1}\right)^\dag\boldsymbol{\mathcal{K}}_k\boldsymbol{M}^{-1}-\boldsymbol{\Omega}\dot{\boldsymbol{M}}\boldsymbol{M}^{-1} .
\end{align}
This suggests that the coarse-graining procedure and the canonical transformations operation commute (note that it would not be the case if $\boldsymbol{M}$ depended on space since the Laplace operator in $\mathcal{K}$ would then act on it), which we are now going to show by considering the deterministic part of the evolution and the quantum diffusion separately. 

Considering firstly the deterministic part of the dynamics, by applying the canonical transformations $\boldsymbol{M}$ directly to $\bar{\boldsymbol{z}}$, one can show that $\dot{\bar{\boldsymbol{z}}}=\boldsymbol{\Omega}\boldsymbol{\mathcal{K}}_0\bar{\boldsymbol{z}}$ transforms into $\dot{\bar{\boldsymbol{v}}}=\boldsymbol{\Omega}\widetilde{\boldsymbol{\mathcal{K}}}_0\bar{\boldsymbol{v}}$. As a consequence, the Green's matrix $\widetilde{\boldsymbol{G}}$ solving for the latter is related to the Green's matrix $\boldsymbol{G}$ solving for the former through
\bea
\label{eq:canonical:Greenfunction}
	\widetilde{\boldsymbol{G}}(\tau,s)=\boldsymbol{M}(\tau)\boldsymbol{G}(\tau,s)\boldsymbol{M}^{-1}(s) .
\eea
This relation is obtained when the canonical transformation is performed prior to coarse graining. If one considers instead a canonical transformation  $\widetilde{\boldsymbol{\Phi}}(\tau)=\overline{\boldsymbol{M}}(\tau)\boldsymbol{\Phi}(\tau)$ on the coarse-grained variable $\boldsymbol{\Phi}$, the solution of the dynamics of the new canonical variables $\widetilde{\boldsymbol{\Phi}}$ satisfies $\widetilde{\boldsymbol{\Phi}}(\tau)=\widetilde{\boldsymbol{G}}(\tau,s)\widetilde{\boldsymbol{\Phi}}(s)$. Given that $\boldsymbol{G}$ is a symplectic matrix (see footnote~\ref{footnote:symplectic}), canonical transformations are preserved across evolution and one has $\overline{\boldsymbol{M}}(\tau)\boldsymbol{\Phi}(\tau)=\widetilde{\boldsymbol{G}}(\tau,s)\overline{\boldsymbol{M}}(s)\boldsymbol{\Phi}(s)$. This has to be compatible with $\boldsymbol{\Phi}(\tau)={\boldsymbol{G}}(\tau,s)\boldsymbol{\Phi}(s)$, which leads to the same equation~(\ref{eq:canonical:Greenfunction}).

Considering secondly the case of quantum fluctuations, one first notes that by directly applying $\boldsymbol{M}$ to $\boldsymbol{z}_k$, $\dot{\boldsymbol{z}}_k=\boldsymbol{\Omega}\boldsymbol{\mathcal{K}}_k\boldsymbol{z}_k$ transforms into the evolution equations for $\boldsymbol{v}_k$ given below \Eq{eq:canonical:noisetilda}, \ie $\dot{\boldsymbol{v}}_k=\boldsymbol{\Omega}\widetilde{\boldsymbol{\mathcal{K}}}_k\boldsymbol{v}_k$. In addition, canonical transformations preserve the Klein-Gordon product normalising the modes since $-i\int_{\Sigma_\tau}\dd^3x\boldsymbol{v}^\dag_k(\tau,\vec{x})\boldsymbol{\Omega}\boldsymbol{v}_k(\tau,\vec{x})=\delta^3(\vec{k}-\vec{k'})$ transforms into $-i\int_{\Sigma_\tau}\dd^3x\boldsymbol{z}^\dag_k(\tau,\vec{x})\boldsymbol{M}^\dag\boldsymbol{\Omega}\boldsymbol{M}\boldsymbol{z}_k(\tau,\vec{x})=\delta^3(\vec{k}-\vec{k'}) = -i\int_{\Sigma_\tau}\dd^3x\boldsymbol{z}^\dag_k(\tau,\vec{x})\boldsymbol{\Omega}\boldsymbol{z}_k(\tau,\vec{x})$, since $\boldsymbol{M}$ is a symplectic matrix. This leads to a simple transformation rule for the correlators of the noise, namely
\bea
\label{eq:canonical_transf_Xi}
	\widetilde{\boldsymbol{\Xi}}(\tau)=\boldsymbol{M}(\tau){\boldsymbol{\Xi}}(\tau)\boldsymbol{M}^\dag(\tau).
\eea
Again, this relation is obtained when the canonical transformation is performed prior to coarse graining. If one considers instead a canonical transformation $\widetilde{\boldsymbol{\Phi}}(\tau)=\overline{\boldsymbol{M}}(\tau)\boldsymbol{\Phi}(\tau)$ on the coarse-grained variable $\boldsymbol{\Phi}$, one has $\tilde{\boldsymbol{\Phi}}\tilde{\boldsymbol{\Phi}}^\dag = \boldsymbol{M}\boldsymbol{\Phi}\boldsymbol{\Phi}^\dag \boldsymbol{M}^\dag$, which leads to the same equation~(\ref{eq:canonical_transf_Xi}) if one makes use of \Eq{eq:quadmoment:quant} and \Eq{eq:canonical:Greenfunction}. 

Finally, let us determine how the diffusion matrices $\boldsymbol{D}$ and $\widetilde{\boldsymbol{D}}$ are related. The diffusion matrix is the symmetric part of the noise correlator matrix, so it is useful to write the expansion~(\ref{eq:Pauli}) as
\bea
\boldsymbol{\Xi} = \boldsymbol{D}+\frac{i}{2}\left(\Xi_{\phi,\pi}-\Xi_{\pi,\phi}\right)\boldsymbol{J}_y .
\eea
Since the Klein-Gordon product $\Xi_{\phi,\pi}-\Xi_{\pi,\phi}$ is preserved through canonical transformations, the antisymmetric part of the noise correlator matrix is preserved as well, $\widetilde{\boldsymbol{\Xi}} - \widetilde{\boldsymbol{D}} = \boldsymbol{\Xi} - \boldsymbol{D} $. Since this difference is proportional to $\boldsymbol{J}_y=-i\boldsymbol{\Omega}$, and since $\boldsymbol{M\Omega M}^\dag=\boldsymbol{\Omega}$ ($\boldsymbol{M}$ being symplectic), one has $\widetilde{\boldsymbol{\Xi}} - \widetilde{\boldsymbol{D}} =\boldsymbol{M}\left( \boldsymbol{\Xi} - \boldsymbol{D}\right)\boldsymbol{M}^\dag = \boldsymbol{M}\boldsymbol{\Xi}\boldsymbol{M}^\dag  -\boldsymbol{M}\boldsymbol{D}\boldsymbol{M}^\dag = \widetilde{\boldsymbol{\Xi}} -\boldsymbol{M}\boldsymbol{D}\boldsymbol{M}^\dag  $, where \Eq{eq:canonical_transf_Xi} has been used in the last equality. This gives rise to
\bea
\label{eq:transfor_canonique:D}
	\widetilde{\boldsymbol{D}}(\tau)=\boldsymbol{M}(\tau){\boldsymbol{D}}(\tau)\boldsymbol{M}^\dag(\tau),
\eea
which means that the diffusion matrix transforms in the same manner as the correlation matrix of the noise under canonical transformations.

In summary, the dynamics of the coarse-grained variables $\bar{\boldsymbol{z}}$ and the dynamics of the coarse-grained variables $\bar{\boldsymbol{v}}$ are both described by Langevin equations with white noise, and their respective PDF are thus solutions of a Fokker-Planck equation of the form given by \Eq{eq:fokker}. For the $\bar{\boldsymbol{z}}$ (respectively $\bar{\boldsymbol{v}}$) variable, the deterministic part is built from $\boldsymbol{A}\equiv \boldsymbol{\Omega}\boldsymbol{\mathcal{K}}_0$ (respectively $\widetilde{\boldsymbol{A}}\equiv \boldsymbol{\Omega}\widetilde{\mathcal{K}}_0$) and gives rise to the Green function $\boldsymbol{G}$ (respectively $\widetilde{\boldsymbol{G}}$), and the diffusion matrix ${\boldsymbol{D}}$ (respectively ${\widetilde{\boldsymbol{D}}}$) is the symmetric part of $\boldsymbol{\Xi}$ (respectively $\widetilde{\boldsymbol{\Xi}}$). The matrices $\boldsymbol{G}$, $\boldsymbol{\Xi}$ and ${\boldsymbol{D}}$ are related through the canonical transformations~(\ref{eq:canonical:Greenfunction}), (\ref{eq:canonical_transf_Xi}) and (\ref{eq:transfor_canonique:D}), and the coarse-graining procedure commutes with canonical transformations as shown in \Sec{sec:canonical}.
\section{Interaction picture versus orthonormalisation}
\label{app:orthonormalisation}
If one introduces the phase-space basis  $\boldsymbol{e_\phi}=(1, 0)^\mathrm{T}$ and $\boldsymbol{e_\pi}=(0, 1)^\mathrm{T}$, so that any phase-space vector $\boldsymbol{\Phi}$ can be decomposed as $\boldsymbol{\Phi} = \phi \boldsymbol{e_\phi}+\pi \boldsymbol{e_\pi}$, the SR and NSR solutions defined in \Eqs{eq:def:PhiSR} and~(\ref{eq:def:PhiNSR}) are not orthonormal with respect to the standard inner dot product since $\boldsymbol{\Phi}_\sr^\dag  \boldsymbol{\Phi}_\nsr=\phi_\sr \phi_\nsr+\pi_\sr \pi_\nsr \neq0$, $\boldsymbol{\Phi}_\sr^\dag  \boldsymbol{\Phi}_\sr \neq1$ and $\boldsymbol{\Phi}_\nsr^\dag  \boldsymbol{\Phi}_\nsr\neq1$. However, one can define orthonormality in a dynamical sense through the Wronskian, where two vectors $\boldsymbol{u}$ and $\boldsymbol{v}$ form a symplectic basis if $\boldsymbol{u}^\dag\boldsymbol{\Omega}\boldsymbol{v}=1$ and $\boldsymbol{u}^\dag\boldsymbol{\Omega}\boldsymbol{u}=\boldsymbol{v}^\dag\boldsymbol{\Omega}\boldsymbol{v}=0$. In this sense, one can check that both $(\boldsymbol{e_\phi},\boldsymbol{e_\pi})$ and $(\boldsymbol{\Phi}_\sr,\boldsymbol{\Phi}_\nsr)$ are symplectic bases.\footnote{
The fact that $(\boldsymbol{\Phi}_\sr,\boldsymbol{\Phi}_\nsr)$ is a symplectic basis is related to the formulas $c_\sr(\boldsymbol{\Phi}_\sr)=c_\nsr(\boldsymbol{\Phi}_\nsr)=1$ and $c_\sr(\boldsymbol{\Phi}_\nsr)=c_\nsr(\boldsymbol{\Phi}_\sr)=0$, where $c_\sr$ and $c_\nsr$ are defined in \Eqs{eq:def:zsr} and~(\ref{eq:def:znsr}).}
In the interaction picture $\boldsymbol{z}=\boldsymbol{U}^{-1}\boldsymbol{\Phi}$, this discussion is simplified since inner dot orthonormality and dynamical orthonormality are equivalent.

This is why, even though the covariance matrix~(\ref{eq:Sigma:z}) in the interaction picture $\boldsymbol{\Sigma}_{\boldsymbol{z}}$ is diagonal, it should not be interpreted as the diagonalisation of the covariance matrix~(\ref{eq:Sigma:light:lightlimit}) in standard phase-space coordinates $\boldsymbol{\Sigma}_{\boldsymbol{\Phi}}$. Indeed, $\boldsymbol{\Sigma}_{\boldsymbol{\Phi}}$ is symmetric and it can thus be diagonalised as $\boldsymbol{\Sigma}_{\boldsymbol{\Phi}}=\boldsymbol{P}\boldsymbol{\Lambda}_{\boldsymbol{\Phi}}\boldsymbol{P}^{-1}$, with $\boldsymbol{\Lambda}_{\boldsymbol{\Phi}}$ a diagonal matrix and $\boldsymbol{P}$ an orthogonal matrix, \ie such that $\boldsymbol{P}^{-1}=\boldsymbol{P}^\dag$. In \Secs{sec:canonical} and~\ref{sec:InteractionPicture}, it was shown that $\boldsymbol{\Sigma}_{\boldsymbol{\Phi}}=\boldsymbol{U}\boldsymbol{\Sigma}_{\boldsymbol{z}}\boldsymbol{U}^\dag$ with $\boldsymbol{\Sigma}_{\boldsymbol{z}}$ a diagonal matrix, but here, the matrix $\boldsymbol{U}$ is not an orthogonal matrix.

In practice, $\boldsymbol{\Sigma}_{\boldsymbol{\Phi}}$ can be diagonalised as follows. Since it is fully anticorrelated, its two eigenvalues are given by $\lambda_1= {\Sigma}_{\phi,\phi}+ {\Sigma}_{\pi,\pi}$ and $\lambda_2=0$. The first eigenvalue defines a direction where diffusion takes place, while the second eigenvalue stands for a direction without diffusion. Their respective orthonormalised eigenvectors read
\bea
	\boldsymbol{e}_{1}=\frac{1}{\sqrt{ {\Sigma}_{\phi,\phi}+ {\Sigma}_{\pi,\pi}}}\left(\begin{array}{c}
		\sqrt {\Sigma_{\phi,\phi}} \\
		-\sqrt{ {\Sigma}_{\pi,\pi}}
	\end{array}\right)
	&~~~\mathrm{and}~~~ &
	\boldsymbol{e}_{2}=\frac{1}{\sqrt{ {\Sigma}_{\phi,\phi}+ {\Sigma}_{\pi,\pi}}}\left(\begin{array}{c}
		\sqrt{{\Sigma}_{\phi,\phi}} \\
		\sqrt{{\Sigma}_{\pi,\pi}}
	\end{array}\right).
\eea
From the expression of $\Sigma_{\phi,\phi}$ and $\Sigma_{\pi,\pi}$ given in \Eq{eq:Sigma:light:lightlimit}, one can check that $\boldsymbol{e}_1\propto\boldsymbol{\Phi}_\sr$, which is consistent with the fact that diffusion occurs along the classical attractor direction only. The direction $\boldsymbol{e}_2$ corresponds to the vanishing eigenvalue and is orthogonal to $\boldsymbol{e}_1$ in the sense of the inner-dot product. As such, it is not aligned with the anti-attractor branch, $\boldsymbol{\Phi}_\nsr$. This may seem slightly counter-intuitive but as we shall now show, the absence of stochasticity along $\boldsymbol{e}_2$ does not imply the presence of stochasticity along $\boldsymbol{\Phi}_\nsr$. At any given time indeed, a solution of the stochastic dynamics can be written in the basis $(\boldsymbol{e}_1,\boldsymbol{e}_2)$ as $\boldsymbol{\Phi}=\widehat{y}_1\boldsymbol{e}_1+\widehat{y}_2\boldsymbol{e}_2$, where $\widehat{y}_1$ and $\widehat{y}_2$ are time-dependent, stochastic (as denoted by the hats) variables. Since diffusion does not occur along $\boldsymbol{e}_2$, $\widehat{y}_2$ is in fact not stochastic and one can write $\boldsymbol{\Phi}=\widehat{y}_1\boldsymbol{e}_1+{y}_2\boldsymbol{e}_2$. Similarly, as explained around \Eq{eq:decom:stoch:interaction}, the solution can be decomposed in the attractor and anti-attractor basis according to $\boldsymbol{\Phi}=\widehat{z}_\sr\boldsymbol{\Phi}_\sr+{z}_\nsr\boldsymbol{\Phi}_\nsr$, with $\widehat{z}_\sr$ a stochastic variable, and ${z}_\nsr$ a deterministic one (since there is no diffusive processes in the NSR direction). The variables $\widehat{z}_\sr$ and $z_\nsr$ are related to $\widehat{y}_1$ and $y_2$ through the Wronskian transformation~(\ref{eq:def:zsr}) and~(\ref{eq:def:znsr})
\bea
	&&\widehat{z}_\sr=\boldsymbol{\Phi}^\dag\boldsymbol{\Omega}\boldsymbol{\Phi}_\nsr=\left(\boldsymbol{e}_1^\dag\boldsymbol{\Omega}\boldsymbol{\Phi}_\nsr\right) \widehat{y}_1+\left(\boldsymbol{e}_2\boldsymbol{\Omega}\boldsymbol{\Phi}_\nsr\right) {y}_2, \\
	&&{z}_\nsr=-\boldsymbol{\Phi}^\dag\boldsymbol{\Omega}\boldsymbol{\Phi}_\sr=-\left(\boldsymbol{e}_2^\dag\boldsymbol{\Omega}\boldsymbol{\Phi}_\sr\right) {y}_2,
\eea
where in the second line, we have used that $\boldsymbol{e}_1^\dag \boldsymbol{\Omega}\boldsymbol{\Phi}_\sr=0$ as $\boldsymbol{e}_1$ is aligned with $\boldsymbol{\Phi}_\sr$ and $(\boldsymbol{\Phi}_\sr,\boldsymbol{\Phi}_\nsr)$ forms a symplectic basis. From these expressions, it is clear that stochasticity occurs along the attractor solution only from both viewpoints, \ie from the eigendecomposition of $\boldsymbol{\Sigma}_{\boldsymbol{\Phi}}$ viewpoint and from the interaction picture viewpoint, even though they are not identical.
\section{Coupling the inflaton to metric fluctuations}
\label{App:InflatonMetricFluctuations}
In \Sec{sec:HamiltonianStochastic}, the Langevin equations for a test scalar field were derived, where one can assume that the field is sufficiently decoupled from the metric and other fields perturbations that the latter can be ignored. However, for the inflaton field discussed in \Sec{sec:Inflaton}, this is not the case anymore and one must keep track of the scalar perturbations of the metric~(\ref{eq:line_element}), which in the spatially-flat gauge where the scalar field fluctuation coincides with the Mukhanov-Sasaki variable reads 
\bea
 \label{eq:dspert}
\dd s^2=-N^2(\tau)\left[1+2A(\vec{x},\tau)\right]\dd \tau^2+p(\tau)\delta_{ij}\left[\partial^iB(\vec{x},\tau)\right]\dd x^i+p(\tau)\delta_{ij}\dd x^i\dd x^j ,
\eea
where $A\ll1$ and $B\ll1$ are perturbations. In this case, the Hamilton equations~(\ref{eq:dotphigen}) and (\ref{eq:dotpgen}) are given by
\bea
\label{eq:eomphiinf}
\dot\phi&=&\frac{N}{p^{3/2}}\pi_\phi\left(1+A\right)+\left(\partial^iB\right)\left(\partial_i\phi\right)  \\ 
\label{eq:eompiinf}
\dot\pi_\phi&=&-Np^{3/2}V_{,\phi}(\phi)\left(1+A\right)+Np^{1/2}\left[\Delta\phi+\delta^{ij}\partial_i\left(A\partial_j\phi\right)\right]
+\partial_i\left[\left(\partial^iB\right)\pi_\phi\right] .
\eea
In the above, all terms containing $A$ or $B$ are at least of order one in fluctuations. Following \Sec{sec:Langevin}, one then introduces a time-dependent comoving scale $k_\sigma(\tau)$ separating the coarse-grained scalar field, $(\bar\phi,\bar\pi)$, from its quantum fluctuations, $(\phi_Q,\pi_Q)$, written as a weighted sum over modes $k>k_\sigma(\tau)$, see \Eqs{eq:phiq} and~(\ref{eq:piq}).

In principle, one should also coarse grain the functions describing metric fluctuations, i.e. $A=\bar{A}+A_Q$ with $\smash{A_Q\propto\int \dd^3kW(k/k_\sigma)A_k(\tau)e^{-i\vec{k}\cdot\vec{x}}}$ (and similarly for $B$). However, $\bar{A}$ and $\bar{B}$ can be set to zero for the following reasons. First, they are considered as background quantities which are reabsorbed into the lapse function and the shift vector of the homogeneous and isotropic sector. Since the lapse and the shift are pure gauge choices, imposing the form~(\ref{eq:line_element:flat}) at large scales leads to $\bar{A}=\bar{B}=0$. Second, the form of the line element~(\ref{eq:dspert}) implicitly assumes that the gravitational degrees of freedom have been split into an homogeneous and isotropic sector, $N$ and $p$, and an inhomogeneous and perturbative sector, $A$ and $B$. There is thus a priori no reason to introduce additional homogeneous degrees of freedom in the gravitational sector [note that this split is not yet performed in the scalar field sector in \Eqs{eq:eomphiinf} and~(\ref{eq:eompiinf})]. Third and finally, metric perturbations are sourced by field fluctuations. Since field fluctuations are restricted to wavenumbers $k>k_\sigma$, there is no source to metric fluctuations at larger scales. Let us stress however that this does not mean that there is no stochasticity in the gravitational dynamics. In the spatially-flat gauge, all fluctuations (hence all sources of stochasticity) are contained in the scalar field, and the stochasticity of the coarse-grained scalar field directly sources the one of the gravitational dynamics through the Friedmann equation~(\ref{eq:Friedman}). 

Linearising \Eqs{eq:eomphiinf} and~(\ref{eq:eompiinf}) with $\bar{A}=\bar{B}=0$, and neglecting gradients of the coarse-grained part of the field, one obtains
\bea
\label{eq:phicginf}
\dot{\bar\phi}&=&\frac{N}{p^{3/2}}\bar{\pi}-\dot\phi_Q+\frac{N}{p^{3/2}}{\pi}_Q+\frac{N}{p^{3/2}}A_Q\bar{\pi} ,  \\
\label{eq:picginf} 
\dot{\bar\pi}&=&-Np^{3/2}V_{,\phi}(\bar\phi)-\dot{\pi}_Q+Np^{1/2}\Delta\phi_Q
-Np^{3/2}V_{,\phi\phi}(\bar\phi)\phi_Q
\nonumber \\ & &
-Np^{3/2}A_QV_{,\phi}(\bar\phi) 
+\bar\pi\partial_i\partial^iB_Q ,
\eea
which generalise \Eqs{eq:Hamilton:split:phi} and~(\ref{eq:Hamilton:split:pi}). These expressions can be simplified by introducing the equation of motion of the quantum modes, which can be derived from the constraint equations. These are obtained by introducing the quantum fluctuations $\phi_Q$ and $\pi_Q$ in the Hamiltonian and expanding it up to second order in $\phi_Q$ and $\pi_Q$. From the scalar field sector of the Hamiltonian, one obtains
\bea
C_\phi\propto\displaystyle\int \dd^3x \left[N\left(\mathcal{C}^{(2)}+A\mathcal{C}^{(1)}\right)+\left(\partial^i B\right)\mathcal{C}^{(1)}_i\right]
\eea
with
%\bea
%	\mathcal{C}^{(2)}&=&\frac{\delta\pi^2}{2p^{3/2}}+\frac{p^{1/2}}{2}\delta^{ij}\left(\partial_i\delta\phi\right)\left(\partial_j\delta\phi\right)+\frac{p^{3/2}}{2}V_{,\phi\phi}\left(\bar\phi\right)\delta\phi, \\
%	\mathcal{C}^{(1)}&=&\frac{1}{p^{3/2}}\bar{\pi}\delta\pi+p^{1/2}\delta^{ij}\left(\partial_i\bar\phi\right)\left(\partial_j\delta\phi\right)+p^{3/2}V_{,\phi}\left(\bar\phi\right)\delta\phi, \\
%	\mathcal{C}^{(1)}_i&=&\delta\pi\partial_i\bar\phi+\bar\pi\partial_i\delta\phi .
%\eea
\bea
	\mathcal{C}^{(2)}&=&\frac{\delta\pi^2}{2p^{3/2}}+\frac{p^{1/2}}{2}\delta^{ij}\left(\partial_i\delta\phi\right)\left(\partial_j\delta\phi\right)+\frac{p^{3/2}}{2}V_{,\phi\phi}\left(\bar\phi\right)\delta\phi^2, \\
	\mathcal{C}^{(1)}&=&\frac{1}{p^{3/2}}\bar{\pi}\delta\pi+p^{3/2}V_{,\phi}\left(\bar\phi\right)\delta\phi, \\
	\mathcal{C}^{(1)}_i&=&\bar\pi\partial_i\delta\phi .
\eea
In the above, $\bar\phi$ and $\bar\pi$ stands for homogeneous degrees of freedom and their spatial gradients have thus been dropped out. The Hamilton equation~(\ref{eq:Hamilton:constraint}) then gives rise to
\bea
\label{eq:eom:fromConstraint:phik}
	\dot\phi_k&=&\frac{N}{p^{3/2}}{\pi}_k+\frac{N}{p^{3/2}}A_k\bar{\pi} , \\
	\dot\pi_k&=&-Np^{3/2}V_{,\phi\phi}(\bar\phi)\phi_k-Np^{1/2}k^2\phi_k-Np^{3/2}A_kV_{,\phi}(\bar\phi)-\bar\pi k^2 B_k ,
\label{eq:eom:fromConstraint:pik}
\eea
where we have introduced the Fourier modes of the quantum fluctuations $\phi_k$ and $\pi_k$.  Introducing the weighted sum decomposition over Fourier modes of $\phi_Q$, $\pi_Q$, $A_Q$ and $B_Q$ into \Eqs{eq:phicginf} and~(\ref{eq:picginf}), \Eqs{eq:eom:fromConstraint:phik} and~(\ref{eq:eom:fromConstraint:pik}) give rise to \Eqs{eq:eombarphi} and~(\ref{eq:eombarpi}) with the quantum noises given by \Eqs{eq:noisephi} and~(\ref{eq:noisepi}), \ie the coarse-grained fields follow the same Langevin equation as in the uncoupled case. 

The only difference lies in the equation of motion of the quantum modes. If one derives the constraint equation $\mathcal{C}_G$ and $\mathcal{C}_i^G$ from the gravitational sector of the Hamiltonian indeed, one can relate $A_k$ and $B_k$ to $\phi_k$, and one obtains~\cite{Langlois:1994ec, Gordon:2000hv, Lyth:2001nq, Levasseur:2013tja},
\bea
 \label{eq:eommetric}
\ddot\phi_k+3H\dot{\phi}_k+\frac{k^2}{a^2}\phi_k+\left[V_{,\phi\phi}(\bar\phi)-\frac{1}{a^3\Mp^2}\frac{\dd}{\dd t}\left(\frac{a^3}{H}\dot{\bar\phi}^2\right)\right]\phi_k=0 ,
\eea
which is written in cosmic time and coincides with Mukhanov-Sasaki equation in spatially-flat gauge. Compared to the equation of motion followed by quantum modes \emph{un}coupled to the metric fluctuations (see \Sec{sec:Langevin}), which reads $\ddot\phi_k+3H\dot{\phi}_k+\frac{k^2}{a^2}\phi_k+V_{,\phi\phi}(\bar\phi)\phi_k=0$, one can check that the only effect of the coupling to the metric fluctuations is to shift the effective mass according to \Eq{eq:meff:inflaton}.

\bibliographystyle{JHEP}
\bibliography{StochaAttractor}
\end{document}

%% file: newcommands.tex
\newcommand{\ie}{{i.e.~}}

\newcommand{\eg}{\textsl{e.g.~}}

\newcommand{\order}[1]{\mathcal{O}\!\left(#1\right)}

\DeclareMathOperator{\erf}{erf}

\newcommand{\dd}{\mathrm{d}}
\newcommand{\ee}{e}

\newcommand{\sss}[1]{{\scriptscriptstyle{#1}}}

\newcommand{\uPl}{\mathrm{Pl}}
\newcommand{\uin}{\mathrm{in}}

\newcommand{\ueff}{\mathrm{eff}}

\newcommand{\uS}{\mathrm{S}}

\newcommand{\usssS}{\sss{\uS}}

\newcommand{\usssPl}{\sss{\uPl}}

\newcommand{\nS}{n_\usssS}

\newcommand{\alphaS}{\alpha_\usssS}

\newcommand{\calP}{\mathcal{P}}

\newcommand{\Mp}{M_\usssPl}

\newcommand{\sr}{{{}_\mathrm{SR}}}
\newcommand{\nsr}{{{}_\mathrm{NSR}}}

\newcommand{\efolds}{$e$-folds}
\newcommand{\efold}{$e$-fold}

\newcommand{\beq}{\begin{equation}}
\newcommand{\eeq}{\end{equation}}
\newcommand{\bea}{\begin{eqnarray}}
\newcommand{\eea}{\end{eqnarray}}
\newlength{\wsingfig}
\newlength{\wdblefig}
\newlength{\wquadfig}
\newlength{\wtriplefig}
\newcommand{\Eq}[1]{Eq.~(\ref{#1})}
\newcommand{\Eqs}[1]{Eqs.~(\ref{#1})}
\newcommand{\Fig}[1]{Fig.~{\ref{#1}}}

\newcommand{\Refc}[1]{Ref.~{\cite{#1}}}

\newcommand{\Sec}[1]{Sec.~\ref{#1}}
\newcommand{\Secs}[1]{Secs.~\ref{#1}}
\newcommand{\App}[1]{Appendix~\ref{#1}}